\documentclass[10pt,aps,prb,twocolumn,floatfix,amsmath,amssymb,superscriptaddress]{revtex4-2}
\usepackage{graphicx}
\usepackage{color}
\usepackage[sort&compress]{natbib}
\usepackage{hyperref}
\usepackage{newtxtext}
\usepackage{newtxmath}
\usepackage{bm}
\usepackage[normalem]{ulem}

\hypersetup{
    colorlinks=true,
    linkcolor=blue,
    citecolor=blue,
    filecolor=magenta,      
    urlcolor=blue}

\begin{document}
\title{Quantum magnetism of iron-based ladders: blocks, spirals, and spin flux}

\author{Maksymilian {\'S}roda}
\affiliation{Department of Theoretical Physics, Faculty of Fundamental Problems of Technology, Wroc\l aw University of Science and Technology, 50-370 Wroc\l aw, Poland}

\author{Elbio Dagotto}
\affiliation{Department of Physics and Astronomy, University of Tennessee, Knoxville, Tennessee 37996, USA}
\affiliation{Materials Science and Technology Division, Oak Ridge National Laboratory, Oak Ridge, Tennessee 37831, USA}

\author{Jacek Herbrych}
\affiliation{Department of Theoretical Physics, Faculty of Fundamental Problems of Technology, Wroc\l aw University of Science and Technology, 50-370 Wroc\l aw, Poland}

\date{\today}

\begin{abstract}
    Motivated by increasing experimental evidence of exotic magnetism in low-dimensional iron-based materials, we present a comprehensive theoretical analysis of magnetic states of the multiorbital Hubbard ladder in the orbital-selective Mott phase (OSMP). The model we used is relevant for iron-based compounds of the AFe$_2$X$_3$ family (where A${}={}$Cs, Rb, Ba, K are alkali metals and X${}={}$S, Se are chalcogenides). To~reduce computational effort, and obtain almost exact numerical results in the ladder geometry, we utilize a low-energy description of the Hubbard model in the OSMP---the generalized Kondo-Heisenberg Hamiltonian. Our main result is the doping vs interaction magnetic phase diagram. We reproduce the experimental findings on the AFe$_2$X$_3$ materials, especially the exotic block magnetism of BaFe$_2$Se$_3$ (antiferromagnetically coupled $2\times 2$ ferromagnetic islands of the $\uparrow\uparrow\downarrow\downarrow$ form). As in recent  studies of the chain geometry, we also unveil block magnetism beyond the $2 \times 2$ pattern (with block sizes varying as a function of the electron doping) and also an interaction-induced frustrated block-spiral state (a~spiral order of rigidly rotating ferromagnetic islands). Moreover, we predict new phases beyond the one-dimensional system: a~robust regime of phase separation close to half-filling, incommensurate antiferromagnetism for weak interaction, and a~quantum spin-flux phase of staggered plaquette spin currents at intermediate doping. Finally, exploiting the bonding/antibonding band occupations, we provide an intuitive physical picture giving insight into the structure of the phase diagram.
\end{abstract}
\maketitle

\section{Introduction}
The lattice geometry plays an important role in quantum many-body systems, especially if the problem is reduced to one (1D) or two (2D) dimensions. For example, the crossover from 1D chains to 2D planes of the spin-$1/2$ Heisenberg model shows that the system behaves fundamentally different if one considers an even or odd number of coupled chains \cite{dogottoSurprises1996}. Consequently, in the last three decades, there was a tremendous effort devoted to understanding the physics of quantum ladders, i.e., the systems at the crossroads between 1D and 2D worlds. Furthermore, while unbiased analytical or numerical calculations are often not possible in 2D, the 1D chains and quasi-1D ladders---due to the possibility of an accurate treatment via quasi-exact numerical simulations---have become a playground for condensed-matter physicists to test various theoretical scenarios.

The interest in the physics of the ladder systems goes beyond a toy model investigation. There are many materials whose lattice structure is of the ladder geometry. The unique  interplay between theory and experiment in low-dimensional systems allows for an in-depth understanding of various complex phenomena. For example, within cuprates the so-called {\it telephone-number} two-leg ladder compounds (La,Sr,Ca)$_{14}$Cu$_{24}$O$_{41}$ were extensively studied motivated by the presence of pressure-induced high-critical-temperature superconductivity \cite{UeharaSCladder1996,MotoyamaTelephon2002,MaekawaLadder1996}. Interestingly, the latter was numerically predicted \cite{DagottoLadder1992,RiceLadder1994}, showing the power of theoretical investigation of low-dimensional systems. Another series of cuprate materials, Sr$_x$Cu$_y$O$_z$, allows to study the differences between various lattice geometries, from chains (Sr$_2$CuO$_3$), through two- (SrCu$_2$O$_3$) and three-leg (Sr$_2$Cu$_2$O$_5$) ladders, to 2D planes (SrCuO$_2$). The first of these compounds is one of the best realizations of a 1D system, with the intrachain exchange integral being four orders of magnitude larger than the interchain one \cite{MotoyamaChain1996}. Despite that the hole doping necessary for superconductivity is hard to achieve, the next two exhibit a large contribution of magnons \cite{HlubekBalisitc2010,hessThermal2019} to the thermal conductivity, in agreement \cite{SteinigewegIntegrable2016,KarraschLadders2015} with the thermal current being a constant of motion of 1D quantum spin systems.

Iron-based ladders are far less explored, especially from the theoretical perspective. Recent experimental investigations have shown that the two-leg ladder materials from the so-called 123 family, i.e., AFe$_2$X$_3$ where A are alkali metals and X chalcogenides, become superconducting under pressure \cite{TakahashiSC2015,YingSC2017,WuBaFeX2018}, as in the Cu-based equivalents. Canonical $(\pi,0)$ order, i.e., staggered antiferromagnetic (AFM) ordering along the legs and ferromagnetic (FM) along the rungs, was identified in (Ba,K)Fe$_2$S$_3$ \cite{WangBaFe2S32017} and (Cs,Rb)Fe$_2$Se$_3$ \cite{HawaiBaFe2Se32015,WangRbFe2Se32016,ChiCsFe2Se32016}. More recent measurements on CsFe$_2$Se$_3$ \cite{MuraseCsFe2Se32020} suggest that an incommensurate order emerges in this compound instead of the AFM.

Interestingly, the magnetic orders identified in AFe$_2$X$_3$ ladders display more variety than those found in cuprates. In a series of experiments on the BaFe$_2$Se$_3$ compound, an exotic block-magnetic order was reported, with the spins forming FM islands which are then AFM coupled $\uparrow\uparrow\downarrow\downarrow\uparrow\uparrow\downarrow\downarrow$ (on the ladder this takes the form of $2\times2$ FM blocks which are AFM coupled). This unusual magnetic state was identified with the help of inelastic neutron scattering (INS) \cite{MourigalINS2015}, X-ray \cite{WuDiff2019}, muon \cite{WuDiff2019}, and neutron powder diffraction \cite{CaronNPD2011,NambuNPD2012,WuDiff2019}. Remarkably, yet again, the block magnetic order was predicted by numerical calculations \cite{RinconBlock2014}. It can be argued that the spin arrangement of the BaFe$_2$Se$_3$ ladder is a low-dimensional equivalent of the magnetic state found in 2D iron-based systems, i.e., the double stripe or staggered dimer ordering found in FeSe \cite{GlasbrennerFeSe2015}, the $\sqrt{5}\times\sqrt{5}$ iron vacancies ordering in (K,Rb)$_{0.8}$Fe$_{1.6}$Se$_2$ \cite{WangReFeSe2015,BaoKFeSe2011,YouKFeSe2011,YuKFeSe2011}, or the block-like magnetism found in the family of 245 iron-based superconductors (K,Rb)$_2$Fe$_4$Se$_5$ \cite{GuKFe2Se22010,YeKFe2Se22011}. Also, similar block magnetism was predicted in a 1D iron selenide compound Na$_2$FeSe$_2$ \cite{PhysRevB.102.035149}.

The theoretical analysis of iron-based systems is a challenging task due to their multiorbital nature. While the single-orbital Hubbard model is often sufficient to describe the Cu-based parent compounds (with the charge density close to one electron per site), the Fe-based materials need (in principle) five orbitals filled with six electrons, i.e., they have to be described by the multiorbital Hubbard model with intra- and interorbital interactions treated on an equal footing. As a consequence, exact-diagonalization many-body calculations are challenging to achieve due to the exponential growth of the Hilbert space of the Hamiltonian---$\mathrm{dim}(\mathcal{H})=4^{\Gamma L}$ with $\Gamma$ the number of active orbitals and $L$ the number of sites in the system. In order to study the physics of such systems, we must rely on some form of approximations. For example, the full five-orbital Hubbard model was investigated via the mean-field Hartree-Fock analysis \cite{LuoHF2011,LuoDFT2013,LuoMHF2014,LuoHF2014}, revealing a complex filling-Hund/Hubbard interaction magnetic phase diagram with many competing phases. Many of such phases were also confirmed by density functional theory \cite{YinDFT2012,LuoDFT2013,DongDFT2014,ZhangDFT2017,ZhangDFT2018,ZhangDFT2019,PhysRevB.101.144417}. Moreover, the electronic properties of the multiorbital Hubbard model were extensively investigated via the dynamical mean-field theory \cite{MediciDMFT2009,GeorgesDMFT2013,IsidoriDMFT2019}, especially the orbital-selective Mott phase (OSMP), namely the possibility of the localization of a fraction of the conduction electrons (on one or more orbitals) \cite{YiOSMP2017,CaronOSMP2012,YamauchiSC2015,CracoOSMP2020}. The latter phase is regarded as a promising candidate for the parent state of iron-based superconductors \cite{yiObservation2013,hardyEvidence2013,yiObservation2015,zhuBand2010} and, most relevantly, of the 123-family ladders \cite{CaronOSMP2012,ChiCsFe2Se32016,TakuboBaFeSe2017,materneBandwidthControlledInsulatormetal2019,patelFingerprints2019,CracoOSMP2020}.

Despite their value, the aforementioned theoretical approaches are limited in that they cannot properly incorporate the effects of quantum fluctuations over long distances. This issue is particularly important for low-dimensional systems, where it is well known that quantum fluctuations must be treated accurately, thereby requiring full many-body calculations. In order to facilitate the latter, an alternative route has to be taken, such as decreasing the number of considered orbitals. For instance, it was shown \cite{Daghofer302010} that the three-orbital Hubbard model can accurately describe the physics of iron-based materials. In the latter, the $e_g$ orbitals ($d_{x^2-y^2}$ and $d_{z^2}$) are far enough from the Fermi level to be neglected, rendering only the $t_{2g}$ orbitals ($d_{xy}$, $d_{xz}$, $d_{yz}$) active. Importantly, the three-orbital model was used to predict \cite{RinconBlock2014,RinconBlock22014} and confirm \cite{HerbrychBSq2018} the INS result on BaFe$_2$Se$_3$ \cite{MourigalINS2015} related to the block magnetic order, while also tracing its origin to the presence of the OSMP. Nevertheless, it should be noted that accurate many-body simulations of three-orbital systems are mostly restricted to the chain geometry, with ladders being largely out of reach. Recently, it was realized that one may further reduce the number of degrees of freedom captured within minimal models by noting that the $d_{yz}$ and $d_{xz}$ orbitals are close to being degenerate in tetragonal systems of the 123 family \cite{TakuboBaFeSe2017,CracoOSMP2020}. As a result, two-orbital models were designed, which, within the OSMP, were found to correctly reproduce both the static \cite{HerbrychBlock2019,herbrychBlockSpiral2020} and dynamic \cite{HerbrychBlock2020} properties of the three-orbital chains.

In this work, we use such a minimal approach to go beyond the chain geometry---bridging the gap between theory and experiment---and perform a comprehensive analysis of the magnetic phases within the OSMP of a multiorbital Hubbard ladder. To facilitate numerically exact many-body calculations, we focus on a two-orbital model, which we further map onto an accurate low-energy description, the generalized Kondo-Heisenberg Hamiltonian. We unveil a rich variety of exotic magnetic phases [see Figs.~\ref{fig1}(d)--\ref{fig1}(i) for sketches], summarized in our \emph{central result}: the doping vs interaction magnetic phase diagram. In particular, we reproduce the experimental finding on BaFe$_2$Se$_3$, i.e., the $\genfrac{}{}{0pt}{}{\uparrow\uparrow\downarrow\downarrow}{\uparrow\uparrow\downarrow\downarrow}$ block phase, and predict the possibility of experimentally realizing larger blocks, e.g., $\genfrac{}{}{0pt}{}{\uparrow\uparrow\uparrow\downarrow\downarrow\downarrow}{\uparrow\uparrow\uparrow\downarrow\downarrow\downarrow}$, by doping this or related compounds. Furthermore, we report a highly unusual block-spiral state (with the blocks rigidly rotating throughout the system), discovered first using a chain geometry \cite{herbrychBlockSpiral2020}, and predict this spiral to be stable also on the experimentally relevant ladder. Surprisingly, we reveal that the ladder supports also phases absent in its chain counterpart. For example, in the vicinity of half-filling, we discover incommensurate AFM order as well as a robust regime of phase separation (relating our effort to previous works on cuprates and manganites, respectively). Last but not least, we report the emergence of a novel quantum spin flux state at intermediate doping, with staggered spin currents circulating around $2 \times 2$ plaquettes. Our magnetic phase survey is supplemented by an intuitive physical picture involving the bonding/antibonding ladder bands, which explains the observed magnetic tendencies and generalizes our conclusions to models with more orbitals.

The paper is organized as follows. In Sec.~\ref{model}, we introduce the two-orbital Hubbard ladder relevant for the AFe$_2$X$_3$ compounds and simplify this formalism into the generalized Kondo-Heisenberg Hamiltonian. Then, we describe the computational method used to solve the many-body problem. In Sec.~\ref{results}, we present the main result: the doping vs interaction magnetic phase diagram. Each reported phase is discussed in detail within three subsections \ref{block}, \ref{phasesep}, \ref{flux}, addressing the cases of large, low, and intermediate doping, respectively. Finally, in Sec. \ref{conclusions}, we give a summary and draw conclusions. In the Appendix, we discuss additional details regarding the computational accuracy.

\section{Model and method}\label{model}

\begin{figure}[!t]
    \includegraphics{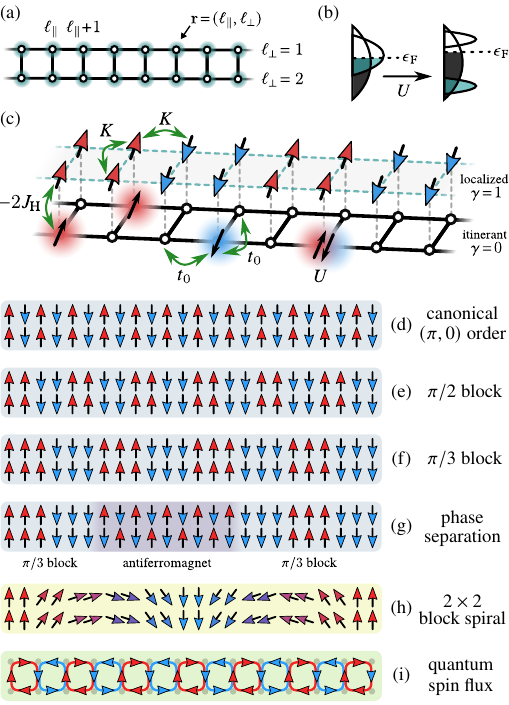}
    \caption{Schematic representation of: (a) the ladder geometry, (b) density of states in the orbital-selective Mott phase, (c) the generalized Kondo-Heisenberg model, (d)-(i) the unveiled exotic magnetic orders.}
    \label{fig1}
\end{figure}

We aim to establish the magnetic properties, within the OSMP, of a two-orbital Hubbard model on a two-leg ladder. In the generic SU(2)-symmetric form, the Hamiltonian reads
\begin{equation}
    \begin{aligned}
        H_\text{H} &= \!\sum_{\gamma\langle \mathbf{r}\,\mathbf{m} \rangle \sigma} \!\! t_{\gamma}\, c^\dagger_{\gamma \mathbf{r} \sigma} c^{\phantom{\dagger}}_{\gamma \mathbf{m} \sigma} + \sum_{\gamma\mathbf{r}} \Delta_\gamma n_{\gamma\mathbf{r}} \\
        &+ U \sum_{\gamma\mathbf{r}} n_{\gamma \mathbf{r} \uparrow} n_{\gamma \mathbf{r} \downarrow} + (U - 5 J_\text{H}/2)\sum_\mathbf{r} n_{0\mathbf{r}} n_{1\mathbf{r}} \\
        &- 2J_\text{H} \sum_{\mathbf{r}} \mathbf{S}_{0\mathbf{r}} \cdot \mathbf{S}_{1\mathbf{r}} + J_\text{H} \sum_\mathbf{r} (P^\dagger_{0\mathbf{r}} P_{1\mathbf{r}} + \text{H.c.}).
    \end{aligned}
\label{hubbard_ham}
\end{equation}
Here, $c^\dagger_{\gamma \mathbf{r} \sigma}$ ($c_{\gamma \mathbf{r} \sigma}$) creates (annihilates) an electron with spin $\sigma=\{\uparrow,\downarrow\}$ at orbital $\gamma=\{0,1\}$ of site $\mathbf{r}=(\ell_\parallel,\ell_\perp)$, where $\ell_\parallel=\{1,\ldots,L_\parallel\}$ and $\ell_\perp=\{1,2\}$ enumerate the sites in directions parallel and perpendicular to the legs, respectively. The total number of sites is $L= 2 \times L_\parallel$. The $\langle \mathbf{r}\,\mathbf{m}\rangle$ brackets indicate summation over nearest-neighbor (NN) sites in the ladder geometry [see the sketch in Fig.~\ref{fig1}(a)]. The first two terms of the Hamiltonian constitute the kinetic energy part, with $t_{\gamma}$ denoting the hopping matrix elements, $\Delta_\gamma$ denoting the crystal-field splitting, and $n_{\gamma\mathbf{r}} = \sum_\sigma n_{\gamma\mathbf{r}\sigma} = \sum_\sigma c^\dagger_{\gamma \mathbf{r} \sigma} c^{\phantom{\dagger}}_{\gamma \mathbf{r} \sigma}$ being the total electron density at $(\gamma,\mathbf{r})$. The remaining four terms form the interaction part: the first is the standard intraorbital Hubbard repulsion $U > 0$, the second is the interorbital repulsion $U-5 J_\text{H}/2$, the third is the ferromagnetic Hund exchange $J_\text{H}$ (which couples spins $\mathbf{S}_{\gamma\mathbf{r}}$ on different orbitals $\gamma$), and the fourth is the interorbital pair hopping ($P_{\gamma\mathbf{r}}=c_{\gamma \mathbf{r} \uparrow}c_{\gamma \mathbf{r} \downarrow}$). Note that all the interaction terms follow directly from the matrix elements of the fundamental $1/r$ Coulomb interaction \cite{kanamoriElectronCorrelationFerromagnetism1963,olesAntiferromagnetismCorrelationElectrons1983,dagottoColossalMagnetoresistantMaterials2001}.

We adopt the following set of hopping amplitudes (eV units): $t_{0} = 0.5$ and $t_{1} = 0.15$. The interorbital hybridization is neglected, as it was shown that a realistically small hybridization leaves the overall physics unaffected \cite{HerbrychBlock2019}. Moreover, here, we choose equal hoppings along the legs and the rungs, i.e., $t^\parallel_{\gamma}=t^\perp_{\gamma}=t_{\gamma}$, although density-functional theory \cite{LuoDFT2013} and spin-wave theory \cite{MourigalINS2015} analyses suggest that this is only an approximation for real materials. Nevertheless, below we shall argue that such a choice does not compromise the generality of our results. The crystal-field splittings are assumed as (eV units): $\Delta_0=0$, $\Delta_1=1.6$, where the latter is taken large enough to energetically separate the two orbitals. The rationale behind the above values of $t_{\gamma}$ and $\Delta_\gamma$ is to reproduce the essential feature of the band structure of the 123-family materials---the coexistence of nondegenerate wide and narrow orbitals \cite{MediciDMFT2009,RinconBlock2014,LuoDFT2013,YiOSMP2017,patelFingerprints2019,Daghofer302010}---and, in this sense, these values are generic. The total kinetic-energy bandwidth $W=3.55$ eV is here the energy unit throughout the paper. To further reduce the number of free parameters in the model, we also fix the Hund exchange to $J_\text{H}=U/4$, a value widely accepted to be experimentally relevant for iron-based materials \cite{hauleCoherence2009,yinKinetic2011,ferberLDA2012,luoNeutron2010,daiMagnetism2012}. Finally, we note that our choice of model parameters ensures that for a wide region of electronic fillings, 2 < $n_\text{H}$ < 3, and Hubbard interaction strengths, $U \gtrsim W$, the ground state is in the OSMP \cite{HerbrychBlock2019,HerbrychBlock2020}, where the narrow ($\gamma=1$) orbital undergoes Mott localization while the wide ($\gamma=0$) orbital remains itinerant [see Fig.~\ref{fig1}(b)]. In the following, we shall vary both $n_\text{H}$ and $U$ to produce a rich variety of magnetic phases [see Figs.~\ref{fig1}(d)--\ref{fig1}(i)].

\begin{figure}[!t]
    \includegraphics{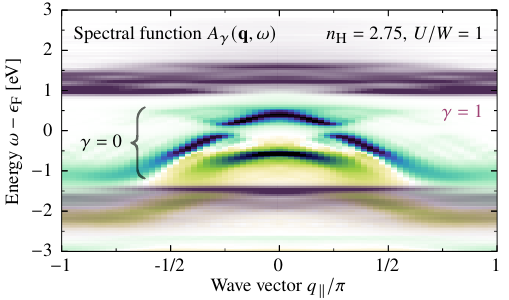}
    \caption{Single-particle spectral function $A_\gamma(\mathbf{q},\omega)$ of the two-orbital Hubbard model \eqref{hubbard_ham} in the vicinity of the Fermi level $\epsilon_\text{F}$. The itinerant ($\gamma=0$) orbital is presented as  blue-green color, whereas the localized ($\gamma=1$) as dark purple.  Both $q_\perp=0,\,\pi$ components of each orbital are displayed. The frequency resolution was chosen to be $\Delta\omega=0.02$~eV with the broadening $\eta=2\Delta\omega$. The results were obtained for a ladder of $L=72$ sites, filling $n_\text{H}=2.75$, and interaction $U/W=1$.}
    \label{fig1-5}
\end{figure}

The selective localization implies that the charge degrees of freedom in the narrow orbital are essentially frozen out and should no longer play a role in low-energy processes. Let us consider the single-particle spectral function of the two-orbital Hubbard ladder \eqref{hubbard_ham} defined as
\begin{equation}
    \begin{aligned}
        A_\gamma(\mathbf{q},\omega) = &-\frac{1}{\pi\sqrt{L}}\sum_{\mathbf{r}}\mathrm{e}^{i\mathbf{q}(\mathbf{r}-\mathbf{c})} \operatorname{Im}\,\langle c^{\dagger}_{\gamma\mathbf{r}}\frac{1}{\omega^{+}+(H-\epsilon_{\text{GS}})} c^{\phantom{\dagger}}_{\gamma\mathbf{c}}\rangle\\
        &-\frac{1}{\pi\sqrt{L}}\sum_{\mathbf{r}}\mathrm{e}^{i\mathbf{q}(\mathbf{r}-\mathbf{c})}\operatorname{Im}\,\langle c^{\phantom{\dagger}}_{\gamma \mathbf{r}}\frac{1}{\omega^{+}-(H-\epsilon_{\text{GS}})} c^{\dagger}_{\gamma\mathbf{c}}\rangle\,,
    \end{aligned}
\label{specakw}
\end{equation}
where $c^{\dagger}_{\gamma\mathbf{r}} = \sum_\sigma c^{\dagger}_{\gamma\mathbf{r}\sigma}$, $\mathbf{c}=(L_\parallel/2,1)$, $\mathbf{q}=(q_\parallel,q_\perp)$, $\omega^{+}=\omega+i\eta$, and $\langle \cdot\rangle \equiv \langle\text{GS}|\cdot|\text{GS}\rangle$ with $|\text{GS}\rangle$ being the ground-state vector with energy $\epsilon_{\text{GS}}$. In Fig.~\ref{fig1-5}, as an example we show the case of $A_\gamma(\mathbf{q},\omega)$ for $\gamma=0,1$, $n_\text{H}=2.75$ and $U/W=1$. Several conclusions can be drawn from the results. (i) The interaction $U$ heavily modifies the dispersion relation which in the $U\to 0$ limit would have a simple cosine form (see also the discussion in the next section). (ii) As expected in the OSMP regime, already at \mbox{$U/W\simeq 1$} the electrons at the  \mbox{$\gamma=1$} orbital localize, which can be deduced from the flat (momentum-independent) spectral function $A_1(\mathbf{q},\omega)$. The two modes of the latter, separated by a wide charge gap, resemble the lower and upper Hubbard subbands of a Mott insulator. Similar properties of the OSMP were also identified in 1D systems \cite{HerbrychBlock2020,herbrychBlockSpiral2020}.

The charge gap of the localized orbital is robust enough to result in vanishing charge fluctuations already for $U\simeq W$. Correspondingly, the double occupancy of the latter orbital can be traced out via the standard Schrieffer-Wolff transformation \cite{schriefferRelation1966}, leaving only the spin degrees of freedom active. Such a procedure results \cite{HerbrychBlock2019} in the generalized Kondo-Heisenberg (gKH) Hamiltonian 
\begin{equation}
    \begin{aligned}
        H_\text{K} &= t_{0} \!\!\sum_{\langle \mathbf{r}\,\mathbf{m} \rangle \sigma} \!\! c^\dagger_{0 \mathbf{r} \sigma} c_{0 \mathbf{m} \sigma} + U \sum_{\mathbf{r}} n_{0 \mathbf{r} \uparrow} n_{0 \mathbf{r} \downarrow} \\
        &+ K \!\sum_{\langle \mathbf{r}\,\mathbf{m} \rangle} \mathbf{S}_{1\mathbf{r}}\cdot\mathbf{S}_{1\mathbf{m}} - 2J_\text{H} \sum_{\mathbf{r}} \mathbf{S}_{0\mathbf{r}} \cdot \mathbf{S}_{1\mathbf{r}},
    \end{aligned}
\label{kondo_ham}
\end{equation}
where $K=4t_{1}^2/U$ is a Heisenberg-like exchange between the localized $\gamma=1$ spins. See Fig.~\ref{fig1}(c) for a graphical representation of the Hamiltonian. The electronic filling $n_\text{K}$ of the gKH model is obtained from the original filling $n_\text{H}$ by subtracting the occupancy of the $\gamma=1$ orbital, i.e., $n_\text{K}=n_\text{H}-1$. However, it is noteworthy that due to the particle-hole symmetry of \eqref{kondo_ham}, one could equivalently choose $n_\text{K}=3-n_\text{H}$. The effective description \eqref{kondo_ham} reveals that the OSMP naturally favors exotic magnetism due to the coexistence of itinerant electrons and well-developed local magnetic moments. In particular, within the OSMP, the Hund exchange induces a remarkably complex correlated behavior where the total on-site magnetic moment $\langle \mathbf{S}_\mathbf{r}^2 \rangle$ ($\mathbf{S}_\mathbf{r} = \sum_\gamma \mathbf{S}_{\gamma\mathbf{r}}$) is completely maximized \cite{HerbrychBlock2019,RinconBlock2014,RinconBlock22014} as in an insulator, despite the system remaining metallic.

Previous comparisons between models \eqref{hubbard_ham} and \eqref{kondo_ham} concluded that the latter not only qualitatively but also quantitatively reproduces both the static \cite{HerbrychBlock2019, herbrychBlockSpiral2020} and dynamic \cite{HerbrychBlock2020} properties of the former (provided that the system is in the OSMP). Accordingly, hereafter, in our numerical calculations, we exclusively use the model \eqref{kondo_ham}, utilizing its considerably smaller Hilbert space to perform extensive simulations with feasible computational cost. The many-body ground state (temperature $T=0$) of the system is studied via the density matrix renormalization group (DMRG) method within the single-center site approach \cite{whiteDensity2005,alvarezDensity2009}. Throughout the DMRG procedure, we typically keep up to $M=1200$ states and perform 20-30 full sweeps in the finite-size algorithm, maintaining the truncation error below 10${}^{-6}$. We focus on the subspace with zero total spin projection and a fixed particle number $N$, which sets the filling $n_\text{K}=N/L$. Open boundary conditions are assumed. All results are obtained using the \textsc{DMRG++} computer program developed at Oak Ridge National Laboratory \cite{alvarezDensity2009,GonzaloDMRGWeb}, and the input scripts are available online \cite{CorrWro}. Additional details regarding the computational accuracy are discussed in the Appendix.

The key observables used to identify the magnetic orders are the total spin-spin correlation function $\langle \mathbf{S}_\mathbf{r} \cdot \mathbf{S}_{\mathbf{m}} \rangle$ (viewed as a function of distance or on NN bonds) and its Fourier transform---the spin structure factor, defined as $S(\mathbf{q}) = \langle \mathbf{S}_\mathbf{q} \cdot \mathbf{S}_\mathbf{-q} \rangle$, where $\mathbf{S}_\mathbf{q} = (1/\sqrt{L}) \sum_\mathbf{r} \exp(i \mathbf{q} \mathbf{r})\, \mathbf{S}_\mathbf{r}$. These two quantities, albeit very useful, cannot distinguish between all possible magnetic orders. Therefore, we supplement our analysis with the chirality correlation function, which is explicitly defined in the next section. Note that the exotic magnetic patterns we observe are not static (as would be the case for a combination of domain walls or a spin density wave), but exhibit significant quantum fluctuations. For example, in the case of the block pattern $\uparrow\uparrow\downarrow\downarrow$ (whose extended version we report), exact diagonalization studies confirm \cite{HerbrychBSq2018} that the many-body ground state is in at least 50\% of the singlet form $|\!\uparrow\uparrow\downarrow\downarrow\rangle - |\!\downarrow\downarrow\uparrow\uparrow\rangle$. Accordingly, the individual magnetic blocks should be considered as regions with strong FM correlations, as opposed to domains with finite magnetization.

\section{Results}\label{results}

\begin{figure}[!t]
    \includegraphics{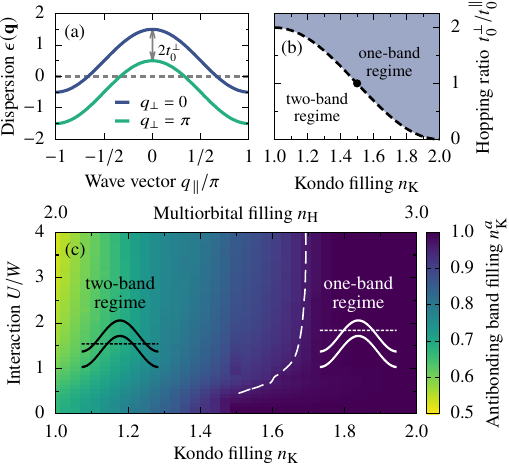}
    \caption{Properties of the band structure. (a) Noninteracting ($U=0$) band structure of \eqref{kondo_ham} for $t_{0}^\perp=t_{0}^\parallel$. The dispersion concerns only the itinerant $\gamma=0$ orbital, as the $\gamma=1$ orbital is completely localized. The dashed line marks the Fermi level $\epsilon_\text{F}$ at half-filling, $n_\text{K}=1$. (b) $n_\text{K}$-$t_{0}^\perp$ phase diagram of the noninteracting ladder. The dashed line marks the point where the Fermi level touches the tip of the antibonding ($q_\perp=\pi$) band, while the dot marks the phase boundary between the one- and two-band regimes for $t_{0}^\perp=t_{0}^\parallel$. (c) DMRG results for the antibonding band filling $n_\text{K}^a$ vs the interaction $U$ and the total filling $n_\text{K}$ at fixed $t_{0}^\perp=t_{0}^\parallel$. The plot is composed of $37 \times 40$ data points obtained for the generalized Kondo-Heisenberg ladder of $L=72$ sites. The sketches show the one-band ($n_\text{K}^a \simeq 1$) and two-band (both bands fractionally occupied) regimes. The dashed line is a contour at $n_\text{K}^a \simeq 1$.}
    \label{fig2}
\end{figure}

To better understand the general structure of the magnetic phase diagram reported below, it is instructive to recall the properties of a noninteracting ($U=0$) ladder system. In such a case, the Hamiltonian \eqref{kondo_ham} retains only the kinetic-energy term which can be easily diagonalized by first introducing the bonding and antibonding (symmetric and antisymmetric, respectively) combinations of the rung states and then Fourier transforming along the leg direction (here, we assume periodic boundary conditions). In the general case of unequal leg and rung hoppings, one obtains the dispersion relation \mbox{$\epsilon(\mathbf{q}) = 2t_{0}^\parallel \cos(q_\parallel) + t_{0}^\perp \cos(q_\perp)$}, consisting of two bands (bonding $q_\perp=0$ and antibonding $q_\perp=\pi$) separated by the energy $2t_{0}^\perp$ [see Fig.~\ref{fig2}(a)]. The respective fillings are denoted by $n^b_\text{K}$, $n^a_\text{K}$. Since these bands can host at most $2L_\parallel$ electrons, the maximum possible filling is $\max\{n^b_\text{K}\}=\max\{n^a_\text{K}\}=1$, and thus $n^b_\text{K}$, $n^a_\text{K} \in [0,1]$, consistent with the relation $n_\text{K} = n^b_\text{K}+n^a_\text{K}$. Note that this dispersion corresponds only to the $\gamma=0$ orbital, as the $\gamma=1$ orbital is completely localized within the model \eqref{kondo_ham}. To avoid any confusion, hereafter, we reserve the term \emph{band} to denote the latter bonding/antibonding bands and not the underlying orbitals.

Owing to the band structure, the behavior of the ladder system is nontrivial even in the noninteracting case. The Fermi level $\epsilon_\text{F}$ can cross either one or both bands [see the sketches in Fig.~\ref{fig2}(c)], giving rise to qualitatively different Fermi ``surfaces'' with two ($\pm k_\text{F}^b$) or four ($\pm k_\text{F}^b$, $\pm k_\text{F}^a$) Fermi points, respectively. Whenever convenient, we will use the abbreviated notation $\mathbf{k}_\text{F}=\{k^b_\text{F}\,,k^a_\text{F}\}$ to collectively refer to both wave vectors. To tune between the one- and two-band regimes, one may use both the filling $n_\text{K}$ (to shift the Fermi level) and/or the rung hopping $t_{0}^\perp$ (to vary the band separation). This is summarized in the $n_\text{K}$-$t_{0}^\perp$ phase diagram \cite{noackGround1996}, Fig.~\ref{fig2}(b), where one clearly recognizes the complementary role of the two parameters in deciding whether one or two bands are fractionally occupied. 

The picture of one- and two-band regimes can be extended also beyond the $U=0$ case. Here, although finite $U$ inevitably renormalizes the band fillings, the latter retain their physical meaning and can be calculated in a straightforward manner. In Fig.~\ref{fig2}(c), we show the antibonding band filling $n^a_\text{K}$ (with $n^a_\text{K} = n_\text{K} - n^b_\text{K}$) as a function of the total filling $n_\text{K}$ and the interaction strength $U$ at fixed $t^\perp_{0}=t^\parallel_{0}$. We observe that there exists a robust region where $n^a_\text{K} \simeq 1$, i.e., the antibonding band is completely filled. This condition provides a convenient definition of the one-band regime for a general $U \neq 0$. Starting from $U=0$, the boundary between the one- and two-band regimes occurs at three-quarter filling $n_\text{K}=1.5$ [in agreement with Fig.~\ref{fig2}(b)] and shifts rightwards with increasing $U$. Notably, although the width of the one-band regime decreases with the interaction, it does not vanish up to the largest considered $U/W=4$.

The significance of the above discussion lies in the fact that the block-magnetism of the gKH chain was shown to be controlled by the Fermi wave vector of the itinerant orbital \cite{HerbrychBlock2019}, even though $U \simeq W$. In the following, we shall see that this insight remains meaningful also on the ladder, where the distinct Fermi surfaces of the one- and two-band regimes will necessarily come into play. In particular, we are already in a position to argue that the main influence of varying $t^\perp_{0}$ on the magnetic properties of our system should come precisely from tuning between the one- and two-band regimes. Consider first the one-band regime. Here, as long as $t^\perp_{0}$ is varied in a range that will not push the system into the two-band regime, there is only one Fermi wave vector $k_\text{F}^b$ available, whose position does not depend on $t^\perp_{0}$. This suggests that the magnetism, which depends on the Fermi wave vector, shall remain mostly unaffected. At a few points within the one-band regime, we checked (not shown) that this indeed holds true, at least for a modest perturbation of the $t^\perp_{0}/t^\parallel_{0}$ ratio (since one expects that for $t^\perp_{0} \gg t^\parallel_{0}$ the system will behave as uncoupled rung dimers and our argument will eventually break). In the two-band regime, the situation becomes more complicated, as here varying $t^\perp_{0}$ at a fixed filling $n_\text{K}$ does change the values of~$\mathbf{k}_\text{F}$. Nevertheless, judging by Fig.~\ref{fig2}(b), it is reasonable to assume that the latter change of Fermi wave vectors---and the resulting impact on magnetism---will be complementary to that achievable by tuning $n_\text{K}$ at fixed $t^\perp_{0}$. In that sense, although in the following we fix $t^\perp_{0}=t^\parallel_{0}$, we do not expect a qualitatively different magnetic phase diagram for other $t^\perp_{0}/t^\parallel_{0}$ ratios, but rather a similar diagram with renormalized magnetic phase boundaries, originating in the renormalization of the one- and two-band regimes. Finally, let us stress that it is the one-band regime where we reproduce the experimentally reported block magnetism, and clearly this is the regime which is least affected by the perturbation of $t^\perp_{0}$.

\begin{figure}[!t]
    \includegraphics{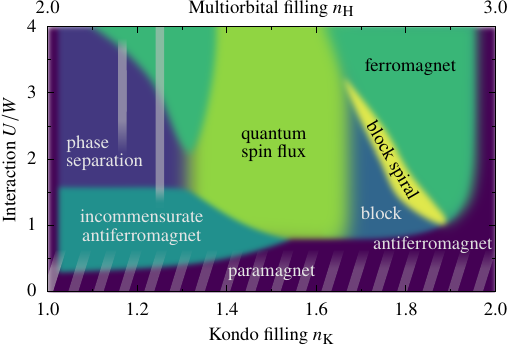}
    \caption{Schematic $n_\text{K}$-$U$ magnetic phase diagram of the generalized Kondo-Heisenberg ladder of $L=72$ sites. The vertical lines within the phase-separation regime mark special fillings $n_\text{K}=1.17,1.25$, where perfect block order is recovered (see the discussion in Sec.~\ref{phasesep}). The phase diagram was inferred from extensive DMRG calculations performed at $37 \times 40$ data points uniformly distributed over the range of the plot. The phase boundaries are necessarily approximate as they cannot be exactly determined from finite-size calculations.}
    \label{fig3}
\end{figure}

The \emph{central result} of this work, shown in Fig.~\ref{fig3}, is the $n_\text{K}$-$U$ magnetic phase diagram of the gKH model on a ladder geometry, relevant for the low-dimensional 123-family iron-based superconductors within the OSMP. The details on each reported magnetic phase are provided in the following three sections: Sec.~\ref{block} discusses the one-band regime, i.e., $n_\text{K} \gtrsim 1.6$, whereas Secs.~\ref{phasesep} and \ref{flux} discuss the two-band regime at low ($n_\text{K} \lesssim 1.3$) and intermediate fillings ($n_\text{K} \sim 1.5$), respectively. Here, let us first focus on a few generic phases. (i) For all considered electronic fillings $n_{\mathrm{K}}$, the system is a paramagnet at small values of the interaction strength $U/W \lesssim 0.5$. Note that in this regime our effective description \eqref{kondo_ham} only approximately depicts the behavior of the full multiorbital Hubbard model. This stems from the fact that the latter is not yet within the OSMP and the magnetic moments $\langle \mathbf{S}_\mathbf{r}^2 \rangle$ are not yet fully developed. (ii) In the other extreme, when $U \gg W$, the system is a ferromagnet for all noninteger fillings, $1<n_{\mathrm{K}}<2$, due to the dominance of the double-exchange mechanism (favored by a large value of the Hund exchange $J_\mathrm{H}$). This phase is also present at moderate interaction strength $U \simeq W$ in the proximity of $n_{\mathrm{K}}=2$. (iii) For special values of the electron density $n_{\mathrm{K}}=1$ and $n_{\mathrm{K}}=2$, i.e., at half-filling and in the case of a band insulator, respectively, the usual $(\pi,\pi)$ (staggered along the legs and rungs) AFM order develops.

\subsection{Block and block-spiral magnetism (one-band regime)}\label{block}

As follows from Fig.~\ref{fig2}(c), the spatially isotropic ($t^\perp_{0}=t^\parallel_{0}$) system is in the one-band regime for \mbox{$n_{\mathrm{K}}\gtrsim 1.6$}. In the rest of this subsection, we shall argue that this is the most experimentally relevant region hosting the block magnetic phase found in BaFe$_2$Se$_3$. It is important to note that the filling $n_{\mathrm{K}}$ of the OSMP effective model \eqref{kondo_ham} does not correspond to the electronic density of the real materials or to the full five-orbital Hubbard model. However, as we will argue below, it is the position of the Fermi wave vectors $\mathbf{k}_\text{F}$ that is crucial for the magnetism within the block phase (as well as strongly influences the behavior of the other phases, even in the two-band regime). This remains true also beyond the noninteracting $U\to 0$ limit where the $\mathbf{k}_\text{F}$ become, in principle, a nontrivial function of the electronic density. As a consequence, we believe that our findings are generic provided that the multiorbital system is in the OSMP and has similar values of the Fermi points $\mathbf{k}_\text{F}$, irrespective of the precise densities necessary to attain them or the number of active orbitals.

\begin{figure}[!b]
    \includegraphics{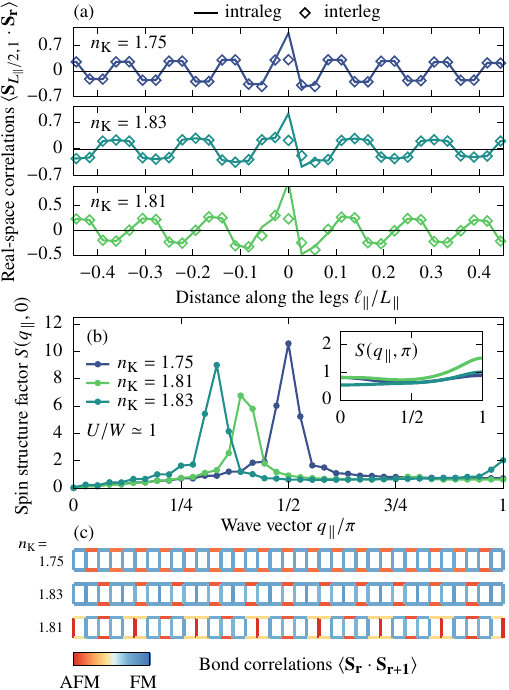}
    \caption{Block magnetic order. (a) Spin-spin correlations $\langle \mathbf{S}_{L_{\parallel}/2,1} \cdot \mathbf{S}_{\mathbf{r}} \rangle$ as a function of distance with $\mathbf{r}=(\ell_\parallel,1)$ (intraleg) or $\mathbf{r}=(\ell_\parallel,2)$ (interleg). Top to bottom: $\pi/2$ block ($n_\text{K}=1.75$, $U/W=1$), $\pi/3$ block ($n_\text{K}=1.83$, $U/W=1.1$), mixed block ($n_\text{K}=1.81$, $U/W=1$). (b) Spin structure factor $S(\mathbf{q})$ being the Fourier transform of the correlations shown in (a). (c) Bond correlations $\langle \mathbf{S}_{\mathbf{r}} \cdot \mathbf{S}_{\mathbf{r+1}} \rangle$ corresponding to (a) and (b). $\mathbf{1}$ connects the nearest-neighbor sites on the ladder. All results were obtained for a generalized Kondo-Heisenberg ladder of $L=72$ sites.}
    \label{fig4}
\end{figure}

Previous efforts \cite{RinconBlock2014,HerbrychBSq2018,HerbrychBlock2019} showed that the magnetic order of the $\uparrow\uparrow\downarrow\downarrow$ form can be stabilized on the chain lattice in the $U\sim {\cal O}(W)$ region of the phase diagram. In such a case, the block magnetism follows twice the Fermi wave vector of the noninteracting limit $2k_\text{F}= \pi (2 - n_\text{K})$ (recall that we work above half-filling, $n_\text{K} > 1$). On the ladder geometry, in the one-band regime, the latter is given by \mbox{$2k_\text{F}^b=\pi(2 - 2 n^b_\text{K})$}, where the additional factor of 2 arises due to $\max\{n^b_\text{K}\} = 1$. Our results shown in Fig.~\ref{fig4} support that the latter predicts also the block magnetic order of the two-leg ladder for $U/W \simeq 1 \to 2$. Namely, in Fig.~\ref{fig4}(a), we present the spin-spin correlation function \mbox{$\langle \mathbf{S}_{L_{\parallel}/2,1} \cdot \mathbf{S}_{\mathbf{r}} \rangle$} between the sites on the same or different legs (lines and symbols, respectively). Clearly, both correlation functions lie on top of each other and exhibit a characteristic step-like pattern. This indicates that the spins are arranged in, e.g., AFM-coupled $2\times2$ FM blocks for $n_\text{K}=1.75$ [sketched in Fig.~\ref{fig1}(e)], i.e., the so-called $\pi/2$-block pattern $\genfrac{}{}{0pt}{}{\uparrow\uparrow\downarrow\downarrow}{\uparrow\uparrow\downarrow\downarrow}$. This unusual magnetic order can be also identified via the spin structure factor $S(\mathbf{q})$, see Fig.~\ref{fig4}(b). Here, the bonding component $S(q_\parallel,0)$ (along the legs) has a well-pronounced maximum at $(2k^b_\text{F},0)$ for all considered fillings $n_\text{K}$. On the other hand, the antibonding component $S(q_\parallel,\pi)$ has only a weak momentum dependence. We again stress that the observed alternating FM block patterns are inferred from the spin-spin correlations and not the static magnetization $\langle {S}^z_{\mathbf{r}} \rangle$.

On our finite lattice of $L=72$ sites, the largest perfect (i.e., AFM-coupled FM) block that we have stabilized is of $3\times2$ size, the so-called $\pi/3$ block $\genfrac{}{}{0pt}{}{\uparrow\uparrow\uparrow\downarrow\downarrow\downarrow}{\uparrow\uparrow\uparrow\downarrow\downarrow\downarrow}$ [sketched in Fig.~\ref{fig1}(f)], present at $n_\text{K}=1.83$. However, it was shown \cite{HerbrychBlock2019} that a small spin anisotropy can be used to stabilize even larger FM islands, possibly accessible here using larger $L$. The block nature of the correlations can be also seen in the NN bond correlations $\langle \mathbf{S}_{\mathbf{r}} \cdot \mathbf{S}_{\mathbf{r+1}} \rangle$ shown in Fig.~\ref{fig4}(c) (here, $\mathbf{1}$ connects the NN sites on the ladder geometry). Interestingly, the block magnetic order is not restricted to perfect blocks of the same size, as those above, but can also involve complicated patterns of differently sized blocks. This is the case of $n_\text{K}=1.81$, for which the real-space and bond correlation functions indicate a repeating motif of a large $5\times2$ magnetic unit cell, within which smaller blocks can be nevertheless still discerned. The unusual periodicity of the latter pattern leads to a strong maximum in $S(q_\parallel,0)$ at $q_\parallel/\pi\simeq0.4$, in agreement with the $2k_\text{F}^b$ prediciton. This finding is consistent with the analysis of the block magnetic orders in 1D systems \cite{HerbrychBlock2019,HerbrychBlock2020}. There, the perfect block order can be found for $2k_\text{F}=\pi/m$ with $m\in \mathbb{Z}$. On the other hand, for $2k_\text{F}\ne\pi/m$, complex block patterns are stabilized. It is important to note that these are not phase-separated regions but true complicated spin arrangements in an overall spatially isotropic system (see also the discussion in the next section).

\begin{figure}[!t]
    \includegraphics{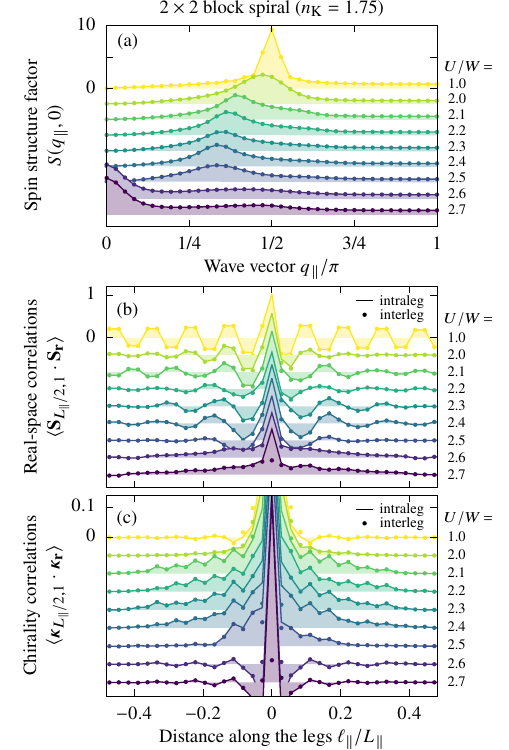}
    \caption{Block-spiral magnetic order. (a) Interaction $U$ evolution of the spin structure factor $S(q_\parallel,0)$ for the $2\times 2$ block spiral ($n_\text{K}=1.75$). (b) Spin-spin correlations $\langle \mathbf{S}_{L_{\parallel}/2,1} \cdot \mathbf{S}_{\mathbf{r}} \rangle$ as a function of distance corresponding to (a). (c) Chirality correlations \mbox{$\langle \bm{\kappa}_{L_{\parallel}/2,1}\cdot\bm{\kappa}_{\mathbf{r}}\rangle$} as a function of distance corresponding to (a). In (a) and (b), both the intraleg [$\mathbf{r}=(\ell_\parallel,1)$] and interleg [$\mathbf{r}=(\ell_\parallel,2)$] components are presented (as lines and symbols, respectively). All results were obtained for a generalized Kondo-Heisenberg ladder of $L=72$ sites.}
    \label{fig5}
\end{figure}

As already discussed, for $U\gg W$, the system orders ferromagnetically due to the double-exchange mechanism dominating for large Hund exchange $J_\text{H}$. Furthermore, it was recently shown in 1D systems \cite{herbrychBlockSpiral2020} that between the block and the FM phases another order exists: the frustrated block-spiral state. In Fig.~\ref{fig5}(a), we show the evolution of the spin structure factor $S(\mathbf{q})$ starting from the block phase at $U/W=1$, for the important special case of $n_\text{K}=1.75$, i.e., the $\pi/2$ block of $\genfrac{}{}{0pt}{}{\uparrow\uparrow\downarrow\downarrow}{\uparrow\uparrow\downarrow\downarrow}$ form. Upon increasing $U$, the maximum of $S(q_\parallel,0)$ smoothly interpolates from $q_\parallel=\pi/2$ towards $q_\parallel\to0$, taking incommensurate values in between. However, the real-space correlations [shown in Fig.~\ref{fig5}(b)] reveal that this order differs significantly from a ``simple'' block pattern. To gain an understanding of this behavior, let us focus on the chirality correlation function along the legs, i.e., $\langle \bm{\kappa}_\mathbf{r}\cdot\bm{\kappa}_{\mathbf{m}}\rangle$ with
\begin{equation}
\bm{\kappa}_\mathbf{r}=\mathbf{S}_\mathbf{r}\times\mathbf{S}_\mathbf{r+1}\,.
\end{equation}
Here, $\mathbf{1}$ connects NN sites along the legs (in Sec.~\ref{flux} we shall generalize it to involve also NN sites along the rungs). Since the above operator is proportional to the angle $\phi$ between NN spins, $\bm{\kappa}_\mathbf{r}\propto\sin(\phi)$, it is evident that if NN bond correlations are of FM ($\phi=0$)  or AFM ($\phi=\pi$) kind, the operator vanishes. On the other hand, if NN spins are rotated by $0<\phi<\pi$, the $\langle \bm{\kappa}_\mathbf{r}\cdot\bm{\kappa}_{\mathbf{m}}\rangle$ correlation can detect the spiral order. In Fig.~\ref{fig5}(c), we present the spatially resolved $\langle \bm{\kappa}_{L_{\parallel}/2,1}\cdot\bm{\kappa}_{\mathbf{r}}\rangle$ vs the interaction strength $U$ for $n_\text{K}=1.75$. As expected, in the block phase ($U/W=1$) the chirality correlation function vanishes. In this phase, spin correlations are alternating between FM and AFM [see Figs.~\ref{fig4}(c) and \ref{fig5}(b)]. Surprisingly, at $U/W\simeq2$, $\langle \bm{\kappa}_\mathbf{r}\cdot\bm{\kappa}_{\mathbf{m}}\rangle$ takes finite values even at distances as long as $L_{\parallel}/2$, and exhibits a zig-zag-like decaying pattern. Such behavior continues until $U/W\simeq2.6$, when the system enters FM phase with $\phi=0$.

\begin{figure}[!t]
    \includegraphics{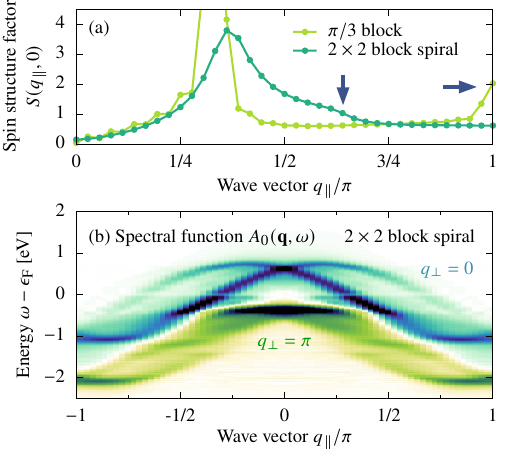}
    \caption{Special features of the block-spiral phase. (a) Secondary Fourier mode (arrows) being a fingerprint of the block and block-spiral order. Shown are the representative cases of $\pi/3$ block ($n_\text{K}=1.83$, $U/W=1.1$) and $2 \times 2$ block spiral ($n_\text{K}=1.75$, $U/W=2.2$). (b) Spectral function of the itinerant orbital $A_0(\mathbf{q},\omega)$ corresponding to the $2 \times 2$ block spiral of (a). The frequency resolution was chosen to be $\Delta\omega=0.02$~eV with the broadening $\eta=2\Delta\omega$. All results were obtained for a generalized Kondo-Heisenberg ladder of $L=72$ sites.}
    \label{fig6}
\end{figure}

The above behavior was identified \cite{herbrychBlockSpiral2020} as the block-spiral phase: upon increasing the strength of the Hubbard interaction $U$, the FM islands of the block phase start to rigidly rotate with respect to each other. The zig-zag (small-large) pattern reflects the fact that within the $\langle \bm{\kappa}_\mathbf{r}\cdot\bm{\kappa}_{\mathbf{m}}\rangle$ correlation the $\bm{\kappa}_\mathbf{r}$ operators act between the blocks (large value) or within the block (small value). Here, we establish that such a phase is also stable on the ladder geometry, where---in the particular case of the $\pi/2$ block at $n_\text{K}=1.75$---all four spins of the block start to rotate $\genfrac{}{}{0pt}{}{\uparrow\uparrow\nearrow\nearrow\rightarrow\rightarrow\searrow\searrow\downarrow\downarrow}{\uparrow\uparrow\nearrow\nearrow\rightarrow\rightarrow\searrow\searrow\downarrow\downarrow}$ [see also the sketch in Fig.~\ref{fig1}(h)]. As marked on the phase diagram, Fig.~\ref{fig3}, the block spiral is not restricted to $n_\text{K}=1.75$, but develops also for the other block patterns at different $n_\text{K}$. The unique block modulation of our spiral is expected to be visible also in the Fourier decomposition, i.e., in the spin structure factor $S(\mathbf{q})$. In Fig.~\ref{fig6}(a), we present a zoom-in plot of $S(\mathbf{q})$ for the $2\times 2$ block spiral. Apart from the standard strong peak at $q_\parallel \simeq \pi/3$ related to the spiral's pitch, there is an additional weaker peak at $q_\parallel \simeq \pi-\pi/3$, which is precisely the fingerprint of the block structure persisting during the spiral rotation \cite{herbrychBlockSpiral2020}. In the same plot, Fig.~\ref{fig6}(a), we point out that the perfect blocks with $2k^b_\text{F}=\pi/m$ also exhibit a unique secondary Fourier peak inherent to their step-like structure \cite{herbrychBlockSpiral2020,HerbrychBlock2020}.

Finally, let us briefly comment on another unique feature of the block-spiral phase which can be observed in the behavior of the itinerant orbital $\gamma=0$. Since the OSMP system is in an overall metallic state, the spiral-like arrangement of the spins heavily modifies the single-particle spectral function $A_0(\mathbf{q},\omega)$, Eq.~\eqref{specakw}. In Fig.~\ref{fig6}(b), we present the bonding ($q_\perp=0$) and antibonding ($q_\perp=\pi$) components of $A_0(\mathbf{q},\omega)$ near the Fermi level $\epsilon_\text{F}$ (evaluated within the gKH model). Both components develop additional two branches which can be associated with parity-breaking quasiparticles, i.e., $q_{\parallel}\to-q_{\parallel}$ changes the character (branch) of the particles. We want to note that this phase was proposed \cite{MuraseCsFe2Se32020} as a possible magnetic order of the CsFe$_2$Se$_3$ ladder compound. Furthermore, a superconducting OSMP system with the parity-breaking quasiparticles was recently predicted \cite{herbrychTopoOSMP2021} to exhibit nontrivial topological properties with Majorana modes emerging at the edges of the system. We refer the interested reader to Refs.~\cite{herbrychBlockSpiral2020} and \cite{herbrychTopoOSMP2021} for details of this exotic phase.

\subsection{Incommensurate antiferromagnet and phase separation (two-band regime at low doping)}\label{phasesep}

We now move to discuss the two-band regime. Here, we find that the four-point Fermi surface makes this regime host qualitatively different magnetic phases than those present in the one-band case. Based on the magnetic phases found, we will split this region into two parts: low and intermediate doping. The latter will be discussed in the next section. 

\begin{figure}[!b]
    \includegraphics{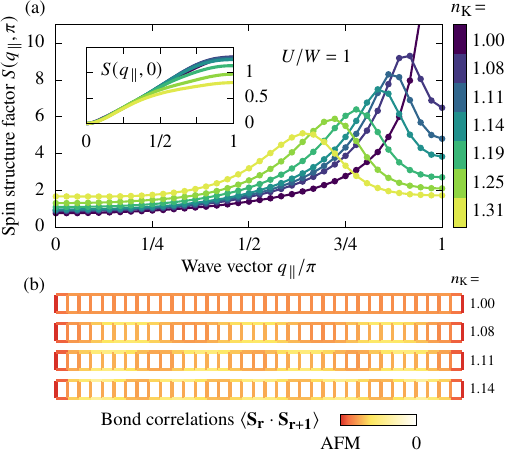}
    \caption{Incommensurate antiferromagnet. (a) The filling $n_\text{K}$ evolution of the spin structure factor $S(\mathbf{q})$ within the two-band regime at $U/W=1$. The main panel (inset) shows the antibonding $S(q_\parallel,\pi)$ [bonding $S(q_\parallel,0)$] component. (b) The filling $n_\text{K}$ evolution of the bond correlations $\langle \mathbf{S}_\mathbf{r} \cdot \mathbf{S}_{\mathbf{r}+\mathbf{1}} \rangle$ corresponding to the structure factors shown in (a). $\mathbf{1}$ connects the nearest-neighbor sites on the ladder. Note the evident amplitude modulation of the AFM correlations. All results were obtained for a generalized Kondo-Heisenberg ladder of $L=72$ sites.}
    \label{fig7}
\end{figure}

\begin{figure*}[!t]
    \includegraphics{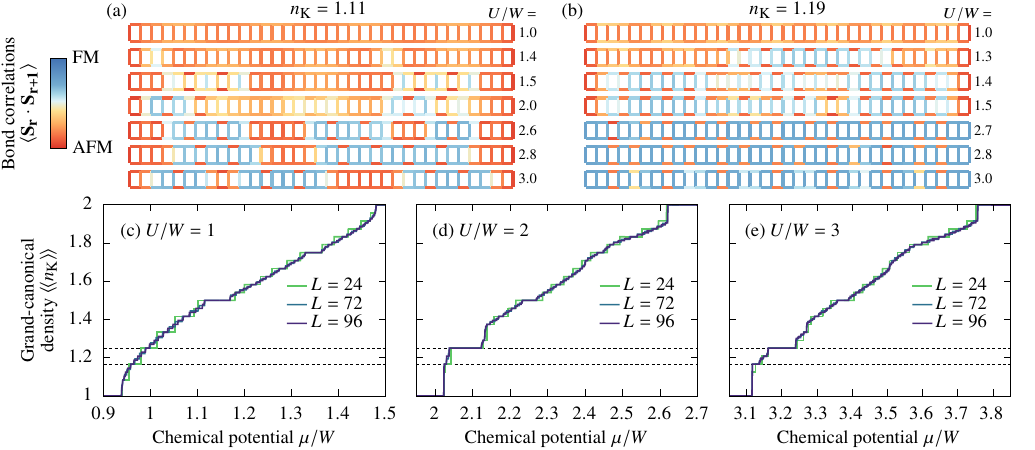}
    \caption{Phase separation. (a), (b) Bond correlations $\langle \mathbf{S}_\mathbf{r} \cdot \mathbf{S}_{\mathbf{r}+\mathbf{1}} \rangle$ for two fillings (a) $n_\text{K}=1.11 \in [1,1.17]$, (b) $n_\text{K}=1.19 \in [1.17,1.25]$ lying within the intervals of unstable densities. $\mathbf{1}$ connects the nearest-neighbor sites on the ladder and the system size is $L=72$ sites. \mbox{(c)-(e)}~Grand-canonical electron density $\langle\!\langle n_\text{K} \rangle\!\rangle$ vs $\mu$ for different values of the interaction $U/W=1,2,3$ and three system sizes $L=24,72,96$. The horizontal lines mark $\langle\!\langle n_\text{K} \rangle\!\rangle = 1.17$ ($\pi/3$ block) and $\langle\!\langle n_\text{K} \rangle\!\rangle = 1.25$ ($\pi/2$ block). All results were obtained for a generalized Kondo-Heisenberg ladder.}
    \label{fig8}
\end{figure*}

In Fig.~\ref{fig7}(a), we show the spin structure factor $S(\mathbf{q})$ for an intermediate interaction strength $U/W = 1$ and a range of fillings $1 < n_\text{K} < 1.3$ in the vicinity of half-filling. Here, in contrast to the block phase of the one-band regime, the antibonding component $S(q_\parallel,\pi)$ (main panel) exhibits a well-defined maximum and dominates the bonding part $S(q_\parallel,0)$ (inset), indicating that the rungs are predominantly AFM coupled. As expected, the half-filled ($n_\text{K}=1$) system is a two-orbital Mott insulator with a $(\pi,\pi)$ AFM order. Upon doping, the $S(q_\parallel,\pi)$ maximum shifts to smaller wave vectors. In contrast to the one-band regime, here, the latter maximum does not correspond to magnetic blocks, as clearly evidenced by the bond correlations [Fig.~\ref{fig7}(b)] which do not show any alternating FM/AFM pattern. Instead, for all $n_\text{K}$ in this region, the bond correlations are AFM with an additional amplitude modulation. Clearly, it is the periodicity of the latter modulation that is responsible for shifting the $S(q_\parallel,\pi)$ peak away from $(\pi,\pi)$. Moreover, the evolution of the $S(q_\parallel,\pi)$ peak also follows the (noninteracting) Fermi wave vectors $\mathbf{k}_\text{F}$ of the itinerant orbital. Being deep in the two-band regime, both Fermi wave vectors play a role and the maximum occurs at $(k_\text{F}^b+k_\text{F}^a,\pi)\simeq(2\pi-\pi n_\text{K},\pi)$, a result recognized already on a single-orbital Hubbard ladder \cite{noackGround1996}. Indeed, this type of exponentially decaying (short-range) incommensurate AFM is not an exclusively multiorbital feature, but relates to the long-standing problem of charge stripes, studied extensively in the doped single-orbital Hubbard model beyond 1D \cite{noackGround1996, martinsDoped2000, whiteStripes2003, hagerStripe2005, ehlersHybridspace2017, zhengStripe2017, jiangSuperconductivity2019, simonscollaborationonthemany-electronproblemAbsence2020}, as relevant in the context of cuprate high-$T_\text{c}$ superconductors \cite{tranquadaEvidence1995, tranquadaNeutronscattering1996, tranquadaCoexistence1997}. These stripes are a combination of codirectional charge-density waves and modulated AFM correlations (or spin-density waves in the case of a symmetry-broken state), wherein the region of maximum charge density coincides with a domain wall in the AFM \cite{zhengStripe2017}. In other words, the AFM correlations experience a $\pi$-phase shift across each charge-density peak, explaining the incommensurate tendencies \cite{martinsDoped2000,ehlersHybridspace2017}. We checked (not shown) that the spin correlations of Fig.~\ref{fig7} are indeed accompanied by striped charge-density waves and exhibit the appropriate phase shift. For completeness, we also verified that the chirality correlation $\langle \bm{\kappa}_\mathbf{r} \cdot \bm{\kappa}_{\mathbf{m}} \rangle$ is zero in this phase. Finally, let us comment that incommensurate AFM was also reported before in the context of the multiorbital Hubbard model \cite{patelPairing2017,pandeyIntertwined2021}.

Intuitively, the single-orbital behavior is recovered in the above since in the vicinity of half-filling the double-exchange mechanism requires larger Hund exchange $J_\text{H}$ to fully develop. Correspondingly, upon increasing $U$ (hence also increasing $J_\text{H}=U/4$), we find that the double exchange starts to play an important role. The bond correlations change drastically: the incommensurate AFM is lost in favor of the returning block formation tendencies, see Figs.~\ref{fig8}(a) and \ref{fig8}(b). At special fillings $n_\text{K}=1.17,1.25$, the system again develops $\pi/3$-block $\genfrac{}{}{0pt}{}{\uparrow\uparrow\uparrow\downarrow\downarrow\downarrow}{\uparrow\uparrow\uparrow\downarrow\downarrow\downarrow}$ and $\pi/2$-block $\genfrac{}{}{0pt}{}{\uparrow\uparrow\downarrow\downarrow}{\uparrow\uparrow\downarrow\downarrow}$ orders, respectively. However, in contrast to the block phase of the one-band regime---where at arbitrary fillings $n_\text{K}$ also other (more complicated) block-magnetic patterns emerged---here, the system coexists in spatially separated regions of the $\pi/3$ block ($n_\text{K}=1.17$), $\pi/2$ block ($n_\text{K}=1.25$), and the $\pi$ AFM ($n_\text{K}=1$) correlations instead. For example, at $n_\text{K}=1.11 \in [1,1.17]$ the system is divided into regions with AFM and $\pi/3$-block-like correlations for all presented values of $U$ [Fig.~\ref{fig8}(a)]. The closer the density is to $n_\text{K}=1.17$, the more the $\pi/3$-block regions grow at the cost of the AFM regions, and vice versa when moving closer to $n_\text{K}=1$. At $n_\text{K}=1.19 \in [1.17,1.25]$, on the other hand, the system shows first AFM--$\pi/2$ separation for $U/W \simeq 1.3$, and then a $\pi/3$--$\pi/2$ separation for $U/W \simeq 2.7$ [Fig.~\ref{fig8}(b)]. Such irregular local correlations are also reflected in an irregular behavior of the structure factor $S(\mathbf{q})$ which does not show any pronounced maximum or shows maxima that appear at seemingly random wave vectors for different values of $U$. 

In order to truly identify the above behavior as phase separation, one usually analyzes whether the compressibility acquires negative values, what signals the system being unstable \cite{dagottoFerromagnetic1998,dagottoNanoscale2003}. However, this can be troublesome, as it involves the evaluation of a second order derivative, which is highly prone to the smallest numerical errors. Therefore, we opt to use another observable---we investigate the $\langle\!\langle n_\text{K} \rangle\!\rangle$ vs $\mu$ curves, where $\langle\!\langle n_\text{K} \rangle\!\rangle$ is the (grand-canonical) electron density at a given chemical potential $\mu$. If $\langle\!\langle n_\text{K} \rangle\!\rangle(\mu)$ exhibits a discontinuity, then there are densities that cannot be stabilized, irrespective of the value of $\mu$. For calculations within the canonical ensemble, as performed here, this means that if the system is initialized with a density in the unstable interval, it will spontaneously separate into two regions of different densities \cite{dagottoFerromagnetic1998,yunokiPhase1998,dagottoNanoscale2003}, i.e., the behavior implied by Figs.~\ref{fig8}(a) and \ref{fig8}(b). 

Although, in principle, $\langle\!\langle n_\text{K} \rangle\!\rangle(\mu)$ needs to be calculated in the grand-canonical ensemble, it is possible to obtain it from the fixed-density DMRG results by searching for the particle number $N$ that minimizes the expectation value $\langle H_\text{K}-\mu \hat{N} \rangle = \epsilon_\text{GS}(N,L) - \mu N$ at each $\mu$ \cite{dagottoFerromagnetic1998, neuberFerromagnetic2006}, where $\hat{N}$ is the total particle number operator and $\epsilon_\text{GS}(N,L)$ is the ground-state energy for a fixed density $n_\text{K}=N/L$. In this way, we are able to study large system sizes, as presented on Figs.~\ref{fig8}(c)--\ref{fig8}(e), enabling us to distinguish a true discontinuity from one being a finite-size effect. One may observe that for $U/W=1$ and the smallest size $L=24$ [Fig.~\ref{fig8}(c)], the $\langle\!\langle n_\text{K} \rangle\!\rangle(\mu)$ curve is manifestly discrete. However, this discreteness disappears for larger sizes $L=72, 96$, where the $\langle\!\langle n_\text{K} \rangle\!\rangle(\mu)$ curves collapse and become smooth. This is the standard non-separated behavior. Contrarily, for $U/W=2$ [Fig.~\ref{fig8}(d)], there are several clear discontinuities at $1 < \langle\!\langle n_\text{K} \rangle\!\rangle \lesssim 1.4$, which persist despite the increasing system size. In particular, there is a discontinuity between $\langle\!\langle n_\text{K} \rangle\!\rangle=1$ and $\langle\!\langle n_\text{K} \rangle\!\rangle=1.17$ (AFM--$\pi/3$ separation), and another between $\langle\!\langle n_\text{K} \rangle\!\rangle=1.17$ and $\langle\!\langle n_\text{K} \rangle\!\rangle=1.25$ ($\pi/3$--$\pi/2$ separation), in perfect agreement with the bond correlation results [Figs.~\ref{fig8}(a) and \ref{fig8}(b)]. Figure~\ref{fig8}(e) leads to the same conclusions, but for $U/W=3$. Therefore, we conclude that a clear tendency to phase-separate exists in our model for relatively low fillings $n_\text{K}$, close to half-filling. 

Finally, from Figs.~\ref{fig8}(c)--\ref{fig8}(e) it also follows that the phase separation is absent in the one-band regime; in particular, even the complicated block patterns [e.g., the one shown in Fig.~\ref{fig4}(c) at $n_\text{K}=1.81$] are robust uniform phases with no phase separation. Curiously, here, in the two-band regime, the blocks with complicated unit cells are in fact entirely absent, as we only see phase separation between AFM, $\pi/3$ and $\pi/2$ blocks. It is possible that larger system sizes would need to be accessed to find separation between the blocks with unusual periodicities. We do, however, find the signatures of block spirals at the special points $n_\text{K}=1.17,1.25$, which appear before FM for excessively large values of $U$. We have also checked (not shown) that phase separation is not present in the chain geometry. This is consistent with the picture that the chain can approximate the ladder well but only in the one-band regime, while the two-band regime cannot be captured (as the Fermi surface is closer to 2D). It is also worth noting that phase separation tendencies were reported experimentally in a layered iron superconductor K$_{0.8}$Fe$_{1.6}$Se$_2$ \cite{ricciNanoscale2011}, albeit they concern separation between magnetic and nonmagnetic regions of different lattice constants.

\subsection{Spin flux (two-band regime at intermediate doping)}\label{flux}
Let us now describe the last region of our phase diagram---the two-band regime at intermediate doping, i.e., in the vicinity of three-quarter filling $n_\text{K} \sim 1.5$. We find this region to behave surprisingly very different from the low-doping case, despite the same (noninteracting) Fermi surface. We shall attribute this difference to the strong renormalization of the Fermi surface due to the simultaneous competition of all energy scales at this intermediate parameter regime.

\begin{figure}[!t]
    \includegraphics{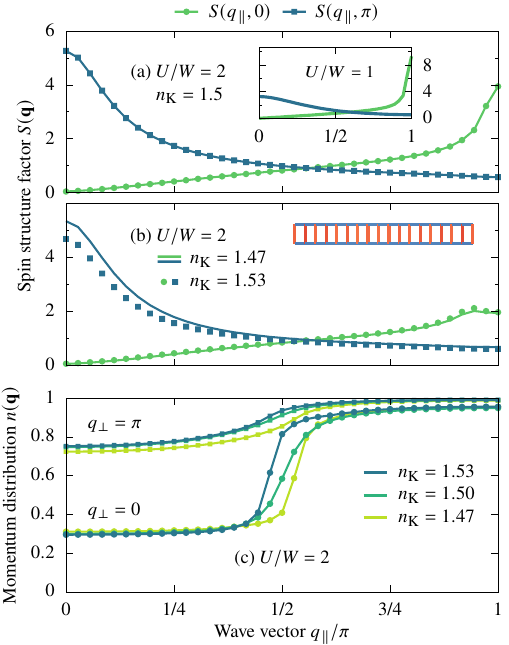}
    \caption{Spin structure factor and momentum distribution function in the spin-flux region. (a) Spin structure factor $S(\mathbf{q})$ at $n_\text{K}=1.5$ and $U/W=1$ (inset), $U/W=2$ (main panel). (b) The same as in (a) but at $n_\text{K}=1.47$ (lines), $n_\text{K}=1.53$ (symbols) and $U/W=2$. The inset shows the bond correlations $\langle \mathbf{S}_\mathbf{r} \cdot \mathbf{S}_{\mathbf{r}+\mathbf{1}} \rangle$ corresponding to $n_\text{K}=1.47$. Here, red (blue) color marks AFM (FM) bonds and $\mathbf{1}$ connects the nearest-neighbor sites on the ladder. (c) Momentum distribution function $n(\mathbf{q})$ at $n_\text{K}=1.47,\, 1.5,\, 1.53$ and $U/W=2$. All results were obtained for a generalized Kondo-Heisenberg ladder of $L=72$ sites.}
    \label{fig9}
\end{figure}

In the inset of Fig.~\ref{fig9}(a), we present the spin structure factor $S(\mathbf{q})$ at $n_\text{K} = 1.5$ and at an intermediate interaction strength $U/W=1$. We observe a dominant maximum in the bonding component $S(q_\parallel,0)$ at wave vector $(\pi,0)$ and a rather structureless antibonding part $S(q_\parallel,\pi)$. This result corresponds to the case of FM rungs and AFM legs, $\genfrac{}{}{0pt}{}{\uparrow\downarrow\uparrow\downarrow}{\uparrow\downarrow\uparrow\downarrow}$, i.e., the canonical magnetic order found experimentally in several iron-based ladders (see the introduction) and also widely believed to be the parent state of 2D iron-based superconductors \cite{GlasbrennerFeSe2015,KreiselReview2020}. Surprisingly, with increasing $U$, we find that this order is suppressed (albeit does not vanish) in favor of a maximum in the antibonding $S(q_\parallel,\pi)$ developing at $(0,\pi)$ [main panel of Fig.~\ref{fig9}(a)]. Similar behavior was recently reported in Ref.~\cite{pandeyIntertwined2021}, which studied the pairing-related properties of the two-leg ladder BaFe$_2$S$_3$. Figure~\ref{fig9}(b) shows that this behavior is not restricted to $n_\text{K} = 1.5$, but occurs consistently at other fillings in the entire $1.3 \lesssim n_\text{K} \lesssim 1.6$ interval and also in a wider range of interactions $1 \lesssim U/W \lesssim 4$. The dominant peak at $(0,\pi)$ leads to the bond correlations taking now the form of AFM rungs and FM legs [inset of Fig.~\ref{fig9}(b)].

\begin{figure*}[!ht]
    \includegraphics{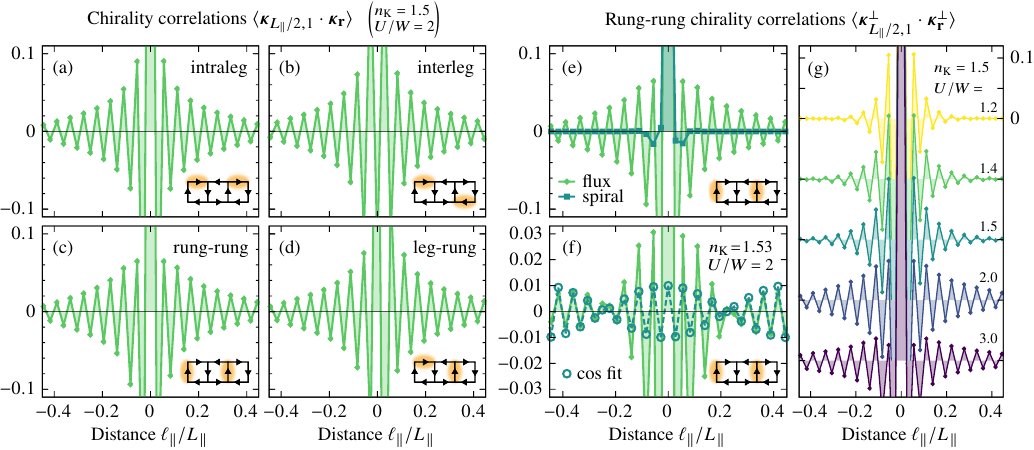}
    \caption{Spin flux. (a)-(d) Chirality correlations $\langle \bm{\kappa}_{L_{\parallel}/2,1}\cdot\bm{\kappa}_{\mathbf{r}}\rangle$ at $n_\text{K}=1.5$, $U/W=2$ as a function of distance. Shown are all possible components: (a) intraleg, (b) interleg, (c) rung-rung, (d) leg-rung. The sketches show the proposed magnetic order and also highlight the bonds which are involved in the calculation. (e) Comparison of the distance-dependent rung-rung chirality $\langle \bm{\kappa}_{L_{\parallel}/2,1}^\perp \cdot \bm{\kappa}_{\mathbf{r}}^\perp \rangle$ between the flux ($n_\text{K}=1.5$, $U/W=2$) and the $2 \times 2$ block spiral ($n_\text{K}=1.75$, $U/W=2.2$). (f) Rung-rung chirality in the flux phase with $n_\text{K}=1.53$, $U/W=2$. The dark symbols correspond to $\cos(2 \tilde{k}_\text{F}^b \,\ell_\parallel)$ fit using the effective Fermi wave vector $\tilde{k}_\text{F}^b$ obtained from Fig.~\ref{fig9}(c). (g) Interaction $U$ evolution of the rung-rung chirality for the spin flux with $n_\text{K}=1.5$. All results were obtained for a generalized Kondo-Heisenberg ladder of $L=72$ sites.}
    \label{fig10}
\end{figure*}

Although in the bond correlations there is no discernible pattern related to the weak $S(q_\parallel,0)$ features present in both Figs.~\ref{fig9}(a) and \ref{fig9}(b), further analysis offers a useful insight. In particular, from Fig.~\ref{fig9}(b) it follows that the structure factor $S(\mathbf{q})$ behaves identically at both $n_\text{K}=1.47$ and $n_\text{K}=1.53$, pointing to a symmetry around the $n_\text{K}=1.5$ point, quite unexpected in the two-band regime. Crucially, here, we are in fact at the crossroads between the one- and two-band regimes, as the latter is enforced solely due to the finite interaction $U$ [Fig.~\ref{fig2}(c)]. In the noninteracting case, we would have two Fermi wave vectors for $n_\text{K} < 1.5$, but only one for $n_\text{K} > 1.5$, meaning that neither the noninteracting $2k_\text{F}^b$ prediction from Sec.~\ref{block} nor the $k_\text{F}^b+k_\text{F}^a$ prediction from Sec.~\ref{phasesep} can be meaningfully applied to the apparent symmetry in the structure factors. 

To elucidate this issue, we investigate the momentum distribution function defined as $n(\mathbf{q})=\langle n_{0\mathbf{q}} \rangle = \langle \sum_\sigma c^\dagger_{0\mathbf{q}\sigma}c_{0\mathbf{q}\sigma} \rangle$, where $c^\dagger_{0\mathbf{q}\sigma}=(1/\sqrt{L})\sum_\mathbf{r} \exp(i\mathbf{q}\mathbf{r})\,c^\dagger_{0\mathbf{r}\sigma}$. The results are presented in Fig.~\ref{fig9}(c) for the three fillings $n_\text{K}=1.47,\, 1.5,\, 1.53$ and the interaction $U/W=2$. One immediately notices that the momentum distribution of the antibonding band ($q_\perp=\pi$) is highly renormalized with respect to the $U=0$ case, where it would be a step function centered at $k_\text{F}^a$. Here, it is strongly flattened instead, and its shape seems to be weakly dependent on the filling $n_\text{K}$. The bonding band ($q_\perp=0$), on the other hand, acts as though it was the only one being filled: Adding (removing) particles shifts its effective Fermi wave vector $\tilde{k}_\text{F}^b$ [taken here as the inflection point of $n(q_\parallel,0)$], and the function $n(q_\parallel,0)$ appears relatively sharp despite the large interaction $U$. Remarkably, the weak peak in $S(q_\parallel,0)$ corresponds to $(2\tilde{k}_\text{F}^b,0)$, explaining why the structure factors at the two fillings $n_\text{K}=1.47,\, 1.53$ are identical [Fig.~\ref{fig9}(b)]. This is precisely due to the symmetric behavior of $n(q_\parallel,0)$ around the $n_\text{K}=1.5$ point. Therefore, although both bands are fractionally filled, as follows from both Figs.~\ref{fig2}(c) and \ref{fig9}(c), the interaction $U$ promotes an emergent one-band behavior. This stands in contrast to the low-doping case (deep within the two-band regime), where such a behavior is absent.

The strong $(0,\pi)$ peak and the bond correlations with FM legs and AFM rungs suggest that this order could be a ladder analogue of 2D patterns argued \cite{GlasbrennerFeSe2015} to be relevant for the layered compounds Fe(Se,Te). However, the analysis of chirality correlations offers a different interpretation. In Fig.~\ref{fig10}(a), we present the intraleg chirality correlation function $\langle \bm{\kappa}_\mathbf{r} \cdot \bm{\kappa}_{\mathbf{m}} \rangle$ at $n_\text{K}=1.5$ and $U/W=2.0$. Remarkably, we observe significant and slowly decaying chirality correlations, indicating that the spins are noncollinear. This is quite unexpected considering the commensurate structure factors shown in Fig.~\ref{fig9}(a). Moreover, the chirality displays an intriguing staggered pattern, which is present not only in the intraleg correlations but also in the interleg case [Fig.~\ref{fig10}(b)], and the highly nontrivial rung-rung and leg-rung cases [Figs.~\ref{fig10}(c)--\ref{fig10}(d)]. It is hard to imagine a (quantum) spin ordering which would lead to all the chirality correlation functions simultaneously showing the same staggered pattern. The solution to this conundrum can be found by noticing that the $z$ component of the chirality correlation function is in fact equal to the spin-current correlation function, $\kappa^z_{\mathbf{r}}=S^x_\mathbf{r}S^y_\mathbf{r+1}-S^y_\mathbf{r}S^x_\mathbf{r+1}= i/2 \, (S^+_\mathbf{r}S^-_\mathbf{r+1}-S^-_\mathbf{r}S^+_\mathbf{r+1})$. Indeed, we checked that the $z$ component has a significant  contribution to the presented chirality values. Therefore, we propose that the system realizes a novel quantum spin-flux phase, wherein the spin currents circulate around $2 \times 2$ plaquettes and are staggered from plaquette to plaquette [see the sketches in Figs.~\ref{fig10}(a)--\ref{fig10}(d) and \ref{fig1}(i)], with no net current flow. 

Since all the chirality functions behave in the same manner, in the following we discuss only the representative rung-rung case $\langle \bm{\kappa}_\mathbf{r}^\perp \cdot \bm{\kappa}_{\mathbf{m}}^\perp \rangle$, which corresponds to spin currents flowing along the rungs. In Fig.~\ref{fig10}(e), we compare the spin flux to the other chiral phase of our model---the block spiral. Clearly, in the latter the rung-rung (and also leg-rung) chirality vanishes, highlighting that it is indeed unique to the flux. Moreover, this unique chirality is not restricted to the $n_\text{K}=1.5$ filling, but appears also for $n_\text{K} \neq 1.5$ in the entire regime being discussed in this section, i.e., for $1.3 \lesssim n_\text{K} \lesssim 1.6$ and $1 \lesssim U/W \lesssim 4$. Variation of the filling introduces an additional modulation of the staggered pattern, as shown in Fig.~\ref{fig10}(f) for $n_\text{K}=1.53$. Furthermore, the latter modulation is controlled by the effective Fermi wave vector $\tilde{k}_\text{F}^b$, i.e., $\langle \bm{\kappa}_\mathbf{r}^\perp \cdot \bm{\kappa}_{\mathbf{r+d}}^\perp \rangle \propto \cos(2 \tilde{k}_\text{F}^b \,d)$ \cite{fjaerestadOrbital2006}. This ``hidden'' periodicity---which is readily seen in the chirality correlations but not in the real-space spin correlations---elucidates the origin of the $(2\tilde{k}_\text{F}^b,0)$ maximum we earlier noted in $S(q_\parallel,0)$, which accompanies the strong $(0,\pi)$ peak [Figs.~\ref{fig9}(a)--\ref{fig9}(b)]. In particular, at $n_\text{K}=1.5$, we have $(2\tilde{k}_\text{F}^b,0)=(\pi,0)$. This understanding is consistent with the results for the 2D FM Kondo lattice of classical spins, where the similar-weight structure factor maxima at $(\pi,0)$ and $(0,\pi)$ were recognized as the hallmark of the spin flux \cite{yamanakaFlux1998,agterbergSpinflux2000,aliagaIsland2001}. The latter spin configuration cannot be specified by only one wave vector \cite{yamanakaFlux1998}. In our case, the spin-flux phase emerges within a fully quantum model and, moreover, it is promoted by the interaction $U$, as follows from Fig.~\ref{fig10}(g). Consistently, the structure factor peak at $(0,\pi)$ [Fig.~\ref{fig9}(a)] acquires significant weight only when the staggered chirality correlations are well-developed. From this perspective, we treat the canonical $(\pi,0)$ order of AFM legs and FM rungs, $\genfrac{}{}{0pt}{}{\uparrow\downarrow\uparrow\downarrow}{\uparrow\downarrow\uparrow\downarrow}$, present at $U/W=1$ [inset of Fig.~\ref{fig9}(a)], as an underdeveloped flux, rather than a separate phase, and we do not mark it individually on the phase diagram, Fig.~\ref{fig3}.

\section{Conclusions}\label{conclusions}

To summarize, using an accurate computational technique we have studied the magnetic phase diagram of the two-leg multiorbital ladder in the orbital-selective Mott phase. Although our effective model, the generalized Kondo-Heisenberg Hamiltonian, describes the electron densities of the iron-based systems in an approximate manner, it properly captures the symmetric and antisymmetric bands (bonding and antibonding, respectively). The latter are crucial to a proper description of the magnetic order. 

The magnetic phase diagram of the ladder OSMP is dominated by tendencies to form magnetic blocks of various shapes and sizes. At large fillings, $n_\text{K}\gtrsim 1.6$, where the antibonding band ($q_\perp=\pi$) is fully filled and only the bonding band ($q_\perp=0$) carries the Fermi wave vectors, the system develops perfect blocks of $\genfrac{}{}{0pt}{}{\uparrow\uparrow\downarrow\downarrow}{\uparrow\uparrow\downarrow\downarrow}$-form at $U \sim W$. Increasing the strength of the interaction $U$ leads to the uniform rotation of the blocks, i.e., to the formation of the exotic block-spiral phase with nontrivial properties. In the opposite limit, close to half-filling $n_\text{K} \sim 1$, the four Fermi wave vectors present in two bands drive the system towards phase separation with (predominantly) $\pi/2$ and $\pi/3$ blocks. Finally, when $n_\text{K} \sim 1.5$, the ladder system develops a quantum spin flux originating in the competing energy scales inherent to the OSMP. This phase can be naively viewed as staggered spin currents circulating within $2 \times 2$ plaquettes (however, no plaquette carries net current due to its quantum nature in a finite system).

Our phase diagram indicates that the magnetism of iron-based ladders, due to the presence of charge, spin, and orbital degrees of freedom, combines phenomena known from cuprates with those found in manganites \cite{PhysRevLett.79.713,PhysRevLett.90.247203}. Namely, at small interaction $U$ and close to half-filling $n_\text{K} \sim 1$, we have found the striped incommensurate antiferromagnetism---the challenging and still not fully understood magnetic order relevant for 2D cuprate superconductors. On the other hand, increasing the interaction strength $U$, one can find the phase-separated region known from the manganites. Most importantly, our results indicate that the family of iron-based AFe$_2$X$_3$ compounds lies within the one-band regime, where the block and block-spiral orders can be found (also experimentally). We believe that our comprehensive study provides motivation and theoretical guidance for crystal growers and experimentalists to discover new iron-based ladder compounds that may display the highly unusual magnetic properties reported here.

\begin{acknowledgments}
We acknowledge fruitful discussions with M.\ M.\ Ma\'{s}ka. M.{\'S}. and J.H. acknowledge grant support by the Polish National Agency of Academic Exchange (NAWA) under contract PPN/PPO/2018/1/00035 and by the National Science Centre (NCN), Poland via project 2019/35/B/ST3/01207. E.D. was supported by the US Department of Energy (DOE), Office of Science, Basic Energy Sciences (BES), Materials Sciences and Engineering Division. Calculations have been carried out using resources provided by Wroclaw Centre for Networking and Supercomputing.
\end{acknowledgments}

\appendix*
\section{Computational accuracy regarding the SU(2) symmetry}\label{SU2}
It is well-known that within DMRG implementations which exploit only the U(1) spin symmetry, and not the full SU(2) symmetry, it is possible to converge to a state with finite local magnetization $\langle{S^z_\mathbf{r}}\rangle$, even though a finite system cannot break the SU(2) symmetry. This effect is more pronounced in simulations beyond 1D and is a recurring issue in the studies of 2D Hubbard and $t$-$J$ models \cite{whiteDensity1998,whiteStripes2003,hagerStripe2005,jiangSuperconductivity2019,jiangGround2020}. Although here we discuss a two-leg ladder, its two-orbital nature makes it effectively a four-leg problem, and in some cases our DMRG indeed ends up in a state with nonvanishing $\langle{S^z_\mathbf{r}}\rangle$. This issue can be mitigated by drastically increasing the number of states kept \cite{whiteStripes2003,jiangSuperconductivity2019,jiangGround2020}, but then the already demanding computational effort would quickly become prohibitively expensive. Still, where feasible, we did verify that increasing $M$ in our DMRG procedure does drive $\langle{S^z_\mathbf{r}}\rangle$ to zero, whilst preserving the spin-spin correlations and introducing a minimal adjustment of the ground-state energy. Moreover, sometimes a slight perturbation of the model parameters (e.g., changing $U$ by as little as 5\%) was enough to tip the algorithm towards a final state which respects the SU(2) symmetry, but appears otherwise unchanged with respect to other quantities. These observations suggest that the finite $\langle{S^z_\mathbf{r}}\rangle$ arises because the DMRG selects a subset of states (with a particular direction of the order parameter) from the macroscopic superposition present within the true ground state \cite{whiteStripes2003,stoudenmireStudying2012}. Since the states in the latter superposition are expected to be (nearly) degenerate, the difference in the final energy is minimal, making it hard to completely converge. Nonetheless, such a spurious ''partial'' symmetry breaking within the final state should not lead to a misrepresentation of the magnetic order existing in our system, nor should it affect the behavior of itinerant carriers doped into those states (see the discussion in Ref.~\cite{whiteStripes2003}). We therefore conclude that the occasional presence of finite $\langle{S^z_\mathbf{r}}\rangle$ is insignificant for our study and does not invalidate our results. That being said, the nondecaying nature of the static magnetization can make the maxima in $S(\mathbf{q})$ appear excessively sharp. To avoid the misinterpretation of $S(\mathbf{q})$, while plotting the latter we discard the fictitious spin-density contribution, i.e., we define $\langle \mathbf{S}_{\mathbf{r}} \cdot \mathbf{S}_{\mathbf{m}} \rangle \equiv \langle \mathbf{S}_{\mathbf{r}} \cdot \mathbf{S}_{\mathbf{m}} \rangle - \langle S^z_{\mathbf{r}} \rangle \langle S^z_{\mathbf{m}} \rangle$. The fact that this does not reduce $\langle \mathbf{S}_{\mathbf{r}} \cdot \mathbf{S}_{\mathbf{m}} \rangle$ to zero confirms that our ground state retains most, if not all, quantum fluctuations on top of the artificial magnetization.

The issue described above is especially troublesome in the intermediate-doping ($n_\text{K} \sim 1.5$) region discussed in Sec.~\ref{flux}, which seems to be the most demanding for the DMRG method. There, all the energy scales are simultaneosly at play, making it hard to fully stabilize the system within current computational limitations. As a consequence, a more detailed analysis focusing solely on the latter region is called for, involving also a systematic study of the $t^\perp_{0}$ influence, which is however beyond the scope of the present, more general, survey of the magnetic phases. Our already interesting findings reported here provide motivation for such a more in-depth study of the intermediate-doping region in the near future.

\bibliographystyle{apsrev4-2}
\bibliography{ref_osmp}

\begin{thebibliography}{104}%
\makeatletter
\providecommand \@ifxundefined [1]{%
 \@ifx{#1\undefined}
}%
\providecommand \@ifnum [1]{%
 \ifnum #1\expandafter \@firstoftwo
 \else \expandafter \@secondoftwo
 \fi
}%
\providecommand \@ifx [1]{%
 \ifx #1\expandafter \@firstoftwo
 \else \expandafter \@secondoftwo
 \fi
}%
\providecommand \natexlab [1]{#1}%
\providecommand \enquote  [1]{``#1''}%
\providecommand \bibnamefont  [1]{#1}%
\providecommand \bibfnamefont [1]{#1}%
\providecommand \citenamefont [1]{#1}%
\providecommand \href@noop [0]{\@secondoftwo}%
\providecommand \href [0]{\begingroup \@sanitize@url \@href}%
\providecommand \@href[1]{\@@startlink{#1}\@@href}%
\providecommand \@@href[1]{\endgroup#1\@@endlink}%
\providecommand \@sanitize@url [0]{\catcode `\\12\catcode `\$12\catcode
  `\&12\catcode `\#12\catcode `\^12\catcode `\_12\catcode `\%12\relax}%
\providecommand \@@startlink[1]{}%
\providecommand \@@endlink[0]{}%
\providecommand \url  [0]{\begingroup\@sanitize@url \@url }%
\providecommand \@url [1]{\endgroup\@href {#1}{\urlprefix }}%
\providecommand \urlprefix  [0]{URL }%
\providecommand \Eprint [0]{\href }%
\providecommand \doibase [0]{https://doi.org/}%
\providecommand \selectlanguage [0]{\@gobble}%
\providecommand \bibinfo  [0]{\@secondoftwo}%
\providecommand \bibfield  [0]{\@secondoftwo}%
\providecommand \translation [1]{[#1]}%
\providecommand \BibitemOpen [0]{}%
\providecommand \bibitemStop [0]{}%
\providecommand \bibitemNoStop [0]{.\EOS\space}%
\providecommand \EOS [0]{\spacefactor3000\relax}%
\providecommand \BibitemShut  [1]{\csname bibitem#1\endcsname}%
\let\auto@bib@innerbib\@empty
\bibitem [{\citenamefont {Dagotto}\ and\ \citenamefont
  {Rice}(1996)}]{dogottoSurprises1996}%
  \BibitemOpen
  \bibfield  {author} {\bibinfo {author} {\bibfnamefont {E.}~\bibnamefont
  {Dagotto}}\ and\ \bibinfo {author} {\bibfnamefont {T.~M.}\ \bibnamefont
  {Rice}},\ }\href {https://doi.org/10.1126/science.271.5249.618} {\bibfield
  {journal} {\bibinfo  {journal} {Science}\ }\textbf {\bibinfo {volume}
  {271}},\ \bibinfo {pages} {618} (\bibinfo {year} {1996})}\BibitemShut
  {NoStop}%
\bibitem [{\citenamefont {Uehara}\ \emph {et~al.}(1996)\citenamefont {Uehara},
  \citenamefont {Nagata}, \citenamefont {Akimitsu}, \citenamefont {Takahashi},
  \citenamefont {M{\^o}ri},\ and\ \citenamefont
  {Kinoshita}}]{UeharaSCladder1996}%
  \BibitemOpen
  \bibfield  {author} {\bibinfo {author} {\bibfnamefont {M.}~\bibnamefont
  {Uehara}}, \bibinfo {author} {\bibfnamefont {T.}~\bibnamefont {Nagata}},
  \bibinfo {author} {\bibfnamefont {J.}~\bibnamefont {Akimitsu}}, \bibinfo
  {author} {\bibfnamefont {H.}~\bibnamefont {Takahashi}}, \bibinfo {author}
  {\bibfnamefont {N.}~\bibnamefont {M{\^o}ri}},\ and\ \bibinfo {author}
  {\bibfnamefont {K.}~\bibnamefont {Kinoshita}},\ }\href
  {https://doi.org/10.1143/JPSJ.65.2764} {\bibfield  {journal} {\bibinfo
  {journal} {J. Phys. Soc. Japan}\ }\textbf {\bibinfo {volume} {65}},\ \bibinfo
  {pages} {2764} (\bibinfo {year} {1996})}\BibitemShut {NoStop}%
\bibitem [{\citenamefont {Motoyama}\ \emph {et~al.}(2002)\citenamefont
  {Motoyama}, \citenamefont {Eisaki}, \citenamefont {Uchida}, \citenamefont
  {Takeshita}, \citenamefont {M{\^o}ri}, \citenamefont {Nakanishi},\ and\
  \citenamefont {Takahashi}}]{MotoyamaTelephon2002}%
  \BibitemOpen
  \bibfield  {author} {\bibinfo {author} {\bibfnamefont {N.}~\bibnamefont
  {Motoyama}}, \bibinfo {author} {\bibfnamefont {H.}~\bibnamefont {Eisaki}},
  \bibinfo {author} {\bibfnamefont {S.}~\bibnamefont {Uchida}}, \bibinfo
  {author} {\bibfnamefont {N.}~\bibnamefont {Takeshita}}, \bibinfo {author}
  {\bibfnamefont {N.}~\bibnamefont {M{\^o}ri}}, \bibinfo {author}
  {\bibfnamefont {T.}~\bibnamefont {Nakanishi}},\ and\ \bibinfo {author}
  {\bibfnamefont {H.}~\bibnamefont {Takahashi}},\ }\href
  {https://doi.org/10.1209/epl/i2002-00414-0} {\bibfield  {journal} {\bibinfo
  {journal} {EPL}\ }\textbf {\bibinfo {volume} {58}},\ \bibinfo {pages} {758}
  (\bibinfo {year} {2002})}\BibitemShut {NoStop}%
\bibitem [{\citenamefont {Maekawa}(1996)}]{MaekawaLadder1996}%
  \BibitemOpen
  \bibfield  {author} {\bibinfo {author} {\bibfnamefont {S.}~\bibnamefont
  {Maekawa}},\ }\href {https://doi.org/10.1126/science.273.5281.1515}
  {\bibfield  {journal} {\bibinfo  {journal} {Science}\ }\textbf {\bibinfo
  {volume} {273}},\ \bibinfo {pages} {1515} (\bibinfo {year}
  {1996})}\BibitemShut {NoStop}%
\bibitem [{\citenamefont {Dagotto}\ \emph {et~al.}(1992)\citenamefont
  {Dagotto}, \citenamefont {Riera},\ and\ \citenamefont
  {Scalapino}}]{DagottoLadder1992}%
  \BibitemOpen
  \bibfield  {author} {\bibinfo {author} {\bibfnamefont {E.}~\bibnamefont
  {Dagotto}}, \bibinfo {author} {\bibfnamefont {J.}~\bibnamefont {Riera}},\
  and\ \bibinfo {author} {\bibfnamefont {D.}~\bibnamefont {Scalapino}},\ }\href
  {https://doi.org/10.1103/PhysRevB.45.5744} {\bibfield  {journal} {\bibinfo
  {journal} {Phys. Rev. B}\ }\textbf {\bibinfo {volume} {45}},\ \bibinfo
  {pages} {5744} (\bibinfo {year} {1992})}\BibitemShut {NoStop}%
\bibitem [{\citenamefont {Rice}\ \emph {et~al.}(1995)\citenamefont {Rice},
  \citenamefont {Gopalan}, \citenamefont {Sigrist},\ and\ \citenamefont
  {Zhang}}]{RiceLadder1994}%
  \BibitemOpen
  \bibfield  {author} {\bibinfo {author} {\bibfnamefont {T.~M.}\ \bibnamefont
  {Rice}}, \bibinfo {author} {\bibfnamefont {S.}~\bibnamefont {Gopalan}},
  \bibinfo {author} {\bibfnamefont {M.}~\bibnamefont {Sigrist}},\ and\ \bibinfo
  {author} {\bibfnamefont {F.~C.}\ \bibnamefont {Zhang}},\ }\href
  {https://doi.org/10.1007/BF00754945} {\bibfield  {journal} {\bibinfo
  {journal} {J. Low Temp. Phys.}\ }\textbf {\bibinfo {volume} {95}},\ \bibinfo
  {pages} {299} (\bibinfo {year} {1995})}\BibitemShut {NoStop}%
\bibitem [{\citenamefont {Motoyama}\ \emph {et~al.}(1996)\citenamefont
  {Motoyama}, \citenamefont {Eisaki},\ and\ \citenamefont
  {Uchida}}]{MotoyamaChain1996}%
  \BibitemOpen
  \bibfield  {author} {\bibinfo {author} {\bibfnamefont {N.}~\bibnamefont
  {Motoyama}}, \bibinfo {author} {\bibfnamefont {H.}~\bibnamefont {Eisaki}},\
  and\ \bibinfo {author} {\bibfnamefont {S.}~\bibnamefont {Uchida}},\ }\href
  {https://doi.org/10.1103/PhysRevLett.76.3212} {\bibfield  {journal} {\bibinfo
   {journal} {Phys. Rev. Lett.}\ }\textbf {\bibinfo {volume} {76}},\ \bibinfo
  {pages} {3212} (\bibinfo {year} {1996})}\BibitemShut {NoStop}%
\bibitem [{\citenamefont {Hlubek}\ \emph {et~al.}(2010)\citenamefont {Hlubek},
  \citenamefont {Ribeiro}, \citenamefont {{Saint-Martin}}, \citenamefont
  {Revcolevschi}, \citenamefont {Roth}, \citenamefont {Behr}, \citenamefont
  {B{\"u}chner},\ and\ \citenamefont {Hess}}]{HlubekBalisitc2010}%
  \BibitemOpen
  \bibfield  {author} {\bibinfo {author} {\bibfnamefont {N.}~\bibnamefont
  {Hlubek}}, \bibinfo {author} {\bibfnamefont {P.}~\bibnamefont {Ribeiro}},
  \bibinfo {author} {\bibfnamefont {R.}~\bibnamefont {{Saint-Martin}}},
  \bibinfo {author} {\bibfnamefont {A.}~\bibnamefont {Revcolevschi}}, \bibinfo
  {author} {\bibfnamefont {G.}~\bibnamefont {Roth}}, \bibinfo {author}
  {\bibfnamefont {G.}~\bibnamefont {Behr}}, \bibinfo {author} {\bibfnamefont
  {B.}~\bibnamefont {B{\"u}chner}},\ and\ \bibinfo {author} {\bibfnamefont
  {C.}~\bibnamefont {Hess}},\ }\href
  {https://doi.org/10.1103/PhysRevB.81.020405} {\bibfield  {journal} {\bibinfo
  {journal} {Phys. Rev. B}\ }\textbf {\bibinfo {volume} {81}},\ \bibinfo
  {pages} {020405(R)} (\bibinfo {year} {2010})}\BibitemShut {NoStop}%
\bibitem [{\citenamefont {Hess}(2019)}]{hessThermal2019}%
  \BibitemOpen
  \bibfield  {author} {\bibinfo {author} {\bibfnamefont {C.}~\bibnamefont
  {Hess}},\ }\href {https://doi.org/10.1016/j.physrep.2019.02.004} {\bibfield
  {journal} {\bibinfo  {journal} {Phys. Rep.}\ }\textbf {\bibinfo {volume}
  {811}},\ \bibinfo {pages} {1} (\bibinfo {year} {2019})}\BibitemShut {NoStop}%
\bibitem [{\citenamefont {Steinigeweg}\ \emph {et~al.}(2016)\citenamefont
  {Steinigeweg}, \citenamefont {Herbrych}, \citenamefont {Zotos},\ and\
  \citenamefont {Brenig}}]{SteinigewegIntegrable2016}%
  \BibitemOpen
  \bibfield  {author} {\bibinfo {author} {\bibfnamefont {R.}~\bibnamefont
  {Steinigeweg}}, \bibinfo {author} {\bibfnamefont {J.}~\bibnamefont
  {Herbrych}}, \bibinfo {author} {\bibfnamefont {X.}~\bibnamefont {Zotos}},\
  and\ \bibinfo {author} {\bibfnamefont {W.}~\bibnamefont {Brenig}},\ }\href
  {https://doi.org/10.1103/PhysRevLett.116.017202} {\bibfield  {journal}
  {\bibinfo  {journal} {Phys. Rev. Lett.}\ }\textbf {\bibinfo {volume} {116}},\
  \bibinfo {pages} {017202} (\bibinfo {year} {2016})}\BibitemShut {NoStop}%
\bibitem [{\citenamefont {Karrasch}\ \emph {et~al.}(2015)\citenamefont
  {Karrasch}, \citenamefont {Kennes},\ and\ \citenamefont
  {{Heidrich-Meisner}}}]{KarraschLadders2015}%
  \BibitemOpen
  \bibfield  {author} {\bibinfo {author} {\bibfnamefont {C.}~\bibnamefont
  {Karrasch}}, \bibinfo {author} {\bibfnamefont {D.~M.}\ \bibnamefont
  {Kennes}},\ and\ \bibinfo {author} {\bibfnamefont {F.}~\bibnamefont
  {{Heidrich-Meisner}}},\ }\href {https://doi.org/10.1103/PhysRevB.91.115130}
  {\bibfield  {journal} {\bibinfo  {journal} {Phys. Rev. B}\ }\textbf {\bibinfo
  {volume} {91}},\ \bibinfo {pages} {115130} (\bibinfo {year}
  {2015})}\BibitemShut {NoStop}%
\bibitem [{\citenamefont {Takahashi}\ \emph {et~al.}(2015)\citenamefont
  {Takahashi}, \citenamefont {Sugimoto}, \citenamefont {Nambu}, \citenamefont
  {Yamauchi}, \citenamefont {Hirata}, \citenamefont {Kawakami}, \citenamefont
  {Avdeev}, \citenamefont {Matsubayashi}, \citenamefont {Du}, \citenamefont
  {Soeda}, \citenamefont {Nakano}, \citenamefont {Uwatoko}, \citenamefont
  {Ueda}, \citenamefont {Sato},\ and\ \citenamefont
  {Ohgushi}}]{TakahashiSC2015}%
  \BibitemOpen
  \bibfield  {author} {\bibinfo {author} {\bibfnamefont {H.}~\bibnamefont
  {Takahashi}}, \bibinfo {author} {\bibfnamefont {A.}~\bibnamefont {Sugimoto}},
  \bibinfo {author} {\bibfnamefont {Y.}~\bibnamefont {Nambu}}, \bibinfo
  {author} {\bibfnamefont {T.}~\bibnamefont {Yamauchi}}, \bibinfo {author}
  {\bibfnamefont {Y.}~\bibnamefont {Hirata}}, \bibinfo {author} {\bibfnamefont
  {T.}~\bibnamefont {Kawakami}}, \bibinfo {author} {\bibfnamefont
  {M.}~\bibnamefont {Avdeev}}, \bibinfo {author} {\bibfnamefont
  {K.}~\bibnamefont {Matsubayashi}}, \bibinfo {author} {\bibfnamefont
  {C.}~\bibnamefont {Du}, \bibfnamefont {F.~Kawashima}}, \bibinfo {author}
  {\bibfnamefont {H.}~\bibnamefont {Soeda}}, \bibinfo {author} {\bibfnamefont
  {S.}~\bibnamefont {Nakano}}, \bibinfo {author} {\bibfnamefont
  {Y.}~\bibnamefont {Uwatoko}}, \bibinfo {author} {\bibfnamefont
  {Y.}~\bibnamefont {Ueda}}, \bibinfo {author} {\bibfnamefont {T.~J.}\
  \bibnamefont {Sato}},\ and\ \bibinfo {author} {\bibfnamefont
  {K.}~\bibnamefont {Ohgushi}},\ }\href {https://doi.org/10.1038/nmat4351}
  {\bibfield  {journal} {\bibinfo  {journal} {Nat. Mater.}\ }\textbf {\bibinfo
  {volume} {14}},\ \bibinfo {pages} {1008} (\bibinfo {year}
  {2015})}\BibitemShut {NoStop}%
\bibitem [{\citenamefont {Ying}\ \emph {et~al.}(2017)\citenamefont {Ying},
  \citenamefont {Lei}, \citenamefont {Petrovic}, \citenamefont {Xiao},\ and\
  \citenamefont {Struzhkin}}]{YingSC2017}%
  \BibitemOpen
  \bibfield  {author} {\bibinfo {author} {\bibfnamefont {J.}~\bibnamefont
  {Ying}}, \bibinfo {author} {\bibfnamefont {H.}~\bibnamefont {Lei}}, \bibinfo
  {author} {\bibfnamefont {C.}~\bibnamefont {Petrovic}}, \bibinfo {author}
  {\bibfnamefont {Y.}~\bibnamefont {Xiao}},\ and\ \bibinfo {author}
  {\bibfnamefont {V.~V.}\ \bibnamefont {Struzhkin}},\ }\href
  {https://doi.org/10.1103/PhysRevB.95.241109} {\bibfield  {journal} {\bibinfo
  {journal} {Phys. Rev. B}\ }\textbf {\bibinfo {volume} {95}},\ \bibinfo
  {pages} {241109(R)} (\bibinfo {year} {2017})}\BibitemShut {NoStop}%
\bibitem [{\citenamefont {Wu}\ \emph {et~al.}(2019{\natexlab{a}})\citenamefont
  {Wu}, \citenamefont {Frandsen}, \citenamefont {Wang}, \citenamefont {Yi},\
  and\ \citenamefont {Birgeneau}}]{WuBaFeX2018}%
  \BibitemOpen
  \bibfield  {author} {\bibinfo {author} {\bibfnamefont {S.}~\bibnamefont
  {Wu}}, \bibinfo {author} {\bibfnamefont {B.~A.}\ \bibnamefont {Frandsen}},
  \bibinfo {author} {\bibfnamefont {M.}~\bibnamefont {Wang}}, \bibinfo {author}
  {\bibfnamefont {M.}~\bibnamefont {Yi}},\ and\ \bibinfo {author}
  {\bibfnamefont {R.}~\bibnamefont {Birgeneau}},\ }\href
  {https://doi.org/10.1007/s10948-019-05304-4} {\bibfield  {journal} {\bibinfo
  {journal} {J. Supercond. Nov. Magn.}\ }\textbf {\bibinfo {volume} {33}},\
  \bibinfo {pages} {143} (\bibinfo {year} {2019}{\natexlab{a}})}\BibitemShut
  {NoStop}%
\bibitem [{\citenamefont {Wang}\ \emph {et~al.}(2017)\citenamefont {Wang},
  \citenamefont {Jin}, \citenamefont {Yi}, \citenamefont {Song}, \citenamefont
  {Jiang}, \citenamefont {Zhang}, \citenamefont {Sun}, \citenamefont {Luo},
  \citenamefont {Christianson}, \citenamefont {{Bourret-Courchesne}},
  \citenamefont {Lee}, \citenamefont {Yao},\ and\ \citenamefont
  {Birgeneau}}]{WangBaFe2S32017}%
  \BibitemOpen
  \bibfield  {author} {\bibinfo {author} {\bibfnamefont {M.}~\bibnamefont
  {Wang}}, \bibinfo {author} {\bibfnamefont {S.~J.}\ \bibnamefont {Jin}},
  \bibinfo {author} {\bibfnamefont {M.}~\bibnamefont {Yi}}, \bibinfo {author}
  {\bibfnamefont {Y.}~\bibnamefont {Song}}, \bibinfo {author} {\bibfnamefont
  {H.~C.}\ \bibnamefont {Jiang}}, \bibinfo {author} {\bibfnamefont {W.~L.}\
  \bibnamefont {Zhang}}, \bibinfo {author} {\bibfnamefont {H.~L.}\ \bibnamefont
  {Sun}}, \bibinfo {author} {\bibfnamefont {H.~Q.}\ \bibnamefont {Luo}},
  \bibinfo {author} {\bibfnamefont {A.~D.}\ \bibnamefont {Christianson}},
  \bibinfo {author} {\bibfnamefont {E.}~\bibnamefont {{Bourret-Courchesne}}},
  \bibinfo {author} {\bibfnamefont {D.~H.}\ \bibnamefont {Lee}}, \bibinfo
  {author} {\bibfnamefont {D.-X.}\ \bibnamefont {Yao}},\ and\ \bibinfo {author}
  {\bibfnamefont {R.~J.}\ \bibnamefont {Birgeneau}},\ }\href
  {https://doi.org/10.1103/PhysRevB.95.060502} {\bibfield  {journal} {\bibinfo
  {journal} {Phys. Rev. B}\ }\textbf {\bibinfo {volume} {95}},\ \bibinfo
  {pages} {060502(R)} (\bibinfo {year} {2017})}\BibitemShut {NoStop}%
\bibitem [{\citenamefont {Hawai}\ \emph {et~al.}(2015)\citenamefont {Hawai},
  \citenamefont {Nambu}, \citenamefont {Ohgushi}, \citenamefont {Du},
  \citenamefont {Hirata}, \citenamefont {Avdeev}, \citenamefont {Uwatoko},
  \citenamefont {Sekine}, \citenamefont {Fukazawa}, \citenamefont {Ma},
  \citenamefont {Chi}, \citenamefont {Ueda}, \citenamefont {Yoshizawa},\ and\
  \citenamefont {Sato}}]{HawaiBaFe2Se32015}%
  \BibitemOpen
  \bibfield  {author} {\bibinfo {author} {\bibfnamefont {T.}~\bibnamefont
  {Hawai}}, \bibinfo {author} {\bibfnamefont {Y.}~\bibnamefont {Nambu}},
  \bibinfo {author} {\bibfnamefont {K.}~\bibnamefont {Ohgushi}}, \bibinfo
  {author} {\bibfnamefont {F.}~\bibnamefont {Du}}, \bibinfo {author}
  {\bibfnamefont {Y.}~\bibnamefont {Hirata}}, \bibinfo {author} {\bibfnamefont
  {M.}~\bibnamefont {Avdeev}}, \bibinfo {author} {\bibfnamefont
  {Y.}~\bibnamefont {Uwatoko}}, \bibinfo {author} {\bibfnamefont
  {Y.}~\bibnamefont {Sekine}}, \bibinfo {author} {\bibfnamefont
  {H.}~\bibnamefont {Fukazawa}}, \bibinfo {author} {\bibfnamefont
  {J.}~\bibnamefont {Ma}}, \bibinfo {author} {\bibfnamefont {S.}~\bibnamefont
  {Chi}}, \bibinfo {author} {\bibfnamefont {Y.}~\bibnamefont {Ueda}}, \bibinfo
  {author} {\bibfnamefont {H.}~\bibnamefont {Yoshizawa}},\ and\ \bibinfo
  {author} {\bibfnamefont {T.~J.}\ \bibnamefont {Sato}},\ }\href
  {https://doi.org/10.1103/PhysRevB.91.184416} {\bibfield  {journal} {\bibinfo
  {journal} {Phys. Rev. B}\ }\textbf {\bibinfo {volume} {91}},\ \bibinfo
  {pages} {184416} (\bibinfo {year} {2015})}\BibitemShut {NoStop}%
\bibitem [{\citenamefont {Wang}\ \emph {et~al.}(2016)\citenamefont {Wang},
  \citenamefont {Yi}, \citenamefont {Jin}, \citenamefont {Jiang}, \citenamefont
  {Song}, \citenamefont {Luo}, \citenamefont {Christianson}, \citenamefont {{de
  la}~Cruz}, \citenamefont {Bourret-Courchesne}, \citenamefont {Yao},
  \citenamefont {Lee},\ and\ \citenamefont {Birgeneau}}]{WangRbFe2Se32016}%
  \BibitemOpen
  \bibfield  {author} {\bibinfo {author} {\bibfnamefont {M.}~\bibnamefont
  {Wang}}, \bibinfo {author} {\bibfnamefont {M.}~\bibnamefont {Yi}}, \bibinfo
  {author} {\bibfnamefont {S.}~\bibnamefont {Jin}}, \bibinfo {author}
  {\bibfnamefont {H.}~\bibnamefont {Jiang}}, \bibinfo {author} {\bibfnamefont
  {Y.}~\bibnamefont {Song}}, \bibinfo {author} {\bibfnamefont {H.}~\bibnamefont
  {Luo}}, \bibinfo {author} {\bibfnamefont {A.~D.}\ \bibnamefont
  {Christianson}}, \bibinfo {author} {\bibfnamefont {C.}~\bibnamefont {{de
  la}~Cruz}}, \bibinfo {author} {\bibfnamefont {E.}~\bibnamefont
  {Bourret-Courchesne}}, \bibinfo {author} {\bibfnamefont {D.-X.}\ \bibnamefont
  {Yao}}, \bibinfo {author} {\bibfnamefont {D.~H.}\ \bibnamefont {Lee}},\ and\
  \bibinfo {author} {\bibfnamefont {R.~J.}\ \bibnamefont {Birgeneau}},\ }\href
  {https://doi.org/10.1103/PhysRevB.94.041111} {\bibfield  {journal} {\bibinfo
  {journal} {Phys. Rev. B}\ }\textbf {\bibinfo {volume} {94}},\ \bibinfo
  {pages} {041111(R)} (\bibinfo {year} {2016})}\BibitemShut {NoStop}%
\bibitem [{\citenamefont {Chi}\ \emph {et~al.}(2016)\citenamefont {Chi},
  \citenamefont {Uwatoko}, \citenamefont {Cao}, \citenamefont {Hirata},
  \citenamefont {Hashizume}, \citenamefont {Aoyama},\ and\ \citenamefont
  {Ohgushi}}]{ChiCsFe2Se32016}%
  \BibitemOpen
  \bibfield  {author} {\bibinfo {author} {\bibfnamefont {S.}~\bibnamefont
  {Chi}}, \bibinfo {author} {\bibfnamefont {Y.}~\bibnamefont {Uwatoko}},
  \bibinfo {author} {\bibfnamefont {H.}~\bibnamefont {Cao}}, \bibinfo {author}
  {\bibfnamefont {Y.}~\bibnamefont {Hirata}}, \bibinfo {author} {\bibfnamefont
  {K.}~\bibnamefont {Hashizume}}, \bibinfo {author} {\bibfnamefont
  {T.}~\bibnamefont {Aoyama}},\ and\ \bibinfo {author} {\bibfnamefont
  {K.}~\bibnamefont {Ohgushi}},\ }\href
  {https://doi.org/10.1103/PhysRevLett.117.047003} {\bibfield  {journal}
  {\bibinfo  {journal} {Phys. Rev. Lett.}\ }\textbf {\bibinfo {volume} {117}},\
  \bibinfo {pages} {047003} (\bibinfo {year} {2016})}\BibitemShut {NoStop}%
\bibitem [{\citenamefont {Murase}\ \emph {et~al.}(2020)\citenamefont {Murase},
  \citenamefont {Okada}, \citenamefont {Kobayashi}, \citenamefont {Hirata},
  \citenamefont {Hashizume}, \citenamefont {Aoyama}, \citenamefont {Ohgushi},\
  and\ \citenamefont {Itoh}}]{MuraseCsFe2Se32020}%
  \BibitemOpen
  \bibfield  {author} {\bibinfo {author} {\bibfnamefont {M.}~\bibnamefont
  {Murase}}, \bibinfo {author} {\bibfnamefont {K.}~\bibnamefont {Okada}},
  \bibinfo {author} {\bibfnamefont {Y.}~\bibnamefont {Kobayashi}}, \bibinfo
  {author} {\bibfnamefont {Y.}~\bibnamefont {Hirata}}, \bibinfo {author}
  {\bibfnamefont {K.}~\bibnamefont {Hashizume}}, \bibinfo {author}
  {\bibfnamefont {T.}~\bibnamefont {Aoyama}}, \bibinfo {author} {\bibfnamefont
  {K.}~\bibnamefont {Ohgushi}},\ and\ \bibinfo {author} {\bibfnamefont
  {M.}~\bibnamefont {Itoh}},\ }\href
  {https://doi.org/10.1103/PhysRevB.102.014433} {\bibfield  {journal} {\bibinfo
   {journal} {Phys. Rev. B}\ }\textbf {\bibinfo {volume} {102}},\ \bibinfo
  {pages} {014433} (\bibinfo {year} {2020})}\BibitemShut {NoStop}%
\bibitem [{\citenamefont {Mourigal}\ \emph {et~al.}(2015)\citenamefont
  {Mourigal}, \citenamefont {Wu}, \citenamefont {Stone}, \citenamefont
  {Neilson}, \citenamefont {Caron}, \citenamefont {McQueen},\ and\
  \citenamefont {Broholm}}]{MourigalINS2015}%
  \BibitemOpen
  \bibfield  {author} {\bibinfo {author} {\bibfnamefont {M.}~\bibnamefont
  {Mourigal}}, \bibinfo {author} {\bibfnamefont {S.}~\bibnamefont {Wu}},
  \bibinfo {author} {\bibfnamefont {M.~B.}\ \bibnamefont {Stone}}, \bibinfo
  {author} {\bibfnamefont {J.~R.}\ \bibnamefont {Neilson}}, \bibinfo {author}
  {\bibfnamefont {J.~M.}\ \bibnamefont {Caron}}, \bibinfo {author}
  {\bibfnamefont {T.~M.}\ \bibnamefont {McQueen}},\ and\ \bibinfo {author}
  {\bibfnamefont {C.~L.}\ \bibnamefont {Broholm}},\ }\href
  {https://doi.org/10.1103/PhysRevLett.115.047401} {\bibfield  {journal}
  {\bibinfo  {journal} {Phys. Rev. Lett.}\ }\textbf {\bibinfo {volume} {115}},\
  \bibinfo {pages} {047401} (\bibinfo {year} {2015})}\BibitemShut {NoStop}%
\bibitem [{\citenamefont {Wu}\ \emph {et~al.}(2019{\natexlab{b}})\citenamefont
  {Wu}, \citenamefont {Yin}, \citenamefont {Smart}, \citenamefont {Acharya},
  \citenamefont {Bull}, \citenamefont {Funnell}, \citenamefont {Forrest},
  \citenamefont {Simutis}, \citenamefont {Khasanov}, \citenamefont {Lewin},
  \citenamefont {Wang}, \citenamefont {Frandsen}, \citenamefont {Jeanloz},\
  and\ \citenamefont {Birgeneau}}]{WuDiff2019}%
  \BibitemOpen
  \bibfield  {author} {\bibinfo {author} {\bibfnamefont {S.}~\bibnamefont
  {Wu}}, \bibinfo {author} {\bibfnamefont {J.}~\bibnamefont {Yin}}, \bibinfo
  {author} {\bibfnamefont {T.}~\bibnamefont {Smart}}, \bibinfo {author}
  {\bibfnamefont {A.}~\bibnamefont {Acharya}}, \bibinfo {author} {\bibfnamefont
  {C.~L.}\ \bibnamefont {Bull}}, \bibinfo {author} {\bibfnamefont {N.~P.}\
  \bibnamefont {Funnell}}, \bibinfo {author} {\bibfnamefont {T.~R.}\
  \bibnamefont {Forrest}}, \bibinfo {author} {\bibfnamefont {G.}~\bibnamefont
  {Simutis}}, \bibinfo {author} {\bibfnamefont {R.}~\bibnamefont {Khasanov}},
  \bibinfo {author} {\bibfnamefont {S.~K.}\ \bibnamefont {Lewin}}, \bibinfo
  {author} {\bibfnamefont {M.}~\bibnamefont {Wang}}, \bibinfo {author}
  {\bibfnamefont {B.~A.}\ \bibnamefont {Frandsen}}, \bibinfo {author}
  {\bibfnamefont {R.}~\bibnamefont {Jeanloz}},\ and\ \bibinfo {author}
  {\bibfnamefont {R.~J.}\ \bibnamefont {Birgeneau}},\ }\href
  {https://doi.org/10.1103/PhysRevB.100.214511} {\bibfield  {journal} {\bibinfo
   {journal} {Phys. Rev. B}\ }\textbf {\bibinfo {volume} {100}},\ \bibinfo
  {pages} {214511} (\bibinfo {year} {2019}{\natexlab{b}})}\BibitemShut
  {NoStop}%
\bibitem [{\citenamefont {Caron}\ \emph {et~al.}(2011)\citenamefont {Caron},
  \citenamefont {Neilson}, \citenamefont {Miller}, \citenamefont {Llobet},\
  and\ \citenamefont {McQueen}}]{CaronNPD2011}%
  \BibitemOpen
  \bibfield  {author} {\bibinfo {author} {\bibfnamefont {J.~M.}\ \bibnamefont
  {Caron}}, \bibinfo {author} {\bibfnamefont {J.~R.}\ \bibnamefont {Neilson}},
  \bibinfo {author} {\bibfnamefont {D.~C.}\ \bibnamefont {Miller}}, \bibinfo
  {author} {\bibfnamefont {A.}~\bibnamefont {Llobet}},\ and\ \bibinfo {author}
  {\bibfnamefont {T.~M.}\ \bibnamefont {McQueen}},\ }\href
  {https://doi.org/10.1103/PhysRevB.84.180409} {\bibfield  {journal} {\bibinfo
  {journal} {Phys. Rev. B}\ }\textbf {\bibinfo {volume} {84}},\ \bibinfo
  {pages} {180409(R)} (\bibinfo {year} {2011})}\BibitemShut {NoStop}%
\bibitem [{\citenamefont {Nambu}\ \emph {et~al.}(2012)\citenamefont {Nambu},
  \citenamefont {Ohgushi}, \citenamefont {Suzuki}, \citenamefont {Du},
  \citenamefont {Avdeev}, \citenamefont {Uwatoko}, \citenamefont {Munakata},
  \citenamefont {Fukazawa}, \citenamefont {Chi}, \citenamefont {Ueda},\ and\
  \citenamefont {Sato}}]{NambuNPD2012}%
  \BibitemOpen
  \bibfield  {author} {\bibinfo {author} {\bibfnamefont {Y.}~\bibnamefont
  {Nambu}}, \bibinfo {author} {\bibfnamefont {K.}~\bibnamefont {Ohgushi}},
  \bibinfo {author} {\bibfnamefont {S.}~\bibnamefont {Suzuki}}, \bibinfo
  {author} {\bibfnamefont {F.}~\bibnamefont {Du}}, \bibinfo {author}
  {\bibfnamefont {M.}~\bibnamefont {Avdeev}}, \bibinfo {author} {\bibfnamefont
  {Y.}~\bibnamefont {Uwatoko}}, \bibinfo {author} {\bibfnamefont
  {K.}~\bibnamefont {Munakata}}, \bibinfo {author} {\bibfnamefont
  {H.}~\bibnamefont {Fukazawa}}, \bibinfo {author} {\bibfnamefont
  {S.}~\bibnamefont {Chi}}, \bibinfo {author} {\bibfnamefont {Y.}~\bibnamefont
  {Ueda}},\ and\ \bibinfo {author} {\bibfnamefont {T.~J.}\ \bibnamefont
  {Sato}},\ }\href {https://doi.org/10.1103/PhysRevB.85.064413} {\bibfield
  {journal} {\bibinfo  {journal} {Phys. Rev. B}\ }\textbf {\bibinfo {volume}
  {85}},\ \bibinfo {pages} {064413} (\bibinfo {year} {2012})}\BibitemShut
  {NoStop}%
\bibitem [{\citenamefont {Rinc{\'o}n}\ \emph
  {et~al.}(2014{\natexlab{a}})\citenamefont {Rinc{\'o}n}, \citenamefont
  {Moreo}, \citenamefont {Alvarez},\ and\ \citenamefont
  {Dagotto}}]{RinconBlock2014}%
  \BibitemOpen
  \bibfield  {author} {\bibinfo {author} {\bibfnamefont {J.}~\bibnamefont
  {Rinc{\'o}n}}, \bibinfo {author} {\bibfnamefont {A.}~\bibnamefont {Moreo}},
  \bibinfo {author} {\bibfnamefont {G.}~\bibnamefont {Alvarez}},\ and\ \bibinfo
  {author} {\bibfnamefont {E.}~\bibnamefont {Dagotto}},\ }\href
  {https://doi.org/10.1103/PhysRevLett.112.106405} {\bibfield  {journal}
  {\bibinfo  {journal} {Phys. Rev. Lett.}\ }\textbf {\bibinfo {volume} {112}},\
  \bibinfo {pages} {106405} (\bibinfo {year} {2014}{\natexlab{a}})}\BibitemShut
  {NoStop}%
\bibitem [{\citenamefont {Glasbrenner}\ \emph {et~al.}(2015)\citenamefont
  {Glasbrenner}, \citenamefont {Mazin}, \citenamefont {Jeschke}, \citenamefont
  {Hirschfeld}, \citenamefont {Fernandes},\ and\ \citenamefont
  {Valent{\'i}}}]{GlasbrennerFeSe2015}%
  \BibitemOpen
  \bibfield  {author} {\bibinfo {author} {\bibfnamefont {J.~K.}\ \bibnamefont
  {Glasbrenner}}, \bibinfo {author} {\bibfnamefont {I.~I.}\ \bibnamefont
  {Mazin}}, \bibinfo {author} {\bibfnamefont {H.~O.}\ \bibnamefont {Jeschke}},
  \bibinfo {author} {\bibfnamefont {P.~J.}\ \bibnamefont {Hirschfeld}},
  \bibinfo {author} {\bibfnamefont {R.~M.}\ \bibnamefont {Fernandes}},\ and\
  \bibinfo {author} {\bibfnamefont {R.}~\bibnamefont {Valent{\'i}}},\ }\href
  {https://doi.org/10.1038/nphys3434} {\bibfield  {journal} {\bibinfo
  {journal} {Nat. Phys.}\ }\textbf {\bibinfo {volume} {11}},\ \bibinfo {pages}
  {953} (\bibinfo {year} {2015})}\BibitemShut {NoStop}%
\bibitem [{\citenamefont {Wang}\ \emph {et~al.}(2015)\citenamefont {Wang},
  \citenamefont {Fang}, \citenamefont {Yao}, \citenamefont {Tan}, \citenamefont
  {Harriger}, \citenamefont {Song}, \citenamefont {Netherton}, \citenamefont
  {Zhang}, \citenamefont {Wang}, \citenamefont {Stone}, \citenamefont {Tian},
  \citenamefont {Hu},\ and\ \citenamefont {Dai}}]{WangReFeSe2015}%
  \BibitemOpen
  \bibfield  {author} {\bibinfo {author} {\bibfnamefont {M.}~\bibnamefont
  {Wang}}, \bibinfo {author} {\bibfnamefont {C.}~\bibnamefont {Fang}}, \bibinfo
  {author} {\bibfnamefont {D.-X.}\ \bibnamefont {Yao}}, \bibinfo {author}
  {\bibfnamefont {G.~T.}\ \bibnamefont {Tan}}, \bibinfo {author} {\bibfnamefont
  {L.~W.}\ \bibnamefont {Harriger}}, \bibinfo {author} {\bibfnamefont
  {Y.}~\bibnamefont {Song}}, \bibinfo {author} {\bibfnamefont {T.}~\bibnamefont
  {Netherton}}, \bibinfo {author} {\bibfnamefont {C.}~\bibnamefont {Zhang}},
  \bibinfo {author} {\bibfnamefont {M.}~\bibnamefont {Wang}}, \bibinfo {author}
  {\bibfnamefont {M.~B.}\ \bibnamefont {Stone}}, \bibinfo {author}
  {\bibfnamefont {W.}~\bibnamefont {Tian}}, \bibinfo {author} {\bibfnamefont
  {J.}~\bibnamefont {Hu}},\ and\ \bibinfo {author} {\bibfnamefont
  {P.}~\bibnamefont {Dai}},\ }\href {https://doi.org/10.1038/ncomms1573}
  {\bibfield  {journal} {\bibinfo  {journal} {Nat. Commun.}\ }\textbf {\bibinfo
  {volume} {2}},\ \bibinfo {pages} {580} (\bibinfo {year} {2015})}\BibitemShut
  {NoStop}%
\bibitem [{\citenamefont {Wei}\ \emph {et~al.}(2011)\citenamefont {Wei},
  \citenamefont {Huang}, \citenamefont {Chen}, \citenamefont {Wang},
  \citenamefont {He},\ and\ \citenamefont {Qiu}}]{BaoKFeSe2011}%
  \BibitemOpen
  \bibfield  {author} {\bibinfo {author} {\bibfnamefont {B.}~\bibnamefont
  {Wei}}, \bibinfo {author} {\bibfnamefont {Q.-Z.}\ \bibnamefont {Huang}},
  \bibinfo {author} {\bibfnamefont {G.-F.}\ \bibnamefont {Chen}}, \bibinfo
  {author} {\bibfnamefont {D.-M.}\ \bibnamefont {Wang}}, \bibinfo {author}
  {\bibfnamefont {J.-B.}\ \bibnamefont {He}},\ and\ \bibinfo {author}
  {\bibfnamefont {Y.-M.}\ \bibnamefont {Qiu}},\ }\href
  {https://doi.org/10.1088/0256-307x/28/8/086104} {\bibfield  {journal}
  {\bibinfo  {journal} {Chin. Phys. Lett.}\ }\textbf {\bibinfo {volume} {28}},\
  \bibinfo {pages} {086104} (\bibinfo {year} {2011})}\BibitemShut {NoStop}%
\bibitem [{\citenamefont {You}\ \emph {et~al.}(2011)\citenamefont {You},
  \citenamefont {Yao},\ and\ \citenamefont {Lee}}]{YouKFeSe2011}%
  \BibitemOpen
  \bibfield  {author} {\bibinfo {author} {\bibfnamefont {Y.-Z.}\ \bibnamefont
  {You}}, \bibinfo {author} {\bibfnamefont {H.}~\bibnamefont {Yao}},\ and\
  \bibinfo {author} {\bibfnamefont {D.-H.}\ \bibnamefont {Lee}},\ }\href
  {https://doi.org/10.1103/PhysRevB.84.020406} {\bibfield  {journal} {\bibinfo
  {journal} {Phys. Rev. B}\ }\textbf {\bibinfo {volume} {84}},\ \bibinfo
  {pages} {020406(R)} (\bibinfo {year} {2011})}\BibitemShut {NoStop}%
\bibitem [{\citenamefont {Yu}\ \emph {et~al.}(2011)\citenamefont {Yu},
  \citenamefont {Goswami},\ and\ \citenamefont {Si}}]{YuKFeSe2011}%
  \BibitemOpen
  \bibfield  {author} {\bibinfo {author} {\bibfnamefont {R.}~\bibnamefont
  {Yu}}, \bibinfo {author} {\bibfnamefont {P.}~\bibnamefont {Goswami}},\ and\
  \bibinfo {author} {\bibfnamefont {Q.}~\bibnamefont {Si}},\ }\href
  {https://doi.org/10.1103/PhysRevB.84.094451} {\bibfield  {journal} {\bibinfo
  {journal} {Phys. Rev. B}\ }\textbf {\bibinfo {volume} {84}},\ \bibinfo
  {pages} {094451} (\bibinfo {year} {2011})}\BibitemShut {NoStop}%
\bibitem [{\citenamefont {Guo}\ \emph {et~al.}(2010)\citenamefont {Guo},
  \citenamefont {Jin}, \citenamefont {Wang}, \citenamefont {Wang},
  \citenamefont {Zhu}, \citenamefont {Zhou}, \citenamefont {He},\ and\
  \citenamefont {Chen}}]{GuKFe2Se22010}%
  \BibitemOpen
  \bibfield  {author} {\bibinfo {author} {\bibfnamefont {J.}~\bibnamefont
  {Guo}}, \bibinfo {author} {\bibfnamefont {S.}~\bibnamefont {Jin}}, \bibinfo
  {author} {\bibfnamefont {G.}~\bibnamefont {Wang}}, \bibinfo {author}
  {\bibfnamefont {S.}~\bibnamefont {Wang}}, \bibinfo {author} {\bibfnamefont
  {K.}~\bibnamefont {Zhu}}, \bibinfo {author} {\bibfnamefont {T.}~\bibnamefont
  {Zhou}}, \bibinfo {author} {\bibfnamefont {M.}~\bibnamefont {He}},\ and\
  \bibinfo {author} {\bibfnamefont {X.}~\bibnamefont {Chen}},\ }\href
  {https://doi.org/10.1103/PhysRevB.82.180520} {\bibfield  {journal} {\bibinfo
  {journal} {Phys. Rev. B}\ }\textbf {\bibinfo {volume} {82}},\ \bibinfo
  {pages} {180520(R)} (\bibinfo {year} {2010})}\BibitemShut {NoStop}%
\bibitem [{\citenamefont {Ye}\ \emph {et~al.}(2011)\citenamefont {Ye},
  \citenamefont {Chi}, \citenamefont {Bao}, \citenamefont {Wang}, \citenamefont
  {Ying}, \citenamefont {Chen}, \citenamefont {Wang}, \citenamefont {Dong},\
  and\ \citenamefont {Fang}}]{YeKFe2Se22011}%
  \BibitemOpen
  \bibfield  {author} {\bibinfo {author} {\bibfnamefont {F.}~\bibnamefont
  {Ye}}, \bibinfo {author} {\bibfnamefont {S.}~\bibnamefont {Chi}}, \bibinfo
  {author} {\bibfnamefont {W.}~\bibnamefont {Bao}}, \bibinfo {author}
  {\bibfnamefont {X.~F.}\ \bibnamefont {Wang}}, \bibinfo {author}
  {\bibfnamefont {J.~J.}\ \bibnamefont {Ying}}, \bibinfo {author}
  {\bibfnamefont {X.~H.}\ \bibnamefont {Chen}}, \bibinfo {author}
  {\bibfnamefont {H.~D.}\ \bibnamefont {Wang}}, \bibinfo {author}
  {\bibfnamefont {C.~H.}\ \bibnamefont {Dong}},\ and\ \bibinfo {author}
  {\bibfnamefont {M.}~\bibnamefont {Fang}},\ }\href
  {https://doi.org/10.1103/PhysRevLett.107.137003} {\bibfield  {journal}
  {\bibinfo  {journal} {Phys. Rev. Lett.}\ }\textbf {\bibinfo {volume} {107}},\
  \bibinfo {pages} {137003} (\bibinfo {year} {2011})}\BibitemShut {NoStop}%
\bibitem [{\citenamefont {Pandey}\ \emph {et~al.}(2020)\citenamefont {Pandey},
  \citenamefont {Lin}, \citenamefont {Soni}, \citenamefont {Kaushal},
  \citenamefont {Herbrych}, \citenamefont {Alvarez},\ and\ \citenamefont
  {Dagotto}}]{PhysRevB.102.035149}%
  \BibitemOpen
  \bibfield  {author} {\bibinfo {author} {\bibfnamefont {B.}~\bibnamefont
  {Pandey}}, \bibinfo {author} {\bibfnamefont {L.-F.}\ \bibnamefont {Lin}},
  \bibinfo {author} {\bibfnamefont {R.}~\bibnamefont {Soni}}, \bibinfo {author}
  {\bibfnamefont {N.}~\bibnamefont {Kaushal}}, \bibinfo {author} {\bibfnamefont
  {J.}~\bibnamefont {Herbrych}}, \bibinfo {author} {\bibfnamefont
  {G.}~\bibnamefont {Alvarez}},\ and\ \bibinfo {author} {\bibfnamefont
  {E.}~\bibnamefont {Dagotto}},\ }\href
  {https://doi.org/10.1103/PhysRevB.102.035149} {\bibfield  {journal} {\bibinfo
   {journal} {Phys. Rev. B}\ }\textbf {\bibinfo {volume} {102}},\ \bibinfo
  {pages} {035149} (\bibinfo {year} {2020})}\BibitemShut {NoStop}%
\bibitem [{\citenamefont {Luo}\ \emph {et~al.}(2011)\citenamefont {Luo},
  \citenamefont {Nicholson}, \citenamefont {Riera}, \citenamefont {Yao},
  \citenamefont {Moreo},\ and\ \citenamefont {Dagotto}}]{LuoHF2011}%
  \BibitemOpen
  \bibfield  {author} {\bibinfo {author} {\bibfnamefont {Q.}~\bibnamefont
  {Luo}}, \bibinfo {author} {\bibfnamefont {A.}~\bibnamefont {Nicholson}},
  \bibinfo {author} {\bibfnamefont {J.}~\bibnamefont {Riera}}, \bibinfo
  {author} {\bibfnamefont {D.-X.}\ \bibnamefont {Yao}}, \bibinfo {author}
  {\bibfnamefont {A.}~\bibnamefont {Moreo}},\ and\ \bibinfo {author}
  {\bibfnamefont {E.}~\bibnamefont {Dagotto}},\ }\href
  {https://doi.org/10.1103/PhysRevB.84.140506} {\bibfield  {journal} {\bibinfo
  {journal} {Phys. Rev. B}\ }\textbf {\bibinfo {volume} {84}},\ \bibinfo
  {pages} {140506(R)} (\bibinfo {year} {2011})}\BibitemShut {NoStop}%
\bibitem [{\citenamefont {Luo}\ \emph {et~al.}(2013)\citenamefont {Luo},
  \citenamefont {Nicholson}, \citenamefont {Rinc{\'o}n}, \citenamefont {Liang},
  \citenamefont {Riera}, \citenamefont {Alvarez}, \citenamefont {Wang},
  \citenamefont {Ku}, \citenamefont {Samolyuk}, \citenamefont {Moreo},\ and\
  \citenamefont {Dagotto}}]{LuoDFT2013}%
  \BibitemOpen
  \bibfield  {author} {\bibinfo {author} {\bibfnamefont {Q.}~\bibnamefont
  {Luo}}, \bibinfo {author} {\bibfnamefont {A.}~\bibnamefont {Nicholson}},
  \bibinfo {author} {\bibfnamefont {J.}~\bibnamefont {Rinc{\'o}n}}, \bibinfo
  {author} {\bibfnamefont {S.}~\bibnamefont {Liang}}, \bibinfo {author}
  {\bibfnamefont {J.}~\bibnamefont {Riera}}, \bibinfo {author} {\bibfnamefont
  {G.}~\bibnamefont {Alvarez}}, \bibinfo {author} {\bibfnamefont
  {L.}~\bibnamefont {Wang}}, \bibinfo {author} {\bibfnamefont {W.}~\bibnamefont
  {Ku}}, \bibinfo {author} {\bibfnamefont {G.~D.}\ \bibnamefont {Samolyuk}},
  \bibinfo {author} {\bibfnamefont {A.}~\bibnamefont {Moreo}},\ and\ \bibinfo
  {author} {\bibfnamefont {E.}~\bibnamefont {Dagotto}},\ }\href
  {https://doi.org/10.1103/PhysRevB.87.024404} {\bibfield  {journal} {\bibinfo
  {journal} {Phys. Rev. B}\ }\textbf {\bibinfo {volume} {87}},\ \bibinfo
  {pages} {024404} (\bibinfo {year} {2013})}\BibitemShut {NoStop}%
\bibitem [{\citenamefont {Luo}\ \emph {et~al.}(2014)\citenamefont {Luo},
  \citenamefont {Foyevtsova}, \citenamefont {Samolyuk}, \citenamefont
  {Reboredo},\ and\ \citenamefont {Dagotto}}]{LuoMHF2014}%
  \BibitemOpen
  \bibfield  {author} {\bibinfo {author} {\bibfnamefont {Q.}~\bibnamefont
  {Luo}}, \bibinfo {author} {\bibfnamefont {K.}~\bibnamefont {Foyevtsova}},
  \bibinfo {author} {\bibfnamefont {G.~D.}\ \bibnamefont {Samolyuk}}, \bibinfo
  {author} {\bibfnamefont {F.}~\bibnamefont {Reboredo}},\ and\ \bibinfo
  {author} {\bibfnamefont {E.}~\bibnamefont {Dagotto}},\ }\href
  {https://doi.org/10.1103/PhysRevB.90.035128} {\bibfield  {journal} {\bibinfo
  {journal} {Phys. Rev. B}\ }\textbf {\bibinfo {volume} {90}},\ \bibinfo
  {pages} {035128} (\bibinfo {year} {2014})}\BibitemShut {NoStop}%
\bibitem [{\citenamefont {Luo}\ and\ \citenamefont
  {Dagotto}(2014)}]{LuoHF2014}%
  \BibitemOpen
  \bibfield  {author} {\bibinfo {author} {\bibfnamefont {Q.}~\bibnamefont
  {Luo}}\ and\ \bibinfo {author} {\bibfnamefont {E.}~\bibnamefont {Dagotto}},\
  }\href {https://doi.org/10.1103/PhysRevB.89.045115} {\bibfield  {journal}
  {\bibinfo  {journal} {Phys. Rev. B}\ }\textbf {\bibinfo {volume} {89}},\
  \bibinfo {pages} {045115} (\bibinfo {year} {2014})}\BibitemShut {NoStop}%
\bibitem [{\citenamefont {Yin}\ \emph {et~al.}(2012)\citenamefont {Yin},
  \citenamefont {Lin},\ and\ \citenamefont {Ku}}]{YinDFT2012}%
  \BibitemOpen
  \bibfield  {author} {\bibinfo {author} {\bibfnamefont {W.-G.}\ \bibnamefont
  {Yin}}, \bibinfo {author} {\bibfnamefont {C.-H.}\ \bibnamefont {Lin}},\ and\
  \bibinfo {author} {\bibfnamefont {W.}~\bibnamefont {Ku}},\ }\href
  {https://doi.org/10.1103/PhysRevB.86.081106} {\bibfield  {journal} {\bibinfo
  {journal} {Phys. Rev. B}\ }\textbf {\bibinfo {volume} {86}},\ \bibinfo
  {pages} {081106(R)} (\bibinfo {year} {2012})}\BibitemShut {NoStop}%
\bibitem [{\citenamefont {Dong}\ \emph {et~al.}(2014)\citenamefont {Dong},
  \citenamefont {Liu},\ and\ \citenamefont {Dagotto}}]{DongDFT2014}%
  \BibitemOpen
  \bibfield  {author} {\bibinfo {author} {\bibfnamefont {S.}~\bibnamefont
  {Dong}}, \bibinfo {author} {\bibfnamefont {J.-M.}\ \bibnamefont {Liu}},\ and\
  \bibinfo {author} {\bibfnamefont {E.}~\bibnamefont {Dagotto}},\ }\href
  {https://doi.org/10.1103/PhysRevLett.113.187204} {\bibfield  {journal}
  {\bibinfo  {journal} {Phys. Rev. Lett.}\ }\textbf {\bibinfo {volume} {113}},\
  \bibinfo {pages} {187204} (\bibinfo {year} {2014})}\BibitemShut {NoStop}%
\bibitem [{\citenamefont {Zhang}\ \emph {et~al.}(2017)\citenamefont {Zhang},
  \citenamefont {Lin}, \citenamefont {Zhang}, \citenamefont {Dagotto},\ and\
  \citenamefont {Dong}}]{ZhangDFT2017}%
  \BibitemOpen
  \bibfield  {author} {\bibinfo {author} {\bibfnamefont {Y.}~\bibnamefont
  {Zhang}}, \bibinfo {author} {\bibfnamefont {L.}~\bibnamefont {Lin}}, \bibinfo
  {author} {\bibfnamefont {J.-J.}\ \bibnamefont {Zhang}}, \bibinfo {author}
  {\bibfnamefont {E.}~\bibnamefont {Dagotto}},\ and\ \bibinfo {author}
  {\bibfnamefont {S.}~\bibnamefont {Dong}},\ }\href
  {https://doi.org/10.1103/PhysRevB.95.115154} {\bibfield  {journal} {\bibinfo
  {journal} {Phys. Rev. B}\ }\textbf {\bibinfo {volume} {95}},\ \bibinfo
  {pages} {115154} (\bibinfo {year} {2017})}\BibitemShut {NoStop}%
\bibitem [{\citenamefont {Zhang}\ \emph {et~al.}(2018)\citenamefont {Zhang},
  \citenamefont {Lin}, \citenamefont {Zhang}, \citenamefont {Dagotto},\ and\
  \citenamefont {Dong}}]{ZhangDFT2018}%
  \BibitemOpen
  \bibfield  {author} {\bibinfo {author} {\bibfnamefont {Y.}~\bibnamefont
  {Zhang}}, \bibinfo {author} {\bibfnamefont {L.-F.}\ \bibnamefont {Lin}},
  \bibinfo {author} {\bibfnamefont {J.-J.}\ \bibnamefont {Zhang}}, \bibinfo
  {author} {\bibfnamefont {E.}~\bibnamefont {Dagotto}},\ and\ \bibinfo {author}
  {\bibfnamefont {S.}~\bibnamefont {Dong}},\ }\href
  {https://doi.org/10.1103/PhysRevB.97.045119} {\bibfield  {journal} {\bibinfo
  {journal} {Phys. Rev. B}\ }\textbf {\bibinfo {volume} {97}},\ \bibinfo
  {pages} {045119} (\bibinfo {year} {2018})}\BibitemShut {NoStop}%
\bibitem [{\citenamefont {Zhang}\ \emph {et~al.}(2019)\citenamefont {Zhang},
  \citenamefont {Lin}, \citenamefont {Moreo}, \citenamefont {Dong},\ and\
  \citenamefont {Dagotto}}]{ZhangDFT2019}%
  \BibitemOpen
  \bibfield  {author} {\bibinfo {author} {\bibfnamefont {Y.}~\bibnamefont
  {Zhang}}, \bibinfo {author} {\bibfnamefont {L.-F.}\ \bibnamefont {Lin}},
  \bibinfo {author} {\bibfnamefont {A.}~\bibnamefont {Moreo}}, \bibinfo
  {author} {\bibfnamefont {S.}~\bibnamefont {Dong}},\ and\ \bibinfo {author}
  {\bibfnamefont {E.}~\bibnamefont {Dagotto}},\ }\href
  {https://doi.org/10.1103/PhysRevB.100.184419} {\bibfield  {journal} {\bibinfo
   {journal} {Phys. Rev. B}\ }\textbf {\bibinfo {volume} {100}},\ \bibinfo
  {pages} {184419} (\bibinfo {year} {2019})}\BibitemShut {NoStop}%
\bibitem [{\citenamefont {Zhang}\ \emph {et~al.}(2020)\citenamefont {Zhang},
  \citenamefont {Lin}, \citenamefont {Moreo}, \citenamefont {Dong},\ and\
  \citenamefont {Dagotto}}]{PhysRevB.101.144417}%
  \BibitemOpen
  \bibfield  {author} {\bibinfo {author} {\bibfnamefont {Y.}~\bibnamefont
  {Zhang}}, \bibinfo {author} {\bibfnamefont {L.-F.}\ \bibnamefont {Lin}},
  \bibinfo {author} {\bibfnamefont {A.}~\bibnamefont {Moreo}}, \bibinfo
  {author} {\bibfnamefont {S.}~\bibnamefont {Dong}},\ and\ \bibinfo {author}
  {\bibfnamefont {E.}~\bibnamefont {Dagotto}},\ }\href
  {https://doi.org/10.1103/PhysRevB.101.144417} {\bibfield  {journal} {\bibinfo
   {journal} {Phys. Rev. B}\ }\textbf {\bibinfo {volume} {101}},\ \bibinfo
  {pages} {144417} (\bibinfo {year} {2020})}\BibitemShut {NoStop}%
\bibitem [{\citenamefont {{de' Medici}}\ \emph {et~al.}(2009)\citenamefont
  {{de' Medici}}, \citenamefont {Hassan}, \citenamefont {Capone},\ and\
  \citenamefont {Dai}}]{MediciDMFT2009}%
  \BibitemOpen
  \bibfield  {author} {\bibinfo {author} {\bibfnamefont {L.}~\bibnamefont {{de'
  Medici}}}, \bibinfo {author} {\bibfnamefont {S.~R.}\ \bibnamefont {Hassan}},
  \bibinfo {author} {\bibfnamefont {M.}~\bibnamefont {Capone}},\ and\ \bibinfo
  {author} {\bibfnamefont {X.}~\bibnamefont {Dai}},\ }\href
  {https://doi.org/10.1103/PhysRevLett.102.126401} {\bibfield  {journal}
  {\bibinfo  {journal} {Phys. Rev. Lett.}\ }\textbf {\bibinfo {volume} {102}},\
  \bibinfo {pages} {126401} (\bibinfo {year} {2009})}\BibitemShut {NoStop}%
\bibitem [{\citenamefont {Georges}\ \emph {et~al.}(2013)\citenamefont
  {Georges}, \citenamefont {{de' Medici}},\ and\ \citenamefont
  {Mravlje}}]{GeorgesDMFT2013}%
  \BibitemOpen
  \bibfield  {author} {\bibinfo {author} {\bibfnamefont {A.}~\bibnamefont
  {Georges}}, \bibinfo {author} {\bibfnamefont {L.}~\bibnamefont {{de'
  Medici}}},\ and\ \bibinfo {author} {\bibfnamefont {J.}~\bibnamefont
  {Mravlje}},\ }\href
  {https://doi.org/10.1146/annurev-conmatphys-020911-125045} {\bibfield
  {journal} {\bibinfo  {journal} {Annu. Rev. Condens. Matter Phys.}\ }\textbf
  {\bibinfo {volume} {4}},\ \bibinfo {pages} {137} (\bibinfo {year}
  {2013})}\BibitemShut {NoStop}%
\bibitem [{\citenamefont {Isidori}\ \emph {et~al.}(2019)\citenamefont
  {Isidori}, \citenamefont {Berovi{\'c}}, \citenamefont {Fanfarillo},
  \citenamefont {{de' Medici}}, \citenamefont {Fabrizio},\ and\ \citenamefont
  {Capone}}]{IsidoriDMFT2019}%
  \BibitemOpen
  \bibfield  {author} {\bibinfo {author} {\bibfnamefont {A.}~\bibnamefont
  {Isidori}}, \bibinfo {author} {\bibfnamefont {M.}~\bibnamefont
  {Berovi{\'c}}}, \bibinfo {author} {\bibfnamefont {L.}~\bibnamefont
  {Fanfarillo}}, \bibinfo {author} {\bibfnamefont {L.}~\bibnamefont {{de'
  Medici}}}, \bibinfo {author} {\bibfnamefont {M.}~\bibnamefont {Fabrizio}},\
  and\ \bibinfo {author} {\bibfnamefont {M.}~\bibnamefont {Capone}},\ }\href
  {https://doi.org/10.1103/PhysRevLett.122.186401} {\bibfield  {journal}
  {\bibinfo  {journal} {Phys. Rev. Lett.}\ }\textbf {\bibinfo {volume} {122}},\
  \bibinfo {pages} {186401} (\bibinfo {year} {2019})}\BibitemShut {NoStop}%
\bibitem [{\citenamefont {Yi}\ \emph {et~al.}(2017)\citenamefont {Yi},
  \citenamefont {Zhang}, \citenamefont {Shen},\ and\ \citenamefont
  {Lu.}}]{YiOSMP2017}%
  \BibitemOpen
  \bibfield  {author} {\bibinfo {author} {\bibfnamefont {M.}~\bibnamefont
  {Yi}}, \bibinfo {author} {\bibfnamefont {Y.}~\bibnamefont {Zhang}}, \bibinfo
  {author} {\bibfnamefont {Z.-X.}\ \bibnamefont {Shen}},\ and\ \bibinfo
  {author} {\bibfnamefont {D.}~\bibnamefont {Lu.}},\ }\href
  {https://doi.org/10.1038/s41535-017-0059-y} {\bibfield  {journal} {\bibinfo
  {journal} {npj Quant. Mater.}\ }\textbf {\bibinfo {volume} {2}},\ \bibinfo
  {pages} {57} (\bibinfo {year} {2017})}\BibitemShut {NoStop}%
\bibitem [{\citenamefont {Caron}\ \emph {et~al.}(2012)\citenamefont {Caron},
  \citenamefont {Neilson}, \citenamefont {Miller}, \citenamefont {Arpino},
  \citenamefont {Llobet},\ and\ \citenamefont {McQueen}}]{CaronOSMP2012}%
  \BibitemOpen
  \bibfield  {author} {\bibinfo {author} {\bibfnamefont {J.~M.}\ \bibnamefont
  {Caron}}, \bibinfo {author} {\bibfnamefont {J.~R.}\ \bibnamefont {Neilson}},
  \bibinfo {author} {\bibfnamefont {D.~C.}\ \bibnamefont {Miller}}, \bibinfo
  {author} {\bibfnamefont {K.}~\bibnamefont {Arpino}}, \bibinfo {author}
  {\bibfnamefont {A.}~\bibnamefont {Llobet}},\ and\ \bibinfo {author}
  {\bibfnamefont {T.~M.}\ \bibnamefont {McQueen}},\ }\href
  {https://doi.org/10.1103/PhysRevB.85.180405} {\bibfield  {journal} {\bibinfo
  {journal} {Phys. Rev. B}\ }\textbf {\bibinfo {volume} {85}},\ \bibinfo
  {pages} {180405(R)} (\bibinfo {year} {2012})}\BibitemShut {NoStop}%
\bibitem [{\citenamefont {Yamauchi}\ \emph {et~al.}(2015)\citenamefont
  {Yamauchi}, \citenamefont {Hirata}, \citenamefont {Ueda},\ and\ \citenamefont
  {Ohgushi}}]{YamauchiSC2015}%
  \BibitemOpen
  \bibfield  {author} {\bibinfo {author} {\bibfnamefont {T.}~\bibnamefont
  {Yamauchi}}, \bibinfo {author} {\bibfnamefont {Y.}~\bibnamefont {Hirata}},
  \bibinfo {author} {\bibfnamefont {Y.}~\bibnamefont {Ueda}},\ and\ \bibinfo
  {author} {\bibfnamefont {K.}~\bibnamefont {Ohgushi}},\ }\href
  {https://doi.org/10.1103/PhysRevLett.115.246402} {\bibfield  {journal}
  {\bibinfo  {journal} {Phys. Rev. Lett.}\ }\textbf {\bibinfo {volume} {115}},\
  \bibinfo {pages} {246402} (\bibinfo {year} {2015})}\BibitemShut {NoStop}%
\bibitem [{\citenamefont {Craco}\ and\ \citenamefont
  {Leoni}(2020)}]{CracoOSMP2020}%
  \BibitemOpen
  \bibfield  {author} {\bibinfo {author} {\bibfnamefont {L.}~\bibnamefont
  {Craco}}\ and\ \bibinfo {author} {\bibfnamefont {S.}~\bibnamefont {Leoni}},\
  }\href {https://doi.org/10.1103/PhysRevB.101.245133} {\bibfield  {journal}
  {\bibinfo  {journal} {Phys. Rev. B}\ }\textbf {\bibinfo {volume} {101}},\
  \bibinfo {pages} {245133} (\bibinfo {year} {2020})}\BibitemShut {NoStop}%
\bibitem [{\citenamefont {Yi}\ \emph {et~al.}(2013)\citenamefont {Yi},
  \citenamefont {Lu}, \citenamefont {Yu}, \citenamefont {Riggs}, \citenamefont
  {Chu}, \citenamefont {Lv}, \citenamefont {Liu}, \citenamefont {Lu},
  \citenamefont {Cui}, \citenamefont {Hashimoto}, \citenamefont {Mo},
  \citenamefont {Hussain}, \citenamefont {Chu}, \citenamefont {Fisher},
  \citenamefont {Si},\ and\ \citenamefont {Shen}}]{yiObservation2013}%
  \BibitemOpen
  \bibfield  {author} {\bibinfo {author} {\bibfnamefont {M.}~\bibnamefont
  {Yi}}, \bibinfo {author} {\bibfnamefont {D.~H.}\ \bibnamefont {Lu}}, \bibinfo
  {author} {\bibfnamefont {R.}~\bibnamefont {Yu}}, \bibinfo {author}
  {\bibfnamefont {S.~C.}\ \bibnamefont {Riggs}}, \bibinfo {author}
  {\bibfnamefont {J.-H.}\ \bibnamefont {Chu}}, \bibinfo {author} {\bibfnamefont
  {B.}~\bibnamefont {Lv}}, \bibinfo {author} {\bibfnamefont {Z.~K.}\
  \bibnamefont {Liu}}, \bibinfo {author} {\bibfnamefont {M.}~\bibnamefont
  {Lu}}, \bibinfo {author} {\bibfnamefont {Y.-T.}\ \bibnamefont {Cui}},
  \bibinfo {author} {\bibfnamefont {M.}~\bibnamefont {Hashimoto}}, \bibinfo
  {author} {\bibfnamefont {S.-K.}\ \bibnamefont {Mo}}, \bibinfo {author}
  {\bibfnamefont {Z.}~\bibnamefont {Hussain}}, \bibinfo {author} {\bibfnamefont
  {C.~W.}\ \bibnamefont {Chu}}, \bibinfo {author} {\bibfnamefont {I.~R.}\
  \bibnamefont {Fisher}}, \bibinfo {author} {\bibfnamefont {Q.}~\bibnamefont
  {Si}},\ and\ \bibinfo {author} {\bibfnamefont {Z.-X.}\ \bibnamefont {Shen}},\
  }\href {https://doi.org/10.1103/PhysRevLett.110.067003} {\bibfield  {journal}
  {\bibinfo  {journal} {Phys. Rev. Lett.}\ }\textbf {\bibinfo {volume} {110}},\
  \bibinfo {pages} {067003} (\bibinfo {year} {2013})}\BibitemShut {NoStop}%
\bibitem [{\citenamefont {Hardy}\ \emph {et~al.}(2013)\citenamefont {Hardy},
  \citenamefont {B{\"o}hmer}, \citenamefont {Aoki}, \citenamefont {Burger},
  \citenamefont {Wolf}, \citenamefont {Schweiss}, \citenamefont {Heid},
  \citenamefont {Adelmann}, \citenamefont {Yao}, \citenamefont {Kotliar},
  \citenamefont {Schmalian},\ and\ \citenamefont
  {Meingast}}]{hardyEvidence2013}%
  \BibitemOpen
  \bibfield  {author} {\bibinfo {author} {\bibfnamefont {F.}~\bibnamefont
  {Hardy}}, \bibinfo {author} {\bibfnamefont {A.~E.}\ \bibnamefont
  {B{\"o}hmer}}, \bibinfo {author} {\bibfnamefont {D.}~\bibnamefont {Aoki}},
  \bibinfo {author} {\bibfnamefont {P.}~\bibnamefont {Burger}}, \bibinfo
  {author} {\bibfnamefont {T.}~\bibnamefont {Wolf}}, \bibinfo {author}
  {\bibfnamefont {P.}~\bibnamefont {Schweiss}}, \bibinfo {author}
  {\bibfnamefont {R.}~\bibnamefont {Heid}}, \bibinfo {author} {\bibfnamefont
  {P.}~\bibnamefont {Adelmann}}, \bibinfo {author} {\bibfnamefont {Y.~X.}\
  \bibnamefont {Yao}}, \bibinfo {author} {\bibfnamefont {G.}~\bibnamefont
  {Kotliar}}, \bibinfo {author} {\bibfnamefont {J.}~\bibnamefont {Schmalian}},\
  and\ \bibinfo {author} {\bibfnamefont {C.}~\bibnamefont {Meingast}},\ }\href
  {https://doi.org/10.1103/PhysRevLett.111.027002} {\bibfield  {journal}
  {\bibinfo  {journal} {Phys. Rev. Lett.}\ }\textbf {\bibinfo {volume} {111}},\
  \bibinfo {pages} {027002} (\bibinfo {year} {2013})}\BibitemShut {NoStop}%
\bibitem [{\citenamefont {Yi}\ \emph {et~al.}(2015)\citenamefont {Yi},
  \citenamefont {Liu}, \citenamefont {Zhang}, \citenamefont {Yu}, \citenamefont
  {Zhu}, \citenamefont {Lee}, \citenamefont {Moore}, \citenamefont {Schmitt},
  \citenamefont {Li}, \citenamefont {Riggs}, \citenamefont {Chu}, \citenamefont
  {Lv}, \citenamefont {Hu}, \citenamefont {Hashimoto}, \citenamefont {Mo},
  \citenamefont {Hussain}, \citenamefont {Mao}, \citenamefont {Chu},
  \citenamefont {Fisher}, \citenamefont {Si}, \citenamefont {Shen},\ and\
  \citenamefont {Lu}}]{yiObservation2015}%
  \BibitemOpen
  \bibfield  {author} {\bibinfo {author} {\bibfnamefont {M.}~\bibnamefont
  {Yi}}, \bibinfo {author} {\bibfnamefont {Z.-K.}\ \bibnamefont {Liu}},
  \bibinfo {author} {\bibfnamefont {Y.}~\bibnamefont {Zhang}}, \bibinfo
  {author} {\bibfnamefont {R.}~\bibnamefont {Yu}}, \bibinfo {author}
  {\bibfnamefont {J.-X.}\ \bibnamefont {Zhu}}, \bibinfo {author} {\bibfnamefont
  {J.~J.}\ \bibnamefont {Lee}}, \bibinfo {author} {\bibfnamefont {R.~G.}\
  \bibnamefont {Moore}}, \bibinfo {author} {\bibfnamefont {F.~T.}\ \bibnamefont
  {Schmitt}}, \bibinfo {author} {\bibfnamefont {W.}~\bibnamefont {Li}},
  \bibinfo {author} {\bibfnamefont {S.~C.}\ \bibnamefont {Riggs}}, \bibinfo
  {author} {\bibfnamefont {J.-H.}\ \bibnamefont {Chu}}, \bibinfo {author}
  {\bibfnamefont {B.}~\bibnamefont {Lv}}, \bibinfo {author} {\bibfnamefont
  {J.}~\bibnamefont {Hu}}, \bibinfo {author} {\bibfnamefont {M.}~\bibnamefont
  {Hashimoto}}, \bibinfo {author} {\bibfnamefont {S.-K.}\ \bibnamefont {Mo}},
  \bibinfo {author} {\bibfnamefont {Z.}~\bibnamefont {Hussain}}, \bibinfo
  {author} {\bibfnamefont {Z.~Q.}\ \bibnamefont {Mao}}, \bibinfo {author}
  {\bibfnamefont {C.~W.}\ \bibnamefont {Chu}}, \bibinfo {author} {\bibfnamefont
  {I.~R.}\ \bibnamefont {Fisher}}, \bibinfo {author} {\bibfnamefont
  {Q.}~\bibnamefont {Si}}, \bibinfo {author} {\bibfnamefont {Z.-X.}\
  \bibnamefont {Shen}},\ and\ \bibinfo {author} {\bibfnamefont {D.~H.}\
  \bibnamefont {Lu}},\ }\href {https://doi.org/10.1038/ncomms8777} {\bibfield
  {journal} {\bibinfo  {journal} {Nat. Commun.}\ }\textbf {\bibinfo {volume}
  {6}},\ \bibinfo {pages} {7777} (\bibinfo {year} {2015})}\BibitemShut
  {NoStop}%
\bibitem [{\citenamefont {Zhu}\ \emph {et~al.}(2010)\citenamefont {Zhu},
  \citenamefont {Yu}, \citenamefont {Wang}, \citenamefont {Zhao}, \citenamefont
  {Jones}, \citenamefont {Dai}, \citenamefont {Abrahams}, \citenamefont
  {Morosan}, \citenamefont {Fang},\ and\ \citenamefont {Si}}]{zhuBand2010}%
  \BibitemOpen
  \bibfield  {author} {\bibinfo {author} {\bibfnamefont {J.-X.}\ \bibnamefont
  {Zhu}}, \bibinfo {author} {\bibfnamefont {R.}~\bibnamefont {Yu}}, \bibinfo
  {author} {\bibfnamefont {H.}~\bibnamefont {Wang}}, \bibinfo {author}
  {\bibfnamefont {L.~L.}\ \bibnamefont {Zhao}}, \bibinfo {author}
  {\bibfnamefont {M.~D.}\ \bibnamefont {Jones}}, \bibinfo {author}
  {\bibfnamefont {J.}~\bibnamefont {Dai}}, \bibinfo {author} {\bibfnamefont
  {E.}~\bibnamefont {Abrahams}}, \bibinfo {author} {\bibfnamefont
  {E.}~\bibnamefont {Morosan}}, \bibinfo {author} {\bibfnamefont
  {M.}~\bibnamefont {Fang}},\ and\ \bibinfo {author} {\bibfnamefont
  {Q.}~\bibnamefont {Si}},\ }\href
  {https://doi.org/10.1103/PhysRevLett.104.216405} {\bibfield  {journal}
  {\bibinfo  {journal} {Phys. Rev. Lett.}\ }\textbf {\bibinfo {volume} {104}},\
  \bibinfo {pages} {216405} (\bibinfo {year} {2010})}\BibitemShut {NoStop}%
\bibitem [{\citenamefont {Takubo}\ \emph {et~al.}(2017)\citenamefont {Takubo},
  \citenamefont {Yokoyama}, \citenamefont {Wadati}, \citenamefont {Iwasaki},
  \citenamefont {Mizokawa}, \citenamefont {Boyko}, \citenamefont {Sutarto},
  \citenamefont {He}, \citenamefont {Hashizume}, \citenamefont {Imaizumi},
  \citenamefont {Aoyama}, \citenamefont {Imai},\ and\ \citenamefont
  {Ohgushi}}]{TakuboBaFeSe2017}%
  \BibitemOpen
  \bibfield  {author} {\bibinfo {author} {\bibfnamefont {K.}~\bibnamefont
  {Takubo}}, \bibinfo {author} {\bibfnamefont {Y.}~\bibnamefont {Yokoyama}},
  \bibinfo {author} {\bibfnamefont {H.}~\bibnamefont {Wadati}}, \bibinfo
  {author} {\bibfnamefont {S.}~\bibnamefont {Iwasaki}}, \bibinfo {author}
  {\bibfnamefont {T.}~\bibnamefont {Mizokawa}}, \bibinfo {author}
  {\bibfnamefont {T.}~\bibnamefont {Boyko}}, \bibinfo {author} {\bibfnamefont
  {R.}~\bibnamefont {Sutarto}}, \bibinfo {author} {\bibfnamefont
  {F.}~\bibnamefont {He}}, \bibinfo {author} {\bibfnamefont {K.}~\bibnamefont
  {Hashizume}}, \bibinfo {author} {\bibfnamefont {S.}~\bibnamefont {Imaizumi}},
  \bibinfo {author} {\bibfnamefont {T.}~\bibnamefont {Aoyama}}, \bibinfo
  {author} {\bibfnamefont {Y.}~\bibnamefont {Imai}},\ and\ \bibinfo {author}
  {\bibfnamefont {K.}~\bibnamefont {Ohgushi}},\ }\href
  {https://doi.org/10.1103/PhysRevB.96.115157} {\bibfield  {journal} {\bibinfo
  {journal} {Phys. Rev. B}\ }\textbf {\bibinfo {volume} {96}},\ \bibinfo
  {pages} {115157} (\bibinfo {year} {2017})}\BibitemShut {NoStop}%
\bibitem [{\citenamefont {Materne}\ \emph {et~al.}(2019)\citenamefont
  {Materne}, \citenamefont {Bi}, \citenamefont {Zhao}, \citenamefont {Hu},
  \citenamefont {Amig{\'o}}, \citenamefont {Seiro}, \citenamefont {Aswartham},
  \citenamefont {B{\"u}chner},\ and\ \citenamefont
  {Alp}}]{materneBandwidthControlledInsulatormetal2019}%
  \BibitemOpen
  \bibfield  {author} {\bibinfo {author} {\bibfnamefont {P.}~\bibnamefont
  {Materne}}, \bibinfo {author} {\bibfnamefont {W.}~\bibnamefont {Bi}},
  \bibinfo {author} {\bibfnamefont {J.}~\bibnamefont {Zhao}}, \bibinfo {author}
  {\bibfnamefont {M.~Y.}\ \bibnamefont {Hu}}, \bibinfo {author} {\bibfnamefont
  {M.~L.}\ \bibnamefont {Amig{\'o}}}, \bibinfo {author} {\bibfnamefont
  {S.}~\bibnamefont {Seiro}}, \bibinfo {author} {\bibfnamefont
  {S.}~\bibnamefont {Aswartham}}, \bibinfo {author} {\bibfnamefont
  {B.}~\bibnamefont {B{\"u}chner}},\ and\ \bibinfo {author} {\bibfnamefont
  {E.~E.}\ \bibnamefont {Alp}},\ }\href
  {https://doi.org/10.1103/PhysRevB.99.020505} {\bibfield  {journal} {\bibinfo
  {journal} {Phys. Rev. B}\ }\textbf {\bibinfo {volume} {99}},\ \bibinfo
  {pages} {020505(R)} (\bibinfo {year} {2019})}\BibitemShut {NoStop}%
\bibitem [{\citenamefont {Patel}\ \emph {et~al.}(2019)\citenamefont {Patel},
  \citenamefont {Nocera}, \citenamefont {Alvarez}, \citenamefont {Moreo},
  \citenamefont {Johnston},\ and\ \citenamefont
  {Dagotto}}]{patelFingerprints2019}%
  \BibitemOpen
  \bibfield  {author} {\bibinfo {author} {\bibfnamefont {N.~D.}\ \bibnamefont
  {Patel}}, \bibinfo {author} {\bibfnamefont {A.}~\bibnamefont {Nocera}},
  \bibinfo {author} {\bibfnamefont {G.}~\bibnamefont {Alvarez}}, \bibinfo
  {author} {\bibfnamefont {A.}~\bibnamefont {Moreo}}, \bibinfo {author}
  {\bibfnamefont {S.}~\bibnamefont {Johnston}},\ and\ \bibinfo {author}
  {\bibfnamefont {E.}~\bibnamefont {Dagotto}},\ }\href
  {https://doi.org/10.1038/s42005-019-0155-3} {\bibfield  {journal} {\bibinfo
  {journal} {Commun. Phys.}\ }\textbf {\bibinfo {volume} {2}},\ \bibinfo
  {pages} {64} (\bibinfo {year} {2019})}\BibitemShut {NoStop}%
\bibitem [{\citenamefont {Daghofer}\ \emph {et~al.}(2010)\citenamefont
  {Daghofer}, \citenamefont {Nicholson}, \citenamefont {Moreo},\ and\
  \citenamefont {Dagotto}}]{Daghofer302010}%
  \BibitemOpen
  \bibfield  {author} {\bibinfo {author} {\bibfnamefont {M.}~\bibnamefont
  {Daghofer}}, \bibinfo {author} {\bibfnamefont {A.}~\bibnamefont {Nicholson}},
  \bibinfo {author} {\bibfnamefont {A.}~\bibnamefont {Moreo}},\ and\ \bibinfo
  {author} {\bibfnamefont {E.}~\bibnamefont {Dagotto}},\ }\href
  {https://doi.org/10.1103/PhysRevB.81.014511} {\bibfield  {journal} {\bibinfo
  {journal} {Phys. Rev. B}\ }\textbf {\bibinfo {volume} {81}},\ \bibinfo
  {pages} {014511} (\bibinfo {year} {2010})}\BibitemShut {NoStop}%
\bibitem [{\citenamefont {Rinc{\'o}n}\ \emph
  {et~al.}(2014{\natexlab{b}})\citenamefont {Rinc{\'o}n}, \citenamefont
  {Moreo}, \citenamefont {Alvarez},\ and\ \citenamefont
  {Dagotto}}]{RinconBlock22014}%
  \BibitemOpen
  \bibfield  {author} {\bibinfo {author} {\bibfnamefont {J.}~\bibnamefont
  {Rinc{\'o}n}}, \bibinfo {author} {\bibfnamefont {A.}~\bibnamefont {Moreo}},
  \bibinfo {author} {\bibfnamefont {G.}~\bibnamefont {Alvarez}},\ and\ \bibinfo
  {author} {\bibfnamefont {E.}~\bibnamefont {Dagotto}},\ }\href
  {https://doi.org/10.1103/PhysRevB.90.241105} {\bibfield  {journal} {\bibinfo
  {journal} {Phys. Rev. B}\ }\textbf {\bibinfo {volume} {90}},\ \bibinfo
  {pages} {241105(R)} (\bibinfo {year} {2014}{\natexlab{b}})}\BibitemShut
  {NoStop}%
\bibitem [{\citenamefont {Herbrych}\ \emph {et~al.}(2018)\citenamefont
  {Herbrych}, \citenamefont {Kaushal}, \citenamefont {Nocera}, \citenamefont
  {Alvarez}, \citenamefont {Moreo},\ and\ \citenamefont
  {Dagotto}}]{HerbrychBSq2018}%
  \BibitemOpen
  \bibfield  {author} {\bibinfo {author} {\bibfnamefont {J.}~\bibnamefont
  {Herbrych}}, \bibinfo {author} {\bibfnamefont {N.}~\bibnamefont {Kaushal}},
  \bibinfo {author} {\bibfnamefont {A.}~\bibnamefont {Nocera}}, \bibinfo
  {author} {\bibfnamefont {G.}~\bibnamefont {Alvarez}}, \bibinfo {author}
  {\bibfnamefont {A.}~\bibnamefont {Moreo}},\ and\ \bibinfo {author}
  {\bibfnamefont {E.}~\bibnamefont {Dagotto}},\ }\href
  {https://doi.org/10.1038/s41467-018-06181-6} {\bibfield  {journal} {\bibinfo
  {journal} {Nat. Commun.}\ }\textbf {\bibinfo {volume} {9}},\ \bibinfo {pages}
  {3736} (\bibinfo {year} {2018})}\BibitemShut {NoStop}%
\bibitem [{\citenamefont {Herbrych}\ \emph {et~al.}(2019)\citenamefont
  {Herbrych}, \citenamefont {Heverhagen}, \citenamefont {Patel}, \citenamefont
  {Alvarez}, \citenamefont {Daghofer}, \citenamefont {Moreo},\ and\
  \citenamefont {Dagotto}}]{HerbrychBlock2019}%
  \BibitemOpen
  \bibfield  {author} {\bibinfo {author} {\bibfnamefont {J.}~\bibnamefont
  {Herbrych}}, \bibinfo {author} {\bibfnamefont {J.}~\bibnamefont
  {Heverhagen}}, \bibinfo {author} {\bibfnamefont {N.~D.}\ \bibnamefont
  {Patel}}, \bibinfo {author} {\bibfnamefont {G.}~\bibnamefont {Alvarez}},
  \bibinfo {author} {\bibfnamefont {M.}~\bibnamefont {Daghofer}}, \bibinfo
  {author} {\bibfnamefont {A.}~\bibnamefont {Moreo}},\ and\ \bibinfo {author}
  {\bibfnamefont {E.}~\bibnamefont {Dagotto}},\ }\href
  {https://doi.org/10.1103/PhysRevLett.123.027203} {\bibfield  {journal}
  {\bibinfo  {journal} {Phys. Rev. Lett.}\ }\textbf {\bibinfo {volume} {123}},\
  \bibinfo {pages} {027203} (\bibinfo {year} {2019})}\BibitemShut {NoStop}%
\bibitem [{\citenamefont {Herbrych}\ \emph
  {et~al.}(2020{\natexlab{a}})\citenamefont {Herbrych}, \citenamefont
  {Heverhagen}, \citenamefont {Alvarez}, \citenamefont {Daghofer},
  \citenamefont {Moreo},\ and\ \citenamefont
  {Dagotto}}]{herbrychBlockSpiral2020}%
  \BibitemOpen
  \bibfield  {author} {\bibinfo {author} {\bibfnamefont {J.}~\bibnamefont
  {Herbrych}}, \bibinfo {author} {\bibfnamefont {J.}~\bibnamefont
  {Heverhagen}}, \bibinfo {author} {\bibfnamefont {G.}~\bibnamefont {Alvarez}},
  \bibinfo {author} {\bibfnamefont {M.}~\bibnamefont {Daghofer}}, \bibinfo
  {author} {\bibfnamefont {A.}~\bibnamefont {Moreo}},\ and\ \bibinfo {author}
  {\bibfnamefont {E.}~\bibnamefont {Dagotto}},\ }\href
  {https://doi.org/10.1073/pnas.2001141117} {\bibfield  {journal} {\bibinfo
  {journal} {Proc. Natl. Acad. Sci. U.S.A.}\ }\textbf {\bibinfo {volume}
  {117}},\ \bibinfo {pages} {16226} (\bibinfo {year}
  {2020}{\natexlab{a}})}\BibitemShut {NoStop}%
\bibitem [{\citenamefont {Herbrych}\ \emph
  {et~al.}(2020{\natexlab{b}})\citenamefont {Herbrych}, \citenamefont
  {Alvarez}, \citenamefont {Moreo},\ and\ \citenamefont
  {Dagotto}}]{HerbrychBlock2020}%
  \BibitemOpen
  \bibfield  {author} {\bibinfo {author} {\bibfnamefont {J.}~\bibnamefont
  {Herbrych}}, \bibinfo {author} {\bibfnamefont {G.}~\bibnamefont {Alvarez}},
  \bibinfo {author} {\bibfnamefont {A.}~\bibnamefont {Moreo}},\ and\ \bibinfo
  {author} {\bibfnamefont {E.}~\bibnamefont {Dagotto}},\ }\href
  {https://doi.org/10.1103/PhysRevB.102.115134} {\bibfield  {journal} {\bibinfo
   {journal} {Phys. Rev. B}\ }\textbf {\bibinfo {volume} {102}},\ \bibinfo
  {pages} {115134} (\bibinfo {year} {2020}{\natexlab{b}})}\BibitemShut
  {NoStop}%
\bibitem [{\citenamefont
  {Kanamori}(1963)}]{kanamoriElectronCorrelationFerromagnetism1963}%
  \BibitemOpen
  \bibfield  {author} {\bibinfo {author} {\bibfnamefont {J.}~\bibnamefont
  {Kanamori}},\ }\href {https://doi.org/10.1143/PTP.30.275} {\bibfield
  {journal} {\bibinfo  {journal} {Prog. Theor. Phys.}\ }\textbf {\bibinfo
  {volume} {30}},\ \bibinfo {pages} {275} (\bibinfo {year} {1963})}\BibitemShut
  {NoStop}%
\bibitem [{\citenamefont
  {Ole{\'s}}(1983)}]{olesAntiferromagnetismCorrelationElectrons1983}%
  \BibitemOpen
  \bibfield  {author} {\bibinfo {author} {\bibfnamefont {A.~M.}\ \bibnamefont
  {Ole{\'s}}},\ }\href {https://doi.org/10.1103/PhysRevB.28.327} {\bibfield
  {journal} {\bibinfo  {journal} {Phys. Rev. B}\ }\textbf {\bibinfo {volume}
  {28}},\ \bibinfo {pages} {327} (\bibinfo {year} {1983})}\BibitemShut
  {NoStop}%
\bibitem [{\citenamefont {Dagotto}\ \emph {et~al.}(2001)\citenamefont
  {Dagotto}, \citenamefont {Hotta},\ and\ \citenamefont
  {Moreo}}]{dagottoColossalMagnetoresistantMaterials2001}%
  \BibitemOpen
  \bibfield  {author} {\bibinfo {author} {\bibfnamefont {E.}~\bibnamefont
  {Dagotto}}, \bibinfo {author} {\bibfnamefont {T.}~\bibnamefont {Hotta}},\
  and\ \bibinfo {author} {\bibfnamefont {A.}~\bibnamefont {Moreo}},\ }\href
  {https://doi.org/10.1016/S0370-1573(00)00121-6} {\bibfield  {journal}
  {\bibinfo  {journal} {Physics Reports}\ }\textbf {\bibinfo {volume} {344}},\
  \bibinfo {pages} {1} (\bibinfo {year} {2001})}\BibitemShut {NoStop}%
\bibitem [{\citenamefont {Haule}\ and\ \citenamefont
  {Kotliar}(2009)}]{hauleCoherence2009}%
  \BibitemOpen
  \bibfield  {author} {\bibinfo {author} {\bibfnamefont {K.}~\bibnamefont
  {Haule}}\ and\ \bibinfo {author} {\bibfnamefont {G.}~\bibnamefont
  {Kotliar}},\ }\href {https://doi.org/10.1088/1367-2630/11/2/025021}
  {\bibfield  {journal} {\bibinfo  {journal} {New J. Phys.}\ }\textbf {\bibinfo
  {volume} {11}},\ \bibinfo {pages} {025021} (\bibinfo {year}
  {2009})}\BibitemShut {NoStop}%
\bibitem [{\citenamefont {Yin}\ \emph {et~al.}(2011)\citenamefont {Yin},
  \citenamefont {Haule},\ and\ \citenamefont {Kotliar}}]{yinKinetic2011}%
  \BibitemOpen
  \bibfield  {author} {\bibinfo {author} {\bibfnamefont {Z.~P.}\ \bibnamefont
  {Yin}}, \bibinfo {author} {\bibfnamefont {K.}~\bibnamefont {Haule}},\ and\
  \bibinfo {author} {\bibfnamefont {G.}~\bibnamefont {Kotliar}},\ }\href
  {https://doi.org/10.1038/nmat3120} {\bibfield  {journal} {\bibinfo  {journal}
  {Nat. Mater.}\ }\textbf {\bibinfo {volume} {10}},\ \bibinfo {pages} {932}
  (\bibinfo {year} {2011})}\BibitemShut {NoStop}%
\bibitem [{\citenamefont {Ferber}\ \emph {et~al.}(2012)\citenamefont {Ferber},
  \citenamefont {Foyevtsova}, \citenamefont {Valent{\'i}},\ and\ \citenamefont
  {Jeschke}}]{ferberLDA2012}%
  \BibitemOpen
  \bibfield  {author} {\bibinfo {author} {\bibfnamefont {J.}~\bibnamefont
  {Ferber}}, \bibinfo {author} {\bibfnamefont {K.}~\bibnamefont {Foyevtsova}},
  \bibinfo {author} {\bibfnamefont {R.}~\bibnamefont {Valent{\'i}}},\ and\
  \bibinfo {author} {\bibfnamefont {H.~O.}\ \bibnamefont {Jeschke}},\ }\href
  {https://doi.org/10.1103/PhysRevB.85.094505} {\bibfield  {journal} {\bibinfo
  {journal} {Phys. Rev. B}\ }\textbf {\bibinfo {volume} {85}},\ \bibinfo
  {pages} {094505} (\bibinfo {year} {2012})}\BibitemShut {NoStop}%
\bibitem [{\citenamefont {Luo}\ \emph {et~al.}(2010)\citenamefont {Luo},
  \citenamefont {Martins}, \citenamefont {Yao}, \citenamefont {Daghofer},
  \citenamefont {Yu}, \citenamefont {Moreo},\ and\ \citenamefont
  {Dagotto}}]{luoNeutron2010}%
  \BibitemOpen
  \bibfield  {author} {\bibinfo {author} {\bibfnamefont {Q.}~\bibnamefont
  {Luo}}, \bibinfo {author} {\bibfnamefont {G.}~\bibnamefont {Martins}},
  \bibinfo {author} {\bibfnamefont {D.-X.}\ \bibnamefont {Yao}}, \bibinfo
  {author} {\bibfnamefont {M.}~\bibnamefont {Daghofer}}, \bibinfo {author}
  {\bibfnamefont {R.}~\bibnamefont {Yu}}, \bibinfo {author} {\bibfnamefont
  {A.}~\bibnamefont {Moreo}},\ and\ \bibinfo {author} {\bibfnamefont
  {E.}~\bibnamefont {Dagotto}},\ }\href
  {https://doi.org/10.1103/PhysRevB.82.104508} {\bibfield  {journal} {\bibinfo
  {journal} {Phys. Rev. B}\ }\textbf {\bibinfo {volume} {82}},\ \bibinfo
  {pages} {104508} (\bibinfo {year} {2010})}\BibitemShut {NoStop}%
\bibitem [{\citenamefont {Dai}\ \emph {et~al.}(2012)\citenamefont {Dai},
  \citenamefont {Hu},\ and\ \citenamefont {Dagotto}}]{daiMagnetism2012}%
  \BibitemOpen
  \bibfield  {author} {\bibinfo {author} {\bibfnamefont {P.}~\bibnamefont
  {Dai}}, \bibinfo {author} {\bibfnamefont {J.}~\bibnamefont {Hu}},\ and\
  \bibinfo {author} {\bibfnamefont {E.}~\bibnamefont {Dagotto}},\ }\href
  {https://doi.org/10.1038/nphys2438} {\bibfield  {journal} {\bibinfo
  {journal} {Nat. Phys.}\ }\textbf {\bibinfo {volume} {8}},\ \bibinfo {pages}
  {709} (\bibinfo {year} {2012})}\BibitemShut {NoStop}%
\bibitem [{\citenamefont {Schrieffer}\ and\ \citenamefont
  {Wolff}(1966)}]{schriefferRelation1966}%
  \BibitemOpen
  \bibfield  {author} {\bibinfo {author} {\bibfnamefont {J.~R.}\ \bibnamefont
  {Schrieffer}}\ and\ \bibinfo {author} {\bibfnamefont {P.~A.}\ \bibnamefont
  {Wolff}},\ }\href {https://doi.org/10.1103/PhysRev.149.491} {\bibfield
  {journal} {\bibinfo  {journal} {Phys. Rev.}\ }\textbf {\bibinfo {volume}
  {149}},\ \bibinfo {pages} {491} (\bibinfo {year} {1966})}\BibitemShut
  {NoStop}%
\bibitem [{\citenamefont {White}(2005)}]{whiteDensity2005}%
  \BibitemOpen
  \bibfield  {author} {\bibinfo {author} {\bibfnamefont {S.~R.}\ \bibnamefont
  {White}},\ }\href {https://doi.org/10.1103/PhysRevB.72.180403} {\bibfield
  {journal} {\bibinfo  {journal} {Phys. Rev. B}\ }\textbf {\bibinfo {volume}
  {72}},\ \bibinfo {pages} {180403(R)} (\bibinfo {year} {2005})}\BibitemShut
  {NoStop}%
\bibitem [{\citenamefont {Alvarez}(2009)}]{alvarezDensity2009}%
  \BibitemOpen
  \bibfield  {author} {\bibinfo {author} {\bibfnamefont {G.}~\bibnamefont
  {Alvarez}},\ }\href {https://doi.org/10.1016/j.cpc.2009.02.016} {\bibfield
  {journal} {\bibinfo  {journal} {Comput. Phys. Commun.}\ }\textbf {\bibinfo
  {volume} {180}},\ \bibinfo {pages} {1572} (\bibinfo {year}
  {2009})}\BibitemShut {NoStop}%
\bibitem [{Gon()}]{GonzaloDMRGWeb}%
  \BibitemOpen
  \href@noop {} {}\bibinfo {howpublished}
  {\url{https://g1257.github.io/dmrgPlusPlus/}}\BibitemShut {NoStop}%
\bibitem [{Cor()}]{CorrWro}%
  \BibitemOpen
  \href@noop {} {}\bibinfo {howpublished}
  {\url{https://bitbucket.org/herbrychjacek/corrwro/}}\BibitemShut {NoStop}%
\bibitem [{\citenamefont {Noack}\ \emph {et~al.}(1996)\citenamefont {Noack},
  \citenamefont {White},\ and\ \citenamefont {Scalapino}}]{noackGround1996}%
  \BibitemOpen
  \bibfield  {author} {\bibinfo {author} {\bibfnamefont {R.~M.}\ \bibnamefont
  {Noack}}, \bibinfo {author} {\bibfnamefont {S.~R.}\ \bibnamefont {White}},\
  and\ \bibinfo {author} {\bibfnamefont {D.~J.}\ \bibnamefont {Scalapino}},\
  }\href {https://doi.org/10.1016/S0921-4534(96)00515-1} {\bibfield  {journal}
  {\bibinfo  {journal} {Physica C: Supercond.}\ }\textbf {\bibinfo {volume}
  {270}},\ \bibinfo {pages} {281} (\bibinfo {year} {1996})}\BibitemShut
  {NoStop}%
\bibitem [{\citenamefont {Herbrych}\ \emph {et~al.}(2021)\citenamefont
  {Herbrych}, \citenamefont {\'{S}roda}, \citenamefont {Alvarez}, \citenamefont
  {Mierzejewski},\ and\ \citenamefont {Dagotto}}]{herbrychTopoOSMP2021}%
  \BibitemOpen
  \bibfield  {author} {\bibinfo {author} {\bibfnamefont {J.}~\bibnamefont
  {Herbrych}}, \bibinfo {author} {\bibfnamefont {M.}~\bibnamefont {\'{S}roda}},
  \bibinfo {author} {\bibfnamefont {G.}~\bibnamefont {Alvarez}}, \bibinfo
  {author} {\bibfnamefont {M.}~\bibnamefont {Mierzejewski}},\ and\ \bibinfo
  {author} {\bibfnamefont {E.}~\bibnamefont {Dagotto}},\ }\href
  {https://doi.org/10.1038/s41467-021-23261-2} {\bibfield  {journal} {\bibinfo
  {journal} {Nat. Commun.}\ }\textbf {\bibinfo {volume} {12}},\ \bibinfo
  {pages} {2955} (\bibinfo {year} {2021})}\BibitemShut {NoStop}%
\bibitem [{\citenamefont {Martins}\ \emph {et~al.}(2000)\citenamefont
  {Martins}, \citenamefont {Gazza}, \citenamefont {Xavier}, \citenamefont
  {Feiguin},\ and\ \citenamefont {Dagotto}}]{martinsDoped2000}%
  \BibitemOpen
  \bibfield  {author} {\bibinfo {author} {\bibfnamefont {G.~B.}\ \bibnamefont
  {Martins}}, \bibinfo {author} {\bibfnamefont {C.}~\bibnamefont {Gazza}},
  \bibinfo {author} {\bibfnamefont {J.~C.}\ \bibnamefont {Xavier}}, \bibinfo
  {author} {\bibfnamefont {A.}~\bibnamefont {Feiguin}},\ and\ \bibinfo {author}
  {\bibfnamefont {E.}~\bibnamefont {Dagotto}},\ }\href
  {https://doi.org/10.1103/PhysRevLett.84.5844} {\bibfield  {journal} {\bibinfo
   {journal} {Phys. Rev. Lett.}\ }\textbf {\bibinfo {volume} {84}},\ \bibinfo
  {pages} {5844} (\bibinfo {year} {2000})}\BibitemShut {NoStop}%
\bibitem [{\citenamefont {White}\ and\ \citenamefont
  {Scalapino}(2003)}]{whiteStripes2003}%
  \BibitemOpen
  \bibfield  {author} {\bibinfo {author} {\bibfnamefont {S.~R.}\ \bibnamefont
  {White}}\ and\ \bibinfo {author} {\bibfnamefont {D.~J.}\ \bibnamefont
  {Scalapino}},\ }\href {https://doi.org/10.1103/PhysRevLett.91.136403}
  {\bibfield  {journal} {\bibinfo  {journal} {Phys. Rev. Lett.}\ }\textbf
  {\bibinfo {volume} {91}},\ \bibinfo {pages} {136403} (\bibinfo {year}
  {2003})}\BibitemShut {NoStop}%
\bibitem [{\citenamefont {Hager}\ \emph {et~al.}(2005)\citenamefont {Hager},
  \citenamefont {Wellein}, \citenamefont {Jeckelmann},\ and\ \citenamefont
  {Fehske}}]{hagerStripe2005}%
  \BibitemOpen
  \bibfield  {author} {\bibinfo {author} {\bibfnamefont {G.}~\bibnamefont
  {Hager}}, \bibinfo {author} {\bibfnamefont {G.}~\bibnamefont {Wellein}},
  \bibinfo {author} {\bibfnamefont {E.}~\bibnamefont {Jeckelmann}},\ and\
  \bibinfo {author} {\bibfnamefont {H.}~\bibnamefont {Fehske}},\ }\href
  {https://doi.org/10.1103/PhysRevB.71.075108} {\bibfield  {journal} {\bibinfo
  {journal} {Phys. Rev. B}\ }\textbf {\bibinfo {volume} {71}},\ \bibinfo
  {pages} {075108} (\bibinfo {year} {2005})}\BibitemShut {NoStop}%
\bibitem [{\citenamefont {Ehlers}\ \emph {et~al.}(2017)\citenamefont {Ehlers},
  \citenamefont {White},\ and\ \citenamefont {Noack}}]{ehlersHybridspace2017}%
  \BibitemOpen
  \bibfield  {author} {\bibinfo {author} {\bibfnamefont {G.}~\bibnamefont
  {Ehlers}}, \bibinfo {author} {\bibfnamefont {S.~R.}\ \bibnamefont {White}},\
  and\ \bibinfo {author} {\bibfnamefont {R.~M.}\ \bibnamefont {Noack}},\ }\href
  {https://doi.org/10.1103/PhysRevB.95.125125} {\bibfield  {journal} {\bibinfo
  {journal} {Phys. Rev. B}\ }\textbf {\bibinfo {volume} {95}},\ \bibinfo
  {pages} {125125} (\bibinfo {year} {2017})}\BibitemShut {NoStop}%
\bibitem [{\citenamefont {Zheng}\ \emph {et~al.}(2017)\citenamefont {Zheng},
  \citenamefont {Chung}, \citenamefont {Corboz}, \citenamefont {Ehlers},
  \citenamefont {Qin}, \citenamefont {Noack}, \citenamefont {Shi},
  \citenamefont {White}, \citenamefont {Zhang},\ and\ \citenamefont
  {Chan}}]{zhengStripe2017}%
  \BibitemOpen
  \bibfield  {author} {\bibinfo {author} {\bibfnamefont {B.-X.}\ \bibnamefont
  {Zheng}}, \bibinfo {author} {\bibfnamefont {C.-M.}\ \bibnamefont {Chung}},
  \bibinfo {author} {\bibfnamefont {P.}~\bibnamefont {Corboz}}, \bibinfo
  {author} {\bibfnamefont {G.}~\bibnamefont {Ehlers}}, \bibinfo {author}
  {\bibfnamefont {M.-P.}\ \bibnamefont {Qin}}, \bibinfo {author} {\bibfnamefont
  {R.~M.}\ \bibnamefont {Noack}}, \bibinfo {author} {\bibfnamefont
  {H.}~\bibnamefont {Shi}}, \bibinfo {author} {\bibfnamefont {S.~R.}\
  \bibnamefont {White}}, \bibinfo {author} {\bibfnamefont {S.}~\bibnamefont
  {Zhang}},\ and\ \bibinfo {author} {\bibfnamefont {G.~K.-L.}\ \bibnamefont
  {Chan}},\ }\href {https://doi.org/10.1126/science.aam7127} {\bibfield
  {journal} {\bibinfo  {journal} {Science}\ }\textbf {\bibinfo {volume}
  {358}},\ \bibinfo {pages} {1155} (\bibinfo {year} {2017})}\BibitemShut
  {NoStop}%
\bibitem [{\citenamefont {Jiang}\ and\ \citenamefont
  {Devereaux}(2019)}]{jiangSuperconductivity2019}%
  \BibitemOpen
  \bibfield  {author} {\bibinfo {author} {\bibfnamefont {H.-C.}\ \bibnamefont
  {Jiang}}\ and\ \bibinfo {author} {\bibfnamefont {T.~P.}\ \bibnamefont
  {Devereaux}},\ }\href {https://doi.org/10.1126/science.aal5304} {\bibfield
  {journal} {\bibinfo  {journal} {Science}\ }\textbf {\bibinfo {volume}
  {365}},\ \bibinfo {pages} {1424} (\bibinfo {year} {2019})}\BibitemShut
  {NoStop}%
\bibitem [{\citenamefont {Qin}\ \emph {et~al.}(2020)\citenamefont {Qin},
  \citenamefont {Chung}, \citenamefont {Shi}, \citenamefont {Vitali},
  \citenamefont {Hubig}, \citenamefont {Schollw\"ock}, \citenamefont {White},\
  and\ \citenamefont
  {Zhang}}]{simonscollaborationonthemany-electronproblemAbsence2020}%
  \BibitemOpen
  \bibfield  {author} {\bibinfo {author} {\bibfnamefont {M.}~\bibnamefont
  {Qin}}, \bibinfo {author} {\bibfnamefont {C.-M.}\ \bibnamefont {Chung}},
  \bibinfo {author} {\bibfnamefont {H.}~\bibnamefont {Shi}}, \bibinfo {author}
  {\bibfnamefont {E.}~\bibnamefont {Vitali}}, \bibinfo {author} {\bibfnamefont
  {C.}~\bibnamefont {Hubig}}, \bibinfo {author} {\bibfnamefont
  {U.}~\bibnamefont {Schollw\"ock}}, \bibinfo {author} {\bibfnamefont {S.~R.}\
  \bibnamefont {White}},\ and\ \bibinfo {author} {\bibfnamefont
  {S.}~\bibnamefont {Zhang}} (\bibinfo {collaboration} {Simons Collaboration on
  the Many-Electron Problem}),\ }\href
  {https://doi.org/10.1103/PhysRevX.10.031016} {\bibfield  {journal} {\bibinfo
  {journal} {Phys. Rev. X}\ }\textbf {\bibinfo {volume} {10}},\ \bibinfo
  {pages} {031016} (\bibinfo {year} {2020})}\BibitemShut {NoStop}%
\bibitem [{\citenamefont {Tranquada}\ \emph {et~al.}(1995)\citenamefont
  {Tranquada}, \citenamefont {Sternlieb}, \citenamefont {Axe}, \citenamefont
  {Nakamura},\ and\ \citenamefont {Uchida}}]{tranquadaEvidence1995}%
  \BibitemOpen
  \bibfield  {author} {\bibinfo {author} {\bibfnamefont {J.~M.}\ \bibnamefont
  {Tranquada}}, \bibinfo {author} {\bibfnamefont {B.~J.}\ \bibnamefont
  {Sternlieb}}, \bibinfo {author} {\bibfnamefont {J.~D.}\ \bibnamefont {Axe}},
  \bibinfo {author} {\bibfnamefont {Y.}~\bibnamefont {Nakamura}},\ and\
  \bibinfo {author} {\bibfnamefont {S.}~\bibnamefont {Uchida}},\ }\href
  {https://doi.org/10.1038/375561a0} {\bibfield  {journal} {\bibinfo  {journal}
  {Nature}\ }\textbf {\bibinfo {volume} {375}},\ \bibinfo {pages} {561}
  (\bibinfo {year} {1995})}\BibitemShut {NoStop}%
\bibitem [{\citenamefont {Tranquada}\ \emph {et~al.}(1996)\citenamefont
  {Tranquada}, \citenamefont {Axe}, \citenamefont {Ichikawa}, \citenamefont
  {Nakamura}, \citenamefont {Uchida},\ and\ \citenamefont
  {Nachumi}}]{tranquadaNeutronscattering1996}%
  \BibitemOpen
  \bibfield  {author} {\bibinfo {author} {\bibfnamefont {J.~M.}\ \bibnamefont
  {Tranquada}}, \bibinfo {author} {\bibfnamefont {J.~D.}\ \bibnamefont {Axe}},
  \bibinfo {author} {\bibfnamefont {N.}~\bibnamefont {Ichikawa}}, \bibinfo
  {author} {\bibfnamefont {Y.}~\bibnamefont {Nakamura}}, \bibinfo {author}
  {\bibfnamefont {S.}~\bibnamefont {Uchida}},\ and\ \bibinfo {author}
  {\bibfnamefont {B.}~\bibnamefont {Nachumi}},\ }\href
  {https://doi.org/10.1103/PhysRevB.54.7489} {\bibfield  {journal} {\bibinfo
  {journal} {Phys. Rev. B}\ }\textbf {\bibinfo {volume} {54}},\ \bibinfo
  {pages} {7489} (\bibinfo {year} {1996})}\BibitemShut {NoStop}%
\bibitem [{\citenamefont {Tranquada}\ \emph {et~al.}(1997)\citenamefont
  {Tranquada}, \citenamefont {Axe}, \citenamefont {Ichikawa}, \citenamefont
  {Moodenbaugh}, \citenamefont {Nakamura},\ and\ \citenamefont
  {Uchida}}]{tranquadaCoexistence1997}%
  \BibitemOpen
  \bibfield  {author} {\bibinfo {author} {\bibfnamefont {J.~M.}\ \bibnamefont
  {Tranquada}}, \bibinfo {author} {\bibfnamefont {J.~D.}\ \bibnamefont {Axe}},
  \bibinfo {author} {\bibfnamefont {N.}~\bibnamefont {Ichikawa}}, \bibinfo
  {author} {\bibfnamefont {A.~R.}\ \bibnamefont {Moodenbaugh}}, \bibinfo
  {author} {\bibfnamefont {Y.}~\bibnamefont {Nakamura}},\ and\ \bibinfo
  {author} {\bibfnamefont {S.}~\bibnamefont {Uchida}},\ }\href
  {https://doi.org/10.1103/PhysRevLett.78.338} {\bibfield  {journal} {\bibinfo
  {journal} {Phys. Rev. Lett.}\ }\textbf {\bibinfo {volume} {78}},\ \bibinfo
  {pages} {338} (\bibinfo {year} {1997})}\BibitemShut {NoStop}%
\bibitem [{\citenamefont {Patel}\ \emph {et~al.}(2017)\citenamefont {Patel},
  \citenamefont {Nocera}, \citenamefont {Alvarez}, \citenamefont {Moreo},\ and\
  \citenamefont {Dagotto}}]{patelPairing2017}%
  \BibitemOpen
  \bibfield  {author} {\bibinfo {author} {\bibfnamefont {N.~D.}\ \bibnamefont
  {Patel}}, \bibinfo {author} {\bibfnamefont {A.}~\bibnamefont {Nocera}},
  \bibinfo {author} {\bibfnamefont {G.}~\bibnamefont {Alvarez}}, \bibinfo
  {author} {\bibfnamefont {A.}~\bibnamefont {Moreo}},\ and\ \bibinfo {author}
  {\bibfnamefont {E.}~\bibnamefont {Dagotto}},\ }\href
  {https://doi.org/10.1103/PhysRevB.96.024520} {\bibfield  {journal} {\bibinfo
  {journal} {Phys. Rev. B}\ }\textbf {\bibinfo {volume} {96}},\ \bibinfo
  {pages} {024520} (\bibinfo {year} {2017})}\BibitemShut {NoStop}%
\bibitem [{\citenamefont {Pandey}\ \emph {et~al.}()\citenamefont {Pandey},
  \citenamefont {Soni}, \citenamefont {Lin}, \citenamefont {Alvarez},\ and\
  \citenamefont {Dagotto}}]{pandeyIntertwined2021}%
  \BibitemOpen
  \bibfield  {author} {\bibinfo {author} {\bibfnamefont {B.}~\bibnamefont
  {Pandey}}, \bibinfo {author} {\bibfnamefont {R.}~\bibnamefont {Soni}},
  \bibinfo {author} {\bibfnamefont {L.-F.}\ \bibnamefont {Lin}}, \bibinfo
  {author} {\bibfnamefont {G.}~\bibnamefont {Alvarez}},\ and\ \bibinfo {author}
  {\bibfnamefont {E.}~\bibnamefont {Dagotto}},\ }\href@noop {} {\ }\Eprint
  {https://arxiv.org/abs/2103.06407} {arXiv:2103.06407} \BibitemShut {NoStop}%
\bibitem [{\citenamefont {Dagotto}\ \emph {et~al.}(1998)\citenamefont
  {Dagotto}, \citenamefont {Yunoki}, \citenamefont {Malvezzi}, \citenamefont
  {Moreo}, \citenamefont {Hu}, \citenamefont {Capponi}, \citenamefont
  {Poilblanc},\ and\ \citenamefont {Furukawa}}]{dagottoFerromagnetic1998}%
  \BibitemOpen
  \bibfield  {author} {\bibinfo {author} {\bibfnamefont {E.}~\bibnamefont
  {Dagotto}}, \bibinfo {author} {\bibfnamefont {S.}~\bibnamefont {Yunoki}},
  \bibinfo {author} {\bibfnamefont {A.~L.}\ \bibnamefont {Malvezzi}}, \bibinfo
  {author} {\bibfnamefont {A.}~\bibnamefont {Moreo}}, \bibinfo {author}
  {\bibfnamefont {J.}~\bibnamefont {Hu}}, \bibinfo {author} {\bibfnamefont
  {S.}~\bibnamefont {Capponi}}, \bibinfo {author} {\bibfnamefont
  {D.}~\bibnamefont {Poilblanc}},\ and\ \bibinfo {author} {\bibfnamefont
  {N.}~\bibnamefont {Furukawa}},\ }\href
  {https://doi.org/10.1103/PhysRevB.58.6414} {\bibfield  {journal} {\bibinfo
  {journal} {Phys. Rev. B}\ }\textbf {\bibinfo {volume} {58}},\ \bibinfo
  {pages} {6414} (\bibinfo {year} {1998})}\BibitemShut {NoStop}%
\bibitem [{\citenamefont {Dagotto}(2003)}]{dagottoNanoscale2003}%
  \BibitemOpen
  \bibfield  {author} {\bibinfo {author} {\bibfnamefont {E.}~\bibnamefont
  {Dagotto}},\ }\href {https://doi.org/10.1007/978-3-662-05244-0} {\emph
  {\bibinfo {title} {Nanoscale {{Phase Separation}} and {{Colossal
  Magnetoresistance}}: {{The Physics}} of {{Manganites}} and {{Related
  Compounds}}}}},\ Springer {{Series}} in {{Solid}}-{{State Sciences}}\
  (\bibinfo  {publisher} {{Springer-Verlag}},\ \bibinfo {address} {{Berlin
  Heidelberg}},\ \bibinfo {year} {2003})\BibitemShut {NoStop}%
\bibitem [{\citenamefont {Yunoki}\ \emph {et~al.}(1998)\citenamefont {Yunoki},
  \citenamefont {Hu}, \citenamefont {Malvezzi}, \citenamefont {Moreo},
  \citenamefont {Furukawa},\ and\ \citenamefont {Dagotto}}]{yunokiPhase1998}%
  \BibitemOpen
  \bibfield  {author} {\bibinfo {author} {\bibfnamefont {S.}~\bibnamefont
  {Yunoki}}, \bibinfo {author} {\bibfnamefont {J.}~\bibnamefont {Hu}}, \bibinfo
  {author} {\bibfnamefont {A.~L.}\ \bibnamefont {Malvezzi}}, \bibinfo {author}
  {\bibfnamefont {A.}~\bibnamefont {Moreo}}, \bibinfo {author} {\bibfnamefont
  {N.}~\bibnamefont {Furukawa}},\ and\ \bibinfo {author} {\bibfnamefont
  {E.}~\bibnamefont {Dagotto}},\ }\href
  {https://doi.org/10.1103/PhysRevLett.80.845} {\bibfield  {journal} {\bibinfo
  {journal} {Phys. Rev. Lett.}\ }\textbf {\bibinfo {volume} {80}},\ \bibinfo
  {pages} {845} (\bibinfo {year} {1998})}\BibitemShut {NoStop}%
\bibitem [{\citenamefont {Neuber}\ \emph {et~al.}(2006)\citenamefont {Neuber},
  \citenamefont {Daghofer}, \citenamefont {Evertz}, \citenamefont {{von der
  Linden}},\ and\ \citenamefont {Noack}}]{neuberFerromagnetic2006}%
  \BibitemOpen
  \bibfield  {author} {\bibinfo {author} {\bibfnamefont {D.~R.}\ \bibnamefont
  {Neuber}}, \bibinfo {author} {\bibfnamefont {M.}~\bibnamefont {Daghofer}},
  \bibinfo {author} {\bibfnamefont {H.~G.}\ \bibnamefont {Evertz}}, \bibinfo
  {author} {\bibfnamefont {W.}~\bibnamefont {{von der Linden}}},\ and\ \bibinfo
  {author} {\bibfnamefont {R.~M.}\ \bibnamefont {Noack}},\ }\href
  {https://doi.org/10.1103/PhysRevB.73.014401} {\bibfield  {journal} {\bibinfo
  {journal} {Phys. Rev. B}\ }\textbf {\bibinfo {volume} {73}},\ \bibinfo
  {pages} {014401} (\bibinfo {year} {2006})}\BibitemShut {NoStop}%
\bibitem [{\citenamefont {Ricci}\ \emph {et~al.}(2011)\citenamefont {Ricci},
  \citenamefont {Poccia}, \citenamefont {Campi}, \citenamefont {Joseph},
  \citenamefont {Arrighetti}, \citenamefont {Barba}, \citenamefont {Reynolds},
  \citenamefont {Burghammer}, \citenamefont {Takeya}, \citenamefont
  {Mizuguchi}, \citenamefont {Takano}, \citenamefont {Colapietro},
  \citenamefont {Saini},\ and\ \citenamefont {Bianconi}}]{ricciNanoscale2011}%
  \BibitemOpen
  \bibfield  {author} {\bibinfo {author} {\bibfnamefont {A.}~\bibnamefont
  {Ricci}}, \bibinfo {author} {\bibfnamefont {N.}~\bibnamefont {Poccia}},
  \bibinfo {author} {\bibfnamefont {G.}~\bibnamefont {Campi}}, \bibinfo
  {author} {\bibfnamefont {B.}~\bibnamefont {Joseph}}, \bibinfo {author}
  {\bibfnamefont {G.}~\bibnamefont {Arrighetti}}, \bibinfo {author}
  {\bibfnamefont {L.}~\bibnamefont {Barba}}, \bibinfo {author} {\bibfnamefont
  {M.}~\bibnamefont {Reynolds}}, \bibinfo {author} {\bibfnamefont
  {M.}~\bibnamefont {Burghammer}}, \bibinfo {author} {\bibfnamefont
  {H.}~\bibnamefont {Takeya}}, \bibinfo {author} {\bibfnamefont
  {Y.}~\bibnamefont {Mizuguchi}}, \bibinfo {author} {\bibfnamefont
  {Y.}~\bibnamefont {Takano}}, \bibinfo {author} {\bibfnamefont
  {M.}~\bibnamefont {Colapietro}}, \bibinfo {author} {\bibfnamefont {N.~L.}\
  \bibnamefont {Saini}},\ and\ \bibinfo {author} {\bibfnamefont
  {A.}~\bibnamefont {Bianconi}},\ }\href
  {https://doi.org/10.1103/PhysRevB.84.060511} {\bibfield  {journal} {\bibinfo
  {journal} {Phys. Rev. B}\ }\textbf {\bibinfo {volume} {84}},\ \bibinfo
  {pages} {060511(R)} (\bibinfo {year} {2011})}\BibitemShut {NoStop}%
\bibitem [{\citenamefont {Kreisel}\ \emph {et~al.}(2020)\citenamefont
  {Kreisel}, \citenamefont {Hirschfeld},\ and\ \citenamefont
  {Andersen}}]{KreiselReview2020}%
  \BibitemOpen
  \bibfield  {author} {\bibinfo {author} {\bibfnamefont {A.}~\bibnamefont
  {Kreisel}}, \bibinfo {author} {\bibfnamefont {P.~J.}\ \bibnamefont
  {Hirschfeld}},\ and\ \bibinfo {author} {\bibfnamefont {B.~M.}\ \bibnamefont
  {Andersen}},\ }\href {https://doi.org/10.3390/sym12091402} {\bibfield
  {journal} {\bibinfo  {journal} {Symmetry}\ }\textbf {\bibinfo {volume}
  {12}},\ \bibinfo {pages} {1402} (\bibinfo {year} {2020})}\BibitemShut
  {NoStop}%
\bibitem [{\citenamefont {Fj{\ae}restad}\ \emph {et~al.}(2006)\citenamefont
  {Fj{\ae}restad}, \citenamefont {Marston},\ and\ \citenamefont
  {Schollw{\"o}ck}}]{fjaerestadOrbital2006}%
  \BibitemOpen
  \bibfield  {author} {\bibinfo {author} {\bibfnamefont {J.~O.}\ \bibnamefont
  {Fj{\ae}restad}}, \bibinfo {author} {\bibfnamefont {J.~B.}\ \bibnamefont
  {Marston}},\ and\ \bibinfo {author} {\bibfnamefont {U.}~\bibnamefont
  {Schollw{\"o}ck}},\ }\href {https://doi.org/10.1016/j.aop.2005.08.005}
  {\bibfield  {journal} {\bibinfo  {journal} {Ann. Phys. (N. Y.)}\ }\textbf
  {\bibinfo {volume} {321}},\ \bibinfo {pages} {894} (\bibinfo {year}
  {2006})}\BibitemShut {NoStop}%
\bibitem [{\citenamefont {Yamanaka}\ \emph {et~al.}(1998)\citenamefont
  {Yamanaka}, \citenamefont {Koshibae},\ and\ \citenamefont
  {Maekawa}}]{yamanakaFlux1998}%
  \BibitemOpen
  \bibfield  {author} {\bibinfo {author} {\bibfnamefont {M.}~\bibnamefont
  {Yamanaka}}, \bibinfo {author} {\bibfnamefont {W.}~\bibnamefont {Koshibae}},\
  and\ \bibinfo {author} {\bibfnamefont {S.}~\bibnamefont {Maekawa}},\ }\href
  {https://doi.org/10.1103/PhysRevLett.81.5604} {\bibfield  {journal} {\bibinfo
   {journal} {Phys. Rev. Lett.}\ }\textbf {\bibinfo {volume} {81}},\ \bibinfo
  {pages} {5604} (\bibinfo {year} {1998})}\BibitemShut {NoStop}%
\bibitem [{\citenamefont {Agterberg}\ and\ \citenamefont
  {Yunoki}(2000)}]{agterbergSpinflux2000}%
  \BibitemOpen
  \bibfield  {author} {\bibinfo {author} {\bibfnamefont {D.~F.}\ \bibnamefont
  {Agterberg}}\ and\ \bibinfo {author} {\bibfnamefont {S.}~\bibnamefont
  {Yunoki}},\ }\href {https://doi.org/10.1103/PhysRevB.62.13816} {\bibfield
  {journal} {\bibinfo  {journal} {Phys. Rev. B}\ }\textbf {\bibinfo {volume}
  {62}},\ \bibinfo {pages} {13816} (\bibinfo {year} {2000})}\BibitemShut
  {NoStop}%
\bibitem [{\citenamefont {Aliaga}\ \emph {et~al.}(2001)\citenamefont {Aliaga},
  \citenamefont {Normand}, \citenamefont {Hallberg}, \citenamefont {Avignon},\
  and\ \citenamefont {Alascio}}]{aliagaIsland2001}%
  \BibitemOpen
  \bibfield  {author} {\bibinfo {author} {\bibfnamefont {H.}~\bibnamefont
  {Aliaga}}, \bibinfo {author} {\bibfnamefont {B.}~\bibnamefont {Normand}},
  \bibinfo {author} {\bibfnamefont {K.}~\bibnamefont {Hallberg}}, \bibinfo
  {author} {\bibfnamefont {M.}~\bibnamefont {Avignon}},\ and\ \bibinfo {author}
  {\bibfnamefont {B.}~\bibnamefont {Alascio}},\ }\href
  {https://doi.org/10.1103/PhysRevB.64.024422} {\bibfield  {journal} {\bibinfo
  {journal} {Phys. Rev. B}\ }\textbf {\bibinfo {volume} {64}},\ \bibinfo
  {pages} {024422} (\bibinfo {year} {2001})}\BibitemShut {NoStop}%
\bibitem [{\citenamefont {Riera}\ \emph {et~al.}(1997)\citenamefont {Riera},
  \citenamefont {Hallberg},\ and\ \citenamefont
  {Dagotto}}]{PhysRevLett.79.713}%
  \BibitemOpen
  \bibfield  {author} {\bibinfo {author} {\bibfnamefont {J.}~\bibnamefont
  {Riera}}, \bibinfo {author} {\bibfnamefont {K.}~\bibnamefont {Hallberg}},\
  and\ \bibinfo {author} {\bibfnamefont {E.}~\bibnamefont {Dagotto}},\ }\href
  {https://doi.org/10.1103/PhysRevLett.79.713} {\bibfield  {journal} {\bibinfo
  {journal} {Phys. Rev. Lett.}\ }\textbf {\bibinfo {volume} {79}},\ \bibinfo
  {pages} {713} (\bibinfo {year} {1997})}\BibitemShut {NoStop}%
\bibitem [{\citenamefont {Hotta}\ \emph {et~al.}(2003)\citenamefont {Hotta},
  \citenamefont {Moraghebi}, \citenamefont {Feiguin}, \citenamefont {Moreo},
  \citenamefont {Yunoki},\ and\ \citenamefont
  {Dagotto}}]{PhysRevLett.90.247203}%
  \BibitemOpen
  \bibfield  {author} {\bibinfo {author} {\bibfnamefont {T.}~\bibnamefont
  {Hotta}}, \bibinfo {author} {\bibfnamefont {M.}~\bibnamefont {Moraghebi}},
  \bibinfo {author} {\bibfnamefont {A.}~\bibnamefont {Feiguin}}, \bibinfo
  {author} {\bibfnamefont {A.}~\bibnamefont {Moreo}}, \bibinfo {author}
  {\bibfnamefont {S.}~\bibnamefont {Yunoki}},\ and\ \bibinfo {author}
  {\bibfnamefont {E.}~\bibnamefont {Dagotto}},\ }\href
  {https://doi.org/10.1103/PhysRevLett.90.247203} {\bibfield  {journal}
  {\bibinfo  {journal} {Phys. Rev. Lett.}\ }\textbf {\bibinfo {volume} {90}},\
  \bibinfo {pages} {247203} (\bibinfo {year} {2003})}\BibitemShut {NoStop}%
\bibitem [{\citenamefont {White}\ and\ \citenamefont
  {Scalapino}(1998)}]{whiteDensity1998}%
  \BibitemOpen
  \bibfield  {author} {\bibinfo {author} {\bibfnamefont {S.~R.}\ \bibnamefont
  {White}}\ and\ \bibinfo {author} {\bibfnamefont {D.~J.}\ \bibnamefont
  {Scalapino}},\ }\href {https://doi.org/10.1103/PhysRevLett.80.1272}
  {\bibfield  {journal} {\bibinfo  {journal} {Phys. Rev. Lett.}\ }\textbf
  {\bibinfo {volume} {80}},\ \bibinfo {pages} {1272} (\bibinfo {year}
  {1998})}\BibitemShut {NoStop}%
\bibitem [{\citenamefont {Jiang}\ \emph {et~al.}(2020)\citenamefont {Jiang},
  \citenamefont {Zaanen}, \citenamefont {Devereaux},\ and\ \citenamefont
  {Jiang}}]{jiangGround2020}%
  \BibitemOpen
  \bibfield  {author} {\bibinfo {author} {\bibfnamefont {Y.-F.}\ \bibnamefont
  {Jiang}}, \bibinfo {author} {\bibfnamefont {J.}~\bibnamefont {Zaanen}},
  \bibinfo {author} {\bibfnamefont {T.~P.}\ \bibnamefont {Devereaux}},\ and\
  \bibinfo {author} {\bibfnamefont {H.-C.}\ \bibnamefont {Jiang}},\ }\href
  {https://doi.org/10.1103/PhysRevResearch.2.033073} {\bibfield  {journal}
  {\bibinfo  {journal} {Phys. Rev. Research}\ }\textbf {\bibinfo {volume}
  {2}},\ \bibinfo {pages} {033073} (\bibinfo {year} {2020})}\BibitemShut
  {NoStop}%
\bibitem [{\citenamefont {Stoudenmire}\ and\ \citenamefont
  {White}(2012)}]{stoudenmireStudying2012}%
  \BibitemOpen
  \bibfield  {author} {\bibinfo {author} {\bibfnamefont {E.}~\bibnamefont
  {Stoudenmire}}\ and\ \bibinfo {author} {\bibfnamefont {S.~R.}\ \bibnamefont
  {White}},\ }\href {https://doi.org/10.1146/annurev-conmatphys-020911-125018}
  {\bibfield  {journal} {\bibinfo  {journal} {Annu. Rev. Condens. Matter
  Phys.}\ }\textbf {\bibinfo {volume} {3}},\ \bibinfo {pages} {111} (\bibinfo
  {year} {2012})}\BibitemShut {NoStop}%
\end{thebibliography}%
\end{document}